\documentclass[journal]{IEEEtran}

\ifCLASSOPTIONcompsoc
  \usepackage[nocompress]{cite}
\else
  \usepackage{cite}
\fi
\ifCLASSINFOpdf
\else
\fi
\DeclareMathAlphabet{\mathsfbr}{OT1}{cmss}{m}{n}
\SetMathAlphabet{\mathsfbr}{bold}{OT1}{cmss}{bx}{n}
\DeclareRobustCommand{\msf}[1]{%
  \ifcat\noexpand#1\relax\msfgreek{#1}\else\mathsfbr{#1}\fi
}

\makeatletter
\newcommand{\msfgreek}[1]{\csname s\expandafter\@gobble\string#1\endcsname}
\makeatother

\DeclareFontEncoding{LGR}{}{} 
\DeclareSymbolFont{sfgreek}{LGR}{cmss}{m}{n}
\SetSymbolFont{sfgreek}{bold}{LGR}{cmss}{bx}{n}
\DeclareMathSymbol{\salpha}{\mathord}{sfgreek}{`a}
\DeclareMathSymbol{\sbeta}{\mathord}{sfgreek}{`b}
\DeclareMathSymbol{\sgamma}{\mathord}{sfgreek}{`g}
\DeclareMathSymbol{\sdelta}{\mathord}{sfgreek}{`d}
\DeclareMathSymbol{\sepsilon}{\mathord}{sfgreek}{`e}
\DeclareMathSymbol{\szeta}{\mathord}{sfgreek}{`z}
\DeclareMathSymbol{\seta}{\mathord}{sfgreek}{`h}
\DeclareMathSymbol{\stheta}{\mathord}{sfgreek}{`j}
\DeclareMathSymbol{\siota}{\mathord}{sfgreek}{`i}
\DeclareMathSymbol{\skappa}{\mathord}{sfgreek}{`k}
\DeclareMathSymbol{\slambda}{\mathord}{sfgreek}{`l}
\DeclareMathSymbol{\smu}{\mathord}{sfgreek}{`m}
\DeclareMathSymbol{\snu}{\mathord}{sfgreek}{`n}
\DeclareMathSymbol{\sxi}{\mathord}{sfgreek}{`x}
\DeclareMathSymbol{\somicron}{\mathord}{sfgreek}{`o}
\DeclareMathSymbol{\spi}{\mathord}{sfgreek}{`p}
\DeclareMathSymbol{\srho}{\mathord}{sfgreek}{`r}
\DeclareMathSymbol{\ssigma}{\mathord}{sfgreek}{`s}
\DeclareMathSymbol{\stau}{\mathord}{sfgreek}{`t}
\DeclareMathSymbol{\supsilon}{\mathord}{sfgreek}{`u}
\DeclareMathSymbol{\sphi}{\mathord}{sfgreek}{`f}
\DeclareMathSymbol{\schi}{\mathord}{sfgreek}{`q}
\DeclareMathSymbol{\spsi}{\mathord}{sfgreek}{`y}
\DeclareMathSymbol{\somega}{\mathord}{sfgreek}{`w}

\DeclareMathSymbol{\svarsigma}{\mathord}{sfgreek}{`c}

\DeclareMathSymbol{\sGamma}{\mathalpha}{sfgreek}{`G}
\DeclareMathSymbol{\sDelta}{\mathalpha}{sfgreek}{`D}
\DeclareMathSymbol{\sTheta}{\mathalpha}{sfgreek}{`J}
\DeclareMathSymbol{\sLambda}{\mathalpha}{sfgreek}{`L}
\DeclareMathSymbol{\sXi}{\mathalpha}{sfgreek}{`X}
\DeclareMathSymbol{\sPi}{\mathalpha}{sfgreek}{`P}
\DeclareMathSymbol{\sSigma}{\mathalpha}{sfgreek}{`S}
\DeclareMathSymbol{\sUpsilon}{\mathalpha}{sfgreek}{`U}
\DeclareMathSymbol{\sPhi}{\mathalpha}{sfgreek}{`F}
\DeclareMathSymbol{\sPsi}{\mathalpha}{sfgreek}{`Y}
\DeclareMathSymbol{\sOmega}{\mathalpha}{sfgreek}{`W}

\DeclareRobustCommand{\mcal}[1]{%
  \ifcat\noexpand#1\relax\mathnormal{#1}\else\cal{#1}\fi
}
\DeclareRobustCommand{\BM}[1]{%
  \ifcat\noexpand#1\relax\bm{\boldUppercaseItalicGreek{#1}}\else\bm{#1}\fi
}
\makeatletter
\newcommand{\boldUppercaseItalicGreek}[1]{\csname var\expandafter\@gobble\string#1\endcsname}
\makeatother

\newcommand{\diag}[1]{\operatorname{diag}\left\{#1\right\}}

\newcommand{\T}{\mathrm{T}}

\newcommand{\V}[1]{\bm{#1}} 
\newcommand{\Set}[1]{{\mcal{#1}}} 

\newcommand{\E}[1]{\mathbb{E}\left\{#1\right\}}

\newcommand{\avg}[1]{\overline{\left\{#1\right\}}}

\usepackage{bm}
\usepackage{color, courier, comment}
\usepackage{psfrag}
\usepackage{acronym}
\usepackage{amsmath}
\usepackage{amssymb}
\usepackage{amsfonts}
\usepackage[colorlinks = true, allcolors = black]{hyperref}
\usepackage{latexsym}
\usepackage{mathrsfs}
\usepackage{epstopdf}
\usepackage{algorithm}
\usepackage{mathtools}
\usepackage[caption=false, font=scriptsize]{subfig}
\usepackage{algorithmic}
\usepackage{relsize}

\graphicspath{{figures/}}

\DeclareMathOperator*{\argmax}{arg\,max}

\newcommand{\st}{\operatorname{s.t.}\,}

\newtheorem{definition}{Definition}
\newtheorem{lemma}{Lemma}
\newtheorem{proposition}{Proposition}
\newtheorem{theorem}{Theorem}
\newtheorem{corollary}{Corollary}
\newtheorem{remark}{Remark}

\definecolor{green}{rgb}{0, 0.6, 0}
\definecolor{pink}{rgb}{1, 0, 1}

\acrodef{agi}[AgI]{augmented information}
\acrodef{mec}[MEC]{mobile edge computing}
\acrodef{ldp}[LDP]{Lyapunov drift-plus-penalty}
\acrodef{lp}[LP]{linear programming}
\acrodef{cmdp}[CMDP]{constrained MDP}
\acrodef{mdp}[MDP]{Markov decision process}
\acrodef{pdf}[pdf]{probability density function}
\acrodef{ue}[UE]{user equipment}
\acrodef{sfc}[SFC]{service function chain}
\acrodef{bp}[BP]{back-pressure}
\acrodef{rsp}[RSP]{restricted shortest path}
\acrodef{sdn}[SDN]{software defined networking}
\acrodef{nfv}[NFV]{network function virtualization}

\begin{document}

\title{
	Ultra-Reliable Distributed Cloud Network Control with End-to-End Latency Constraints
}

\author{
	\vspace{0.2cm}

	Yang~Cai,~\IEEEmembership{Student~Member,~IEEE},
	Jaime~Llorca,~\IEEEmembership{Member,~IEEE},
	Antonia~M.~Tulino,~\IEEEmembership{Fellow,~IEEE}, and
	Andreas~F.~Molisch,~\IEEEmembership{Fellow,~IEEE}

	\thanks{
	Part of this paper was presented at IEEE ICC 2021 \cite{cai2021delay}.
	
	Y. Cai and A. F. Molisch are with the Department of Electrical Engineering, University of Southern California, Los Angeles, CA 90089, USA (e-mail: yangcai@usc.edu; molisch@usc.edu).

	J. Llorca is with New York University, NY 10012, USA (e-mail: jllorca@nyu.edu).

	A. M. Tulino is with New York University, NY 10012, USA, and also with the University\`{a} degli Studi di Napoli Federico II, Naples 80138, Italy (e-mail: atulino@nyu.edu; antoniamaria.tulino@unina.it).
	}
}

\maketitle

\IEEEpubid{
\begin{minipage}{3\columnwidth}
	\centering
	{\footnotesize
	\vspace{50pt}
	This work has been submitted to the IEEE for possible publication. Copyright may be transferred without notice, after which this version may no longer be accessible.
	}
\end{minipage}
}

\begin{abstract}
We are entering a rapidly unfolding future driven by the delivery of real-time computation services, such as industrial automation and augmented reality, collectively referred to as \ac{agi} services, over highly distributed cloud/edge computing networks. The interaction intensive nature of \ac{agi} services is accelerating the need for networking solutions that provide strict latency guarantees. In contrast to most existing studies that can only characterize average delay performance, we focus on the critical goal of delivering \ac{agi} services ahead of corresponding deadlines on a per-packet basis, while minimizing overall cloud network operational cost. To this end, we design a novel queuing system able to track data packets' lifetime and formalize the {\em delay-constrained least-cost dynamic network control problem}. To address this challenging problem, we first study the setting with average capacity (or resource budget) constraints, for which we characterize the delay-constrained stability region and design a near-optimal control policy leveraging Lyapunov optimization theory on an equivalent virtual network. Guided by the same principle, we tackle the peak capacity constrained scenario by developing the {\em reliable cloud network control} (RCNC) algorithm, which employs a two-way optimization method to make actual and virtual network flow solutions converge in an iterative manner. Extensive numerical results show the superior performance of the proposed control policy compared with the state-of-the-art cloud network control algorithm, and the value of guaranteeing strict end-to-end deadlines for the delivery of next-generation \ac{agi} services.
\end{abstract}


\acresetall


\IEEEpeerreviewmaketitle


\section{Introduction}

\IEEEPARstart{T}{he} so-called automation era or fourth industrial revolution will be driven by the proliferation of compute- and interaction-intensive applications, such as real-time computer vision, autonomous transportation, machine control in Industry 4.0, telepresence, and augmented/virtual reality (AR/VR), which we collectively refer to as {\em \ac{agi} services} \cite{cai2022metaverse,FenLloTulMol:C17,Wel:B16}. In addition to the communication resources needed for the delivery of data streams to corresponding destinations, \ac{agi} services also require a significant amount of computation resources for the real-time processing, via possibly multiple functions, of source and intermediate data streams.

The evolution of \acp{ue} towards increasingly small, lightweight, seamless devices, and their associated limitations in power and computing capabilities, has been pushing the need to offload many computation-intensive tasks to the cloud. However, increased access delays associated with distant centralized cloud data centers are fueling advanced network architectures such as fog and \ac{mec} that push computation resources closer to the end users in order to strike a better balance between resource efficiency and end-to-end delay \cite{Wel:B16,mach17mecsurvey,cai2020mec}. In this work, we refer to the overall wide-area distributed computation network that results from the convergence of telco networks and cloud/edge/\ac{ue} resources as a {\em distributed cloud network}.

Delay and cost are thus two essential metrics when evaluating the performance of \ac{agi} service delivery over a distributed cloud network. From the consumers' perspective, excessive end-to-end delays can significantly impact quality of experience (QoE), especially for delay-sensitive AgI applications (e.g., industrial automation, augmented reality) where packets must be delivered by a strict deadline in order to be effective (i.e., packets delivered after their deadline become irrelevant and/or break application interactivity). In this context, {\em timely throughput}, which measures the rate of effective packet delivery (i.e., within-deadline packet delivery rate), becomes the appropriate performance metric \cite{LasAve:J13,CheHua:J18,sun2021AoI}. On the other hand, network operators care about the overall resource (e.g., computation, communication) consumption needed to support the  dynamic service requests raised by end users.

Both delay and cost will ultimately be dictated by the choice of cloud/edge locations where to execute the various \ac{agi} service functions, the network paths over which to route the service data streams, and the corresponding allocation of computation and communication resources. Therefore, to maximize the benefit of distributed cloud networks for the delivery of \ac{agi} services, two fundamental problems need to be jointly addressed:
{\em \begin{itemize}
	\item where to execute the requested \ac{agi} service functions, and how much computation resource to allocate
	\item how to route and schedule data streams through the appropriate sequence of service functions, and how much communication resource to allocate
\end{itemize}}

In addition, due to the dynamic and unpredictable nature of \ac{agi} service requests, the above placement, processing, routing, and resource allocation problems must be addressed in an online manner, in response to stochastic network conditions and service demands.

\subsection{Related Work}

With the advent of \ac{sdn} and \ac{nfv}, network (and, by extension, \ac{agi}) services can be deployed as a sequence of software functions or \acp{sfc} instantiated over distributed cloud locations. A number of studies have investigated the problem of joint \ac{sfc} placement and routing over multi-hop networks with the objective of either minimizing overall operational cost \cite{BarLloTulRam:C15,BarChoAhmBou:C15,AddBelBouSec:C15,BarCorLloTulVicMor:J16}, or maximizing accepted service requests \cite{huang2021throughput,xu2020placement,yue2021throughput}. Nonetheless, these solutions exhibit two main drawbacks. First, the problem is formulated as a {\em static} optimization problem without considering the dynamic nature of service requests. In addition, when it comes to delay performance, these studies mainly focus on propagation delay \cite{AddBelBouSec:C15,BarCorLloTulVicMor:J16,huang2021throughput,xu2020placement}, while neglecting queuing delay, or using simplified models (e.g., M/M/1) to approximate it \cite{yue2021throughput}. Second, due to the combinatorial nature of the problem, the corresponding formulations typically take the form of (NP-hard) mixed integer linear programs and either heuristic or loose approximation algorithms are developed, compromising the quality of the resulting solution.

More recently, a number of studies have addressed the \ac{sfc} optimization problem in {\em dynamic} scenarios, where one needs to make joint packet processing and routing decisions in an online manner \cite{FenLloTulMol:J18a,cai2021multicast,zhang2021multicast,cai2022multicast_arxiv,cai2022CCC_arXiv}. The works in \cite{FenLloTulMol:J18a,cai2021multicast} employ a generalized cloud network flow model that allows joint control of processing and transmission flows. The works in \cite{zhang2021multicast,cai2022multicast_arxiv,cai2022CCC_arXiv} show that the traffic control problem in distributed cloud networks (involving joint packet processing and routing decisions) can be reduced to a packet routing problem on a properly constructed {\em layered graph} that includes extra edges to characterize the processing operations (i.e., packets pushed through these edges are interpreted as being processed by a service function). By this transformation, many control policies designed for packet routing can be extended to address cloud network control problems (i.e., packet processing and routing), especially those aiming at maximizing network throughput with bounded {\em average} delay performance.

In particular, \ac{bp} \cite{TasEph:J92} is a well-known algorithm for throughput-optimal routing that leverages Lyapunov drift control theory to steer data packets based on the {\em pressure} difference (differential backlog) between neighbor nodes. In addition, the \ac{ldp} control approach \cite{Nee:B10} extends the \ac{bp} algorithm to also minimize network operational cost (e.g., energy expenditure), while preserving throughput optimality. Despite the remarkable advantages of achieving optimal throughput performance via simple local policies without requiring any knowledge of network topology and traffic demands, both \ac{bp} and \ac{ldp} approaches can suffer from poor average delay performance, especially in low congestion scenarios, where packets can take unnecessarily long, and sometimes even cyclic, paths \cite{BuiSriSto:C09}. Average delay reductions were then shown to be obtained in \cite{YinShaRedLiu:J11} by combining \ac{bp} and hop-distance based shortest-path routing, using a more complex \ac{mdp} formulation in \cite{neely2013MDPdelay}, or via the use of source routing to dynamically select acyclic routes for incoming packets, albeit requiring global network information, in \cite{SinMod:J18}.

Going beyond average delay and analyzing per-packet delay performance is a more challenging problem with much fewer known results, even in the context of packet routing and under {\em static arrivals}. In particular, the \ac{rsp} problem, which aims to find the min-cost path for a given source-destination pair subject to an end-to-end delay (or path length) constraint, is known to be NP-hard \cite{GarJoh:B79}. Considering {\em dynamic arrivals} becomes a further obstacle that requires additional attention.
An opportunistic scheduling policy is proposed in \cite{Nee:C11} that trades off worst-case delay and timely throughput, which preserves the delay guarantee when applied to hop-count-limited transmissions. However, it requires a link selection procedure (to meet the hop-count requirement) that weakens its performance in general networks (e.g., mesh topologies); besides, the timely throughput is with respect to the worst-case delay, rather than the deadline imposed by the application, leading to either sub-optimal throughput under stringent deadline constraints, or looser guarantees on the worst-case delay; finally, it treats packet scheduling on different links separately, lacking an end-to-end optimization of the overall delay.
In \cite{singh2018delay}, the authors formulate the problem of timely throughput maximization as an exponential-size \ac{cmdp}, and derive an approximate solution based on solving the {\em optimal single-packet transportation problem} for each packet; in addition, \cite{singh2021delay} addresses the more complex set-up of wireless networks with link interference. While this approach reduces the complexity from exponential (of a general solution that makes joint packet decisions) to polynomial, it requires solving a dynamic programming problem for each packet at every time slot, which can still become computationally expensive.
Furthermore, none of these works takes operational cost minimization into account, an important aspect in modern elastic cloud environments.

\subsection{Contributions}

In this paper, we investigate the problem of multi-hop cloud network control with the goal of delivering \ac{agi} services with strict per-packet deadline constraints, while minimizing overall operational cost. More concretely, we focus on {\bf reliable service delivery}, which requires the timely throughput of each service, i.e., the rate of packets delivered by their {\em deadlines}, to surpass a given level in order to meet a desired QoE. We study the problem in dynamic scenarios, i.e., assuming the service requests are unknown and time-varying.

There are two main challenges that prohibit the use of existing cloud network control methods (e.g., \cite{FenLloTulMol:J18a}) and associated queuing systems for reliable service delivery. In particular, existing queuing systems:
(i) do not take packet deadlines into account and cannot track associated {\em packet lifetimes};
(ii) do not allow {\em packet drops}, which becomes critical in delay-constrained routing, since dropping {\em outdated} packets can benefit cost performance without impacting {\em timely} throughput.

To overcome these drawbacks, we construct a novel queuing system with separate queues for different deadline-driven packet lifetimes, and allow packet drops upon lifetime expiry. In contrast to standard queuing systems, where packets are transmitted to reduce network congestion and keep physical queues stable \cite{Nee:B10}, the new queuing model is fundamentally different: stability of physical queues becomes irrelevant (due to packet drops), and packet transmission is driven by the requirement to deliver packets on time ({\bf reliable service delivery}). The proposed solution is presented in two stages. First, we study a relaxed {\em average-constrained network control} problem and derive an exact solution via a flow matching technique where flow scheduling decisions are driven by an \ac{ldp} solution to an equivalent virtual network control problem. Then, we address the original {\em peak-constrained} problem, with the additional challenge of non-equivalent actual and virtual network formulations, and devise an algorithm that adapts the \ac{ldp} plus flow matching technique via an iterative procedure.

Our contributions can be summarized as follows:

\begin{enumerate}
	\item We develop a novel queuing model that allows tracking data packet lifetimes and dropping outdated packets, and formalize the delay-constrained least-cost dynamic network control problem $\mathscr{P}_0$.
	\item We derive a relaxed problem $\mathscr{P}_1$ targeting the same objective in an average-capacity-constrained network, and characterize its delay-constrained stability region based on a lifetime-driven flow conservation law.
	\item We design a fully distributed near-optimal control policy for $\mathscr{P}_1$ by (i) deriving an equivalent virtual network control problem $\mathscr{P}_2$ that admits an efficient \ac{ldp}-based solution, (ii) proving that $\mathscr{P}_1$ and  $\mathscr{P}_2$ have identical stability region, flow space, and optimal objective value, and (iii) designing a randomized policy for $\mathscr{P}_1$ guided by matching the virtual flow solution to $\mathscr{P}_2$.
	\item We leverage the flow matching technique to develop an algorithm for $\mathscr{P}_0$, referred to as {\em reliable cloud network control} (RCNC), whose solution results from the convergence of actual (for $\mathscr{P}_0$) and virtual (for $\mathscr{P}_2$) flows via an iterative optimization procedure.
\end{enumerate}

The rest of the paper is organized as follows.
In Section \ref{sec:system_model}, we introduce network model and associated queuing system.
In Section \ref{sec:feasible_optimization}, we define the policy space  and formulate the original problem $\mathscr{P}_0$.
In Section \ref{sec:approximate}, we study the relaxed problem $\mathscr{P}_1$ and derive 
an equivalent \ac{ldp}-amenable formulation $\mathscr{P}_2$.
Section \ref{sec:average_solution} presents the algorithm for solving $\mathscr{P}_1$ as well as its performance analysis, which is extended to develop an iterative algorithm for $\mathscr{P}_0$ in Section \ref{sec:peak_solution}.
Numerical results are shown in Section \ref{sec:experiments},
and possible extensions are discussed in Section \ref{sec:extensions}. Finally, we summarize the main conclusions in Section \ref{sec:conclusion}.

\begin{table}[!t]\label{table1}
\footnotesize
\newcommand{\tabincell}[2]{\begin{tabular}{@{}#1@{}}#2\end{tabular}}
\caption{Table of Notations}
\centering
\vspace{-6 pt}
\renewcommand{\arraystretch}{1.2}
\begin{tabular}{l p{5.8 cm}}
\hline
Symbol         & Description \\
\hline
$\Set{G}$; $\Set{V}$, $\Set{E}; d$ & Network graph model; node, edge sets; destination. \\
$\delta_i^-$, $\delta_i^+$ & Sets of incoming/outgoing neighbors of $i$. \\
$C_{ij}$, $e_{ij}$ & Transmission capacity and cost of link $(i, j)$. \\
$l$, $L$, $\Set{L}$ & Lifetime, maximum lifetime, set of lifetimes. \\
$a(t)$, $\lambda$   & Number of arrival packets, arrival rate. \\
$x(t)$; $\nu(t)$, $\mu(t)$ & Flow variable; virtual, actual flows. \\
$\gamma$, $\Lambda$   & Reliability level, network stability region. \\
$\Set{F}$, $\Gamma$ & Feasible policy space, flow space. \\
$Q(t)$, $U(t)$, $R(t)$ & Actual queue, virtual queue, request queue. \\
\hline
\end{tabular}
\label{usedsymbols}
\end{table}

\section{System Model}\label{sec:system_model}

\subsection{Cloud Layered Graph}

The ultimate goal of this work is to design control policies for distributed cloud networks to reliably support multiple delay-sensitive \ac{agi} services, where the network is equipped with computation resources (cloud servers, edge/fog computing nodes, etc.) able to host service functions and execute corresponding computation tasks.

While in traditional packet routing problems, each node treats its {\em neighbor nodes} as outgoing interfaces over which packets can be scheduled for transmission, a key step to address the \ac{agi} service control problem is to treat the co-located {\em computing resources} as an additional outgoing interface over which packets can be scheduled for processing \cite{FenLloTulMol:J18a}. Indeed, as illustrated in \cite{zhang2021multicast}, the \ac{agi} service control problem, involving both packet routing and processing, can be reduced to a packet routing problem on a {\em layered graph} where cross-layer edges represent computation resources.

Motivated by such a connection and for ease of exposition, in this paper, w.l.o.g., we illustrate the developed approach focusing on the {\em single-commodity delay-constrained min-cost packet routing} problem. We remark that (i) it is still an open problem even in traditional communication networks, and (ii) the extension to distributed cloud networks hosting \ac{agi} services is presented in Appendix \ref{apdx:agi}.

\subsection{Network Model}

The considered packet routing network is modeled via a directed graph $\Set{G} = (\Set{V}, \Set{E})$, where edge $(i, j) \in \Set{E}$ represents a network link supporting data transmission from node $i\in\Set{V}$ to $j\in\Set{V}$, and where $\delta_i^-$ and $\delta_i^+$ denote the incoming and outgoing neighbor sets of node $i$, respectively.

Time is divided into equal-sized slots, and the available transmission resources and associated costs at each network link are quantified as:
\begin{itemize}
	\item $C_{ij}$: the transmission capacity, i.e., the maximum number of data units (e.g., packets) that can be transmitted in one time slot, on link $(i, j)$;
	\item $e_{ij}$: the unit transmission cost, i.e., the cost of transmitting one unit of data in one time slot, on link $(i, j)$.
\end{itemize}

We emphasize that in the layered graph, cross-layer edges represent data processing, i.e., data streams pushed through these edges are interpreted as being processed by corresponding service functions, and the capacity and cost of these edges represent the {\em processing capacity} and {\em processing cost} of the associated computation resources (e.g., cloud/edge servers).

\subsection{Arrival Model}

In this work, we focus on a delay-sensitive application, assuming that each packet has a strict deadline by which it must be delivered to the destination $d\in \Set{V}$. In other words, each packet must be delivered within its {\em lifetime}, defined as the number of time slots between the current time and its deadline. A packet is called {\em effective} if its remaining lifetime $l$ is positive, and {\em outdated} otherwise. In addition, we define {\em timely throughput} as the rate of effective packet delivery.

We assume that input packets can originate at any source node of the application, and in general, we assume that the set of source nodes can be any network node except the destination, $\Set{V}\setminus \{d\}$. The packet's initial lifetime $l\in \Set{L} \triangleq \{1, \cdots, L\}$ is determined by the application (based on the sensitivity of the contained information to delay), which can vary from packet to packet, with $L$ denoting the maximum possible lifetime. Denote by $a_i^{(l)}(t)$ the number of exogenous packets (i.e., packets generated externally) of lifetime $l$ arriving at node $i$. We assume that the arrival process is i.i.d. over time, with mean arrival rate $\lambda_i^{(l)} \triangleq \mathbb{E}\big\{a_i^{(l)}(t)\big\}$ and an upper bound of $A_{\max}$; besides, we define the corresponding vectors $\V{a}(t) = \big\{ a_i^{(l)}(t): \forall \, i\in \Set{V},\, l \in \Set{L} \big\}$ and $\V{\lambda} = \E{\V{a}(t)}$.

\subsection{Queuing System}

Since each packet has its own delivery deadline, keeping track of data packets' lifetimes is essential. A key step is to construct a queuing system with distinct queues for packets of different {\em current lifetimes} $l \in \Set{L}$. In particular, we denote by $Q_i^{(l)}(t)$ the queue backlog of lifetime $l$ packets at node $i$ at time slot $t$, and define $\V{Q}(t) = \big\{ Q_i^{(l)}(t): \forall \, i\in \Set{V},\, l \in \Set{L} \big\}$. Let $x_{ij}^{(l)}(t)$ be the {\em actual} number of lifetime $l$ packets transmitted from node $i$ to $j$ at time $t$ (which is different from a widely used assigned flow model, as explained in Remark \ref{remark:q}).

\begin{figure}[t]
	\centering
	\includegraphics[width = .98 \columnwidth]{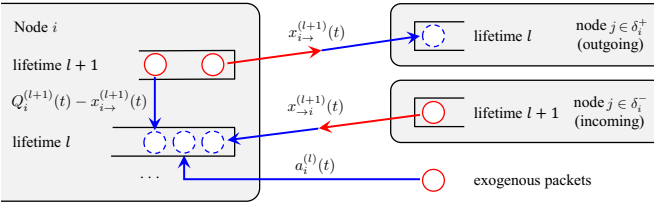}
	\caption{
		Interaction between lifetime queues. Red and blue colors denote packet states and actions during {\em transmitting} and {\em receiving} phases, respectively.
	}
	\label{fig:queuing_dynamics}
\end{figure}

Each time slot is divided into two phases, as illustrated in Fig. \ref{fig:queuing_dynamics}. In the {\em transmitting} phase, each node makes and executes transmission decisions based on observed queuing states. The number of lifetime $l+1$ packets at the end of this phase is given by
\begin{align}
	\breve{Q}_i^{(l+1)}(t) = Q_i^{(l+1)}(t) - x_{i\to}^{(l+1)}(t)
\end{align}
where $x_{i \to}^{(l+1)}(t) \triangleq \sum_{j\in \delta_i^+} x_{ij}^{(l+1)}(t)$ denotes the number of outgoing packets. In the {\em receiving} phase, the incoming packets, including those from neighbor nodes $x_{\to i}^{(l+1)}(t) \triangleq  \sum_{j\in \delta_i^-} x_{ji}^{(l+1)}(t)$ as well as exogenously arriving packets $a_i^{(l)}(t)$, are loaded into the queuing system, and the queuing states are updated as:
\begin{align}
	Q_i^{(l)}(t+1) = \big[ \breve{Q}_i^{(l+1)}(t) + x_{\to i}^{(l+1)}(t) \big] + a_i^{(l)}(t)
\end{align}
where lifetime $l+1$ packets, including those still in the queue as well as those arriving from incoming neighbors during the {\em transmitting} phase of time slot $t$ (i.e., terms in the square bracket) turn into lifetime $l$ packets during the {\em receiving} phase of time slot $t$. In addition, lifetime $l$ exogenous packets, $a_i^{(l)}(t)$, also enter the lifetime $l$ queue during the {\em receiving} phase of slot $t$. All such arriving packets become ready for transmission at the {\em transmitting} phase of slot $t+1$.

To sum up, the queuing dynamics are given by
\begin{align}\label{eq:queue_dynamics_1}
\hspace{-5pt} Q_i^{(l)}(t+1) = Q_i^{(l+1)}(t) - x_{i\to}^{(l+1)}(t) + x_{\to i}^{(l+1)}(t) + a_i^{(l)}(t)
\end{align}
for $\forall\, i \in \Set{V}, l \in \Set{L}$.

In addition, we assume: 1) as the information contained in outdated packets is useless, i.e., outdated packets do not contribute to timely throughput, they are immediately dropped to avoid inefficient use of network resources:
\begin{align}\label{eq:queue_dynamics_2}
Q_i^{(0)}(t) = 0, \quad \forall\, i \in \Set{V},
\end{align}
and 2) for the destination node $d$, every effective packet is consumed as soon as it arrives, and therefore
\begin{align}\label{eq:queue_dynamics_3}
Q_d^{(l)}(t) = 0, \quad \forall\, l \in \Set{L}.
\end{align}

Considering the lifetime reduction over time slots, in general, we do not send packets of lifetime $1$, i.e., $x_{ij}^{(1)}(t) = 0$, since the packets turn outdated at node $j$ at the next time slot. The only exception occurs when $j = d$: we assume that the packets of lifetime $l = 1$ are consumed as soon as the destination node receives them, while they are still effective.

\section{Problem Formulation} \label{sec:feasible_optimization}

In this section, we introduce the admissible policy space, the reliability constraint, and the formalized {\em delay-constrained least-cost dynamic network control} problem.

\subsection{Admissible Policy Space}\label{sec:admissible}

The control policies of interest make packet routing and scheduling decisions at each time slot, which are dictated by the flow variables $\V{x}(t) = \big\{ x_{ij}^{(l)}(t): \forall\, (i, j)\in \Set{E}, l \in \Set{L} \big\}$. In particular, we focus on the space of {\em admissible} control policies with decision flow variables satisfying:
\begin{enumerate}
	\item non-negativity constraint, i.e.,
	\begin{align}
	x_{ij}^{(l)}(t) \geq 0\text{ for }\forall\,(i, j)\in \Set{E},\text{ or }\V{x}(t) \succeq 0;
	\end{align}
	\item peak link capacity constraint, i.e.,
	\begin{align}\label{eq:capacity}
	x_{ij}(t) \triangleq \sum_{l\in \Set{L}} x_{ij}^{(l)}(t) \leq C_{ij},\ \forall\, (i, j)\in \Set{E};
	\end{align}
	\item availability constraint, i.e.,
	\begin{align}\label{eq:availability}
	x_{i \to}^{(l)}(t) \leq Q_i^{(l)}(t),\ \forall\,i\in \Set{V},\, l\in \Set{L}.
	\end{align}
\end{enumerate}

The availability constraint \eqref{eq:availability} requires the total number of (scheduled) outgoing packets to not exceed those in the current queuing system, since we define $\V{x}(t)$ as the actual flow (see Remark \ref{remark:q} for a detailed explanation). As will be shown throughout the paper, it plays an equivalent role to {\em flow conservation} in traditional packet routing formulations.

\subsection{General Network Stability Region} \label{sec:capacity_region}

In addition to the above admissibility constraints, we require the timely throughput achieved by the designed control policy to surpass a given level specified by the application, i.e.,
\begin{align}\label{eq:reliability}
\avg{ \E{ x_{\to d}(t) } } \geq \gamma \| \V{\lambda} \|_1
\end{align}
where $\gamma$ denotes the {\em reliability level}, $\| \V{\lambda} \|_1$ is the total arrival rate, and $\avg{ z(t) } \triangleq \lim_{T\to\infty} \frac{1}{T} \sum_{t=0}^{T-1}{ z(t) }$ denotes the long-term average of random process $\{ z(t): t\geq 0\}$.

The reliability constraint \eqref{eq:reliability} imposes the requirement on the routing policy to provide reliable (delay-constrained) packet delivery. It forces packets to be routed {\em efficiently} and avoid excessive in-network packet drops due to lifetime expiry. The reliability level $\gamma$ characterizes the robustness of the considered service to missing information, i.e., a percentage of up to $(1-\gamma)$ of the packets can be {\em dropped} without causing a significant performance loss. The reliability constraint plays an equivalent role to network stability in traditional packet routing formulations.

\begin{definition} \label{def:general_capacity_region}
For a given capacitated network $\Set{G}$, we define the {\em delay-constrained stability region} as the set of $(f_{\V{a}}, \gamma)$ pairs that can be supported by an admissible policy, i.e., the pairs $(f_{\V{a}}, \gamma)$ such that there exists an admissible policy that satisfies \eqref{eq:reliability} under an arrival process with \ac{pdf} $f_{\V{a}}$.
\end{definition}

Note that via the complete information of the \ac{pdf} $f_{\V{a}}$, the mean arrival vector $\V{\lambda}$ in \eqref{eq:reliability} can be derived, which is employed to characterize the stability region in many existing works (e.g., \cite{Nee:B10, FenLloTulMol:J18a}). However, such first order characterization is not sufficient for the studied problem, as illustrated in Remark \ref{remark:capacity_region}, showing the necessity to include the entire \ac{pdf} information.

\begin{figure}[t]
	\centering
	\subfloat[Single-hop network.]{
	\includegraphics[width = 0.45\linewidth]{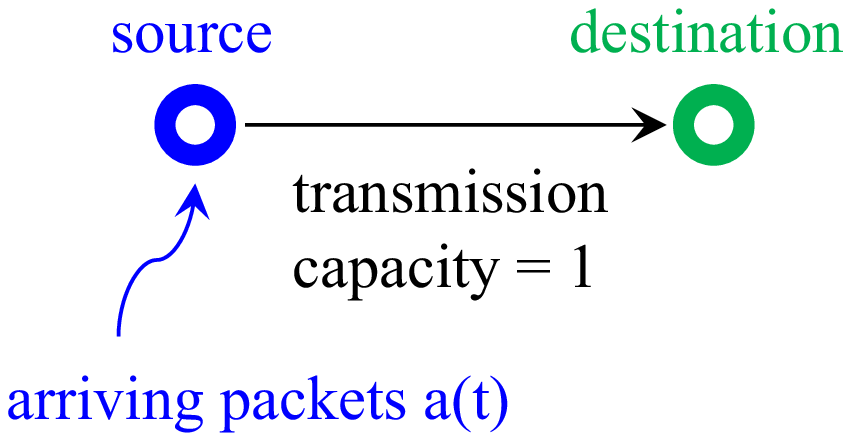}
	\label{fig:example1}
	}
	\subfloat[Arrival processes.]{
	\includegraphics[width = 0.45\linewidth]{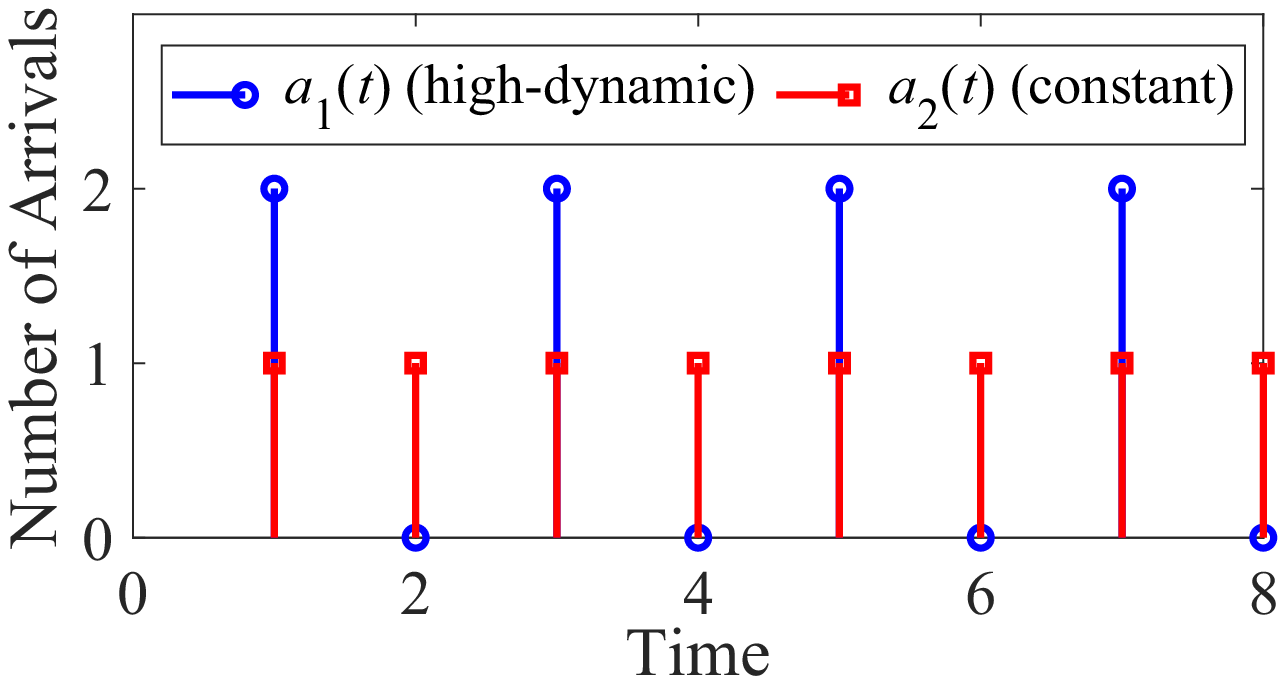}
	\label{fig:example2}
	}
	\caption{
	A one-hop example network. Packets of lifetime $L=1$ arrive at the source according to two arrival processes of equal mean arrival rate $\lambda = 1$.
	}
	\label{fig:example}
\end{figure}

\begin{remark} \label{remark:capacity_region}
Consider the {\bf Example} shown in Fig. \ref{fig:example}, where the initial lifetime of every packet is equal to 1. The achievable reliability level is $\gamma_1 = 50\%$ under a high-dynamic arrival $a_1(t)$, and $\gamma_2 = 100\%$ under the constant arrival $a_2(t)$; while the two arrival processes have the same rate of $1$. This example shows that: in addition to arrival rate, arrival dynamics can also impact the performance in the studied problem.
\end{remark}

\begin{remark}
\label{remark:q}
In the existing literature of stochastic network optimization (e.g., \cite{FenLloTulMol:J18a,Nee:B10,Nee:C11,TasEph:J92}), a key element that has gained widespread adoption to improve tractability is the use of the {\em assigned} flow, which is different from the {\em actual} flow in that it does not need to satisfy the availability constraint \eqref{eq:availability}. Dummy packets are created when there are not sufficient packets in the queue to support the scheduling decision, making the decision variables not constrained by the queuing process. Such formulation, however, is not suitable for delay-constrained routing, where reliable packet delivery is imposed on the {\em actual} packets received by the destination (via constraint \eqref{eq:reliability}).
\end{remark}

\subsection{Problem Formulation}

The goal is to develop an admissible control policy that guarantees reliable packet delivery, while minimizing overall network operational cost. Formally, we aim to find the policy with decisions $\{ \V{x}(t): t\geq 0 \}$ satisfying
\begin{subequations}\label{eq:p0}
\begin{align}
\mathscr{P}_0: \ & \min_{\V{x}(t) \succeq 0}\ \avg{ \E{ h(\V{x}(t)) } } \\
& \hspace{5pt} \st \hspace{6pt} \avg{ \E{ x_{\to d}(t) } } \geq \gamma \| \V{\lambda} \|_1 \label{eq:p0_reliability} \\
& \hspace{32pt} x_{ij}(t) \leq C_{ij},\ \forall \, (i,j) \in \Set{E} \label{eq:p0_capacity} \\
& \hspace{32pt} x_{i\to}^{(l)}(t) \leq Q_i^{(l)}(t),\ \forall \, i\in \Set{V}, l \in \Set{L} \label{eq:p0_availability} \\
& \hspace{32pt} \V{Q}(t) \text{ evolves by \eqref{eq:queue_dynamics_1} -- \eqref{eq:queue_dynamics_3}} \label{eq:p0_queue}
\end{align}
\end{subequations}
where the instantaneous cost of the decision $\V{x}(t)$ is given by
\begin{align}\label{eq:cost}
h(t) = h(\V{x}(t)) = \sum\nolimits_{(i, j)\in \Set{E}} e_{ij} x_{ij}(t) = \langle \V{e}, \V{x}(t) \rangle
\end{align}
with $\langle \cdot, \cdot \rangle$ denoting the inner product of the two vectors.

The above problem belongs to the category of \ac{cmdp}, by defining the queuing vector $\V{Q}(t)$ as the {\em state} and the flow variable $\V{x}(t)$ as the {\em action}. However, note that the dimension of state-action space grows {\em exponentially} with the network dimension, which prohibits the application of the standard solution to this problem \cite{Alt:B99}. Even if we leave out the operational cost minimization aspect, it is still challenging to find an exact efficient solution to the remaining problem of timely throughput maximization, as studied in \cite{singh2018delay}.

On the other hand, note that \eqref{eq:p0} deals with a queuing process, together with long-term average objective and constraints, which is within the scope of Lyapunov drift control \cite{Nee:B10}. However, it cannot be directly applied to solve \eqref{eq:p0} because: 
(i) the related queuing process \eqref{eq:p0_queue} is not of standard form;%
\footnote{
	In the designed lifetime-based queuing system, a packet can traverse queues of reducing lifetimes and eventually get dropped when entering the lifetime $0$ queue. On the other hand, in traditional queuing systems \cite{Nee:B10,FenLloTulMol:J18a}, a packet stays in the same queue until selected for operation; in addition, since there are no packet drops, queue build up contributes to network congestion and creates {\em pressure} driving packet transmission \cite{Nee:B10}.
}
(ii) the decision variables are actual flows and depend on the queuing states \eqref{eq:p0_availability}, which is different from a widely used assigned flow model (see Remark \ref{remark:q}).

\section{The Average Capacity Constrained Problem} \label{sec:approximate}

The goal of this work is to derive an efficient approximate solution to $\mathscr{P}_0$. To this end, we start out with a less restricted setup in which only the average flow is constrained to be below capacity, leading  to the following {\em relaxed} control problem:
\begin{subequations}\label{eq:problem_1}
\begin{align}
\mathscr{P}_1: \ & \min_{\V{x}(t) \succeq 0} \ \avg{ \E{ h(\V{x}(t)) } } \\
& \hspace{5pt} \st \hspace{6pt} \avg{ \E{ x_{\to d}(t) } } \geq \gamma \| \V{\lambda} \|_1 \label{eq:p1_reliability} \\
& \hspace{32pt} \avg{ \E{ x_{ij}(t) } } \leq C_{ij} \label{eq:p1_capacity} \\
& \hspace{32pt} x_{i\to}^{(l)}(t) \leq Q_i^{(l)}(t) \label{eq:p1_availability} \\
& \hspace{32pt} \V{Q}(t) \text{ evolves by \eqref{eq:queue_dynamics_1} -- \eqref{eq:queue_dynamics_3}} \label{eq:p1_queue}
\end{align}
\end{subequations}
which relaxes the peak capacity constraint \eqref{eq:p0_capacity} by the corresponding average capacity constraint \eqref{eq:p1_capacity}.\footnote{We note that such an average-constrained setting may find interesting applications of its own in next-generation virtual networks that allow elastic scaling of network resources \cite{Wel:B16}.}

Mathematically, $\mathscr{P}_1$ is still a \ac{cmdp} problem, making it challenging to solve. Instead of tackling it directly, in the following, we derive a tractable problem $\mathscr{P}_2$ corresponding to a {\em virtual network}, and establish the connection between them by showing that they have identical flow spaces, which allows to address $\mathscr{P}_1$ using the solution to $\mathscr{P}_2$ as a stepping-stone.

\subsection{The Virtual Network} \label{sec:virtual_network}

The virtual network control problem is cast as
\begin{subequations}\label{eq:problem_2}
\begin{align}
\mathscr{P}_2: \ & \min_{\V{x}(t) \succeq 0} \ \avg{ \E{ h(\V{x}(t)) } } \\
& \hspace{5pt} \st \hspace{6pt}  \avg{ \E{ x_{\to d}(t) } } \geq \gamma \| \V{\lambda} \|_1 \label{eq:p2_reliability} \\
& \hspace{32pt} x_{ij}(t) \leq C_{ij} \label{eq:p2_capacity} \\
& \hspace{32pt} \bar{x}_{i\to}^{(\geq l)} \leq \bar{x}_{\to i}^{(\geq l+1)} + \lambda_i^{(\geq l)} \label{eq:p2_causality}
\end{align}
\end{subequations}
where $\bar{x}_{i\to}^{(\geq l)} = \lim_{T\to\infty}\frac{1}{T} \sum_{t=0}^{T-1} \mathbb{E}\big\{ x_{i\to}^{(\geq l)}(t) \big\}$ denotes the average transmission rate of packets with lifetime $\geq l$, with $x_{i\to}^{(\geq l)}(t) = \sum_{\ell = l}^{L}{ x_{i\to}^{(\ell)}(t) }$ (similarly for $\bar{x}_{\to i}^{(\geq l+1)}$).

A crucial difference in the derivation of $\mathscr{P}_2$ is to replace the availability constraint \eqref{eq:p1_availability} by \eqref{eq:p2_causality}, which states the fact that the lifetime of the packets must decrease (by at least $1$) as they traverse any node $i$, and thus is called the {\em causality} constraint. As a consequence, we eliminate the unconventional queuing process \eqref{eq:p1_queue} and the dependency of $\V{x}(t)$ on $\V{Q}(t)$, i.e., the two factors resulting in the failure of employing the \ac{ldp} approach to address $\mathscr{P}_1$. Especially, we will use $\V{\nu}(t)$ (instead of $\V{x}(t)$) to represent the decisions determined in $\mathscr{P}_2$, referred to as the virtual flow.

\subsubsection{Virtual Queue}

Although there is no explicit queuing system  in $\mathscr{P}_2$, it consists of long-term average objective and constraints, which can be addressed via the \ac{ldp} control of a virtual queuing system \cite{Nee:B10}. More concretely, the {\em virtual queuing system} $\V{U}(t) = \{ U_d(t) \} \cup \{ U_i^{(l)}(t): i\in \Set{V}\setminus \{d\},\, l\in \Set{L} \}$ must be stabilized to ensure \eqref{eq:p2_reliability} and \eqref{eq:p2_causality}, defined as
\begin{subequations} \label{eq:virtual_q}
\begin{align}
U_d(t+1) & = \max\big\{ U_d(t) + \gamma A(t) - \nu_{\to d}(t), \, 0 \big\} \label{eq:virtual_sink}, \\
U_i^{(l)}(t+1) & = \max \big\{U_i^{(l)}(t) + \nu_{i\to}^{(\geq l)}(t) - \nu_{\to i}^{(\geq l+1)}(t) \nonumber \\
& \hspace{.5in} - a_i^{(\geq l)}(t), \, 0 \big\}. \label{eq:virtual_intermediate}
\end{align}
\end{subequations}
where $A(t) = \sum_{i\in \Set{V},\, l\in \Set{L}} a_i^{(l)}(t)$ is the total amount of packets arriving in the network at time slot $t$.%
\footnote{
	Here we use $A(t)$ instead of $\| \V{\lambda} \|_1$ as the latter information is usually not available in practice; furthermore, if the arrival information cannot be obtained immediately, delayed information, i.e., $A(t-\tau)$ with $\tau > 0$, can be used as an alternative, which does not impact the result of time average.
}
We refer to \eqref{eq:virtual_sink} and \eqref{eq:virtual_intermediate} as the virtual queues associated with node $d$ and $i$.

To sum up, it is equivalent to cast $\mathscr{P}_2$ as
\begin{subequations}\label{eq:problem_2_eq}
\begin{align}
\hspace{-.1in}\mathscr{P}_2^{\,\text{e}}: \ & \min_{\V{\nu}(t)} \ \avg{ \E{ h(\V{\nu}(t)) } } \\
& \st\ \text{stabilize }\V{U}(t) \text{ evolving by }\eqref{eq:virtual_q} \\
& \hspace{.33in} 0\leq \nu_{ij}(t) \leq C_{ij}\quad \forall\, (i,j)\in \Set{E}.
\end{align}
\end{subequations}

\subsubsection{Physical Interpretation} \label{sec:interpretation}

\begin{figure}[t]
	\centering
	\includegraphics[width = .8 \columnwidth]{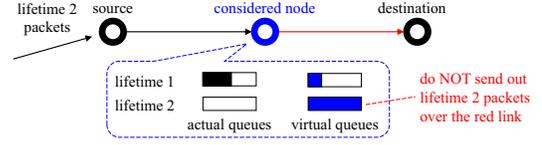}
	\caption{Illustration of the devised virtual system. The source node supplies packets of lifetime $2$, which arrive as lifetime $1$ packets to the actual queue of the considered node. In the virtual system, the considered node is allowed to supply packets of any lifetime to the destination by {\em borrowing} them from the reservoir and building up in the corresponding virtual queue. The virtual queue of lifetime $1$ is stable, since the received lifetime $1$ packets from the source node can compensate the borrowed packets; while the virtual queue of lifetime $2$ builds up, pushing the node to stop sending out more lifetime $2$ packets. The decision derived from the stability of the virtual system is aligned with the desired operation of the actual network, as only lifetime $1$ packets are available for transmission at the considered node.}
	\label{fig:virtual_network}
\end{figure}

When deriving the virtual network control problem $\mathscr{P}_2$, we relax the precedence constraint that imposes that a packet cannot be transmitted from a node before it arrives at the given node. Instead, we assume that each node in the virtual network is a {\em data-reservoir} and has access to abundant (virtual) packets of any lifetime. At every time slot, each node checks packet requests from its outgoing neighbors and supplies such needs using the virtual packets borrowed from the reservoir, which are compensated when it receives packets of the same lifetime (either from incoming neighbors or exogenous arrivals). The virtual queues can be interpreted as the {\em data deficits} (difference between outgoing and incoming packets) of the corresponding data-reservoirs. Specially, in \eqref{eq:virtual_sink}, the destination reservoir {\em sends out} $\gamma A(t)$ packets to the end user (to meet the reliability requirement), while {\em receiving} $\nu_{\to d}(t)$ in return. When \eqref{eq:p2_reliability} and \eqref{eq:p2_causality} are satisfied, or the virtual queues are stabilized, the network nodes do not need to embezzle virtual packets from the reservoirs; since the achieved network flow can be attained by the actual packets, it can serve as guidance for packet steering in the actual network (see Fig. \ref{fig:virtual_network} for illustration).

\subsection{Connections Between $\mathscr{P}_1$ and $\mathscr{P}_2$}

We now describe key connections between the actual and virtual network control problems.

\begin{definition}[Feasible Policy]
For problem $\mathscr{P}_\iota\ (\iota = 1, 2)$, a policy $p$ is called {\em feasible} if it makes decisions satisfying \eqref{eq:p1_reliability} -- \eqref{eq:p1_queue} ($\iota = 1$) or \eqref{eq:p2_reliability} -- \eqref{eq:p2_causality} ($\iota = 2$). The set of feasible policies is called {\em feasible policy space} $\Set{F}_\iota$.
\end{definition}

\begin{definition}[Flow Assignment]
Given a feasible policy $p$ for problem $\mathscr{P}_\iota\ (\iota = 1, 2)$ (with decisions $\{ \V{x}_p(t): t\geq 0 \}$), the achieved {\em flow assignment} is defined as $\V{x}_p = \avg{ \E{ \V{x}_p(t) } }$, i.e., the vector of transmission rates for packets with different lifetimes on all network links. Furthermore, the {\em flow space} is defined as the set of all achievable flow assignments, i.e., $\Gamma_\iota = \big\{ \V{x}_p: p\in \Set{F}_\iota \big\}$.
\end{definition}

\begin{definition}[Stability Region]
For problem $\mathscr{P}_\iota\ (\iota = 1, 2)$, the {\em stability region} $\Lambda_\iota$ is defined as the set of $(\V{\lambda}, \gamma)$ pairs, under which the feasible policy space $\Set{F}_\iota$ is non-empty.
\end{definition}

We make the following clarifications about the above definitions:
(i) we will prove (in Theorem \ref{thm:cap_region}) that the stability region of $\mathscr{P}_1$ only depends on the mean arrival rate $\V{\lambda}$, in contrast to the general {\em Definition \ref{def:general_capacity_region}} which involves the arrival \ac{pdf} $f_{\V{a}}$; and so is that of $\mathscr{P}_2$, which is clear from its definition \eqref{eq:problem_2};
(ii) the feasible policy space and the flow space are associated with a certain point $(\V{\lambda}, \gamma)$ in the stability region;
(iii) since the networks considered in $\mathscr{P}_1$ and $\mathscr{P}_2$ are of the same topology, the flow assignment vectors are of the same dimension.

We then reveal the intimate relationship between the two problems by the following three results.

\begin{proposition}\label{prop:causality_constraint}
The availability constraint \eqref{eq:p1_availability} implies the causality constraint \eqref{eq:p2_causality}.
\end{proposition}

\begin{IEEEproof}
See Appendix \ref{apdx:proof_causality} in supplementary material.
\end{IEEEproof}

\begin{theorem}\label{thm:cap_region}
For a given network, the stability regions of $\mathscr{P}_1$ and $\mathscr{P}_2$ are identical, i.e., $\Lambda_1 = \Lambda_2$. 
In addition, a pair $(\V{\lambda},\gamma)$ is within the stability region $\Lambda_\iota$ ($\iota = 1, 2$) if and only if there exist flow variables $\V{x} = \{ x^{(l)}_{ij} \geq 0: \forall\, (i,j)\in \Set{E}, l \in \Set{L} \}$, such that for $\forall\, i\in \Set{V}$, $(i,j)\in \Set{E}$,\, $l \in \Set{L}$, 
\begin{subequations}\label{eq:capacity_region} \begin{align}
x_{\to d} & \geq \gamma \| \V{\lambda} \|_1 \label{eq:cr_reliability} \\
x_{ij} & \leq C_{ij},\ \forall\,(i,j)\in \Set{E} \label{eq:cr_capacity} \\
x_{\to i}^{(\geq l+1)} + \lambda_i^{(\geq l)} & \geq x_{i\to}^{(\geq l)},\ \forall\,i\in \Set{V}, l\in \Set{L} \label{eq:flow_conserve} \\
x_{ij}^{(0)} = x_{dk}^{(l)} & = 0,\ \forall\,k\in \delta_d^{+},(i, j)\in \Set{E}, l\in \Set{L}.
\end{align} \end{subequations} 
Furthermore, 
$\forall\, (\V{\lambda}, \gamma) \in \Lambda_\iota$, there exists a feasible randomized policy 
that achieves the optimal cost.
\end{theorem}

\begin{IEEEproof}
See Appendix \ref{apdx:cap_p1}, \ref{apdx:cap_p2} in supplementary material.
\end{IEEEproof}

\begin{proposition}\label{prop:flow_space}
For $\forall \,(\V{\lambda}, \gamma) \in \Lambda_1 = \Lambda_2$, the two problems have identical flow spaces, i.e., $\Gamma_1 = \Gamma_2$.
\end{proposition}

\begin{IEEEproof}
By Theorem \ref{thm:cap_region}, $\mathscr{P}_1$ and $\mathscr{P}_2$ have the same stability region, i.e, $\Lambda_1 = \Lambda_2$. Consider a point in the stability region $\Lambda_1=\Lambda_2$. For any flow assignment $\V{x} \in \Gamma_1$, there exists a feasible policy $p_1\in \Set{F}_1$, with decision variables $\{\V{x}_1(t): t\geq 0\}$, that attains flow assignment $\V{x}$, i.e., $\avg{\E{\V{x}_1(t)}} = \V{x}$. Therefore, $\{\V{x}_1(t): t\geq 0\}$ satisfies \eqref{eq:p1_reliability} -- \eqref{eq:p1_queue}, which implies that $\V{x}$ satisfies all the conditions in \eqref{eq:capacity_region} (by Proposition \ref{prop:causality_constraint}). Using the method provided in Appendix \ref{sec:random_2}, we can construct a feasible randomized policy $p_2 \in \Set{F}_2$, with decision variables $\{\V{x}_2(t): t\geq 0\}$, that achieves the same flow assignment, i.e., $\avg{\E{\V{x}_2(t)}} = \V{x}$, for $\mathscr{P}_2$. Therefore, $\V{x} \in \Gamma_2$, and thus $\Gamma_1\subset \Gamma_2$.
The reverse direction $\Gamma_2\subset \Gamma_1$ can be shown via the same argument.
Hence, $\Gamma_1 = \Gamma_2$.
\end{IEEEproof}

The above propositions are explained in the following:
by Proposition \ref{prop:causality_constraint}, the feasible policy spaces satisfy $\Set{F}_1 \nsubseteq \Set{F}_2$ and $\Set{F}_2 \nsubseteq \Set{F}_1$;
while Theorem \ref{thm:cap_region} suggests that they lead to the same stability region, by presenting an explicit, identical characterization \eqref{eq:capacity_region} (where \eqref{eq:flow_conserve} is the generalized lifetime-driven {\em flow conservation} law), which is in the form of a \ac{lp} problem with $L|E|$ variables (and thus of pseudo polynomial complexity); Proposition \ref{prop:flow_space} further shows that $\mathscr{P}_1$ and $\mathscr{P}_2$ share the same flow space (for any point in the stability region), 
which is a crucial property 
since the two metrics of interest, i.e., timely throughput \eqref{eq:reliability} and operational cost \eqref{eq:cost}, are both {\em linear} functions of the flow assignment.

\begin{corollary}\label{coro:opt_value}
$\mathscr{P}_1$ and $\mathscr{P}_2$ have the same optimal value.
\end{corollary}

\begin{IEEEproof}
Consider a feasible policy $p_1\in \Set{F}_1$, whose decisions $\V{x}_1(t)$ attain flow assignment $\V{x}$. According to the Proposition \ref{prop:flow_space}, there exists a feasible policy $p_2 \in \Set{F}_2$ attaining the same flow assignment $\V{x}$ by making decisions $\V{x}_2(t)$. The operational cost satisfies
\begin{align}\begin{split}
\hspace{-.1cm}\avg{ \E{ h( \V{x}_1(t) ) } } & = \avg{ \E{ \langle \V{e}, \V{x}_1(t) \rangle } }
= \langle \V{e}, \avg{ \E{ \V{x}_1(t) } } \rangle \\
& = \langle \V{e}, \V{x} \rangle 
= \langle \V{e}, \avg{ \E{ \V{x}_2(t) } } \rangle \\
& = \avg{ \E{ \langle \V{e}, \V{x}_2(t) \rangle } }
= \avg{ \E{ h( \V{x}_2(t) ) } }.
\end{split}\end{align}
The reverse direction can be shown by the same argument. As a result, they have the same range (when treating the cost as a function of the policy), and thus optimal value.
\end{IEEEproof}

\begin{corollary}\label{coro:opt_solution}
Given a feasible policy to $\mathscr{P}_2$, we can construct a feasible randomized policy for $\mathscr{P}_1$ to achieve the same flow assignment.
\end{corollary}

\begin{IEEEproof}
Suppose $\{ \V{\nu}(t): t\geq 0 \}\in \Set{F}_2$. The associated flow assignment $\V{\nu} = \avg{ \E{ \V{\nu}(t) } }$ satisfies \eqref{eq:capacity_region}, and we can construct a feasible randomized policy for $\mathscr{P}_1$ as follows (see Appendix \ref{sec:random_1} for details): at each time slot, for any packet of lifetime $l\in \Set{L}$ in the queuing system, node $i\in \Set{V}$ selects the outgoing neighbor $j\in \delta_i^+$ for it according to the \ac{pdf}
\begin{align} \label{eq:p1_randomize}
\alpha_i^{(l)}(j) = \nu_{ij}^{(l)} \big/ \big( \nu_{\to i}^{(\geq l+1)} + \lambda_i^{(\geq l)} - \nu_{i\to}^{(\geq l+1)} \big)
\end{align}
otherwise the packet stays in node $i$. It is shown in Appendix \ref{apdx:cap_p1} that this policy achieves flow assignment $\V{\nu}$.
\end{IEEEproof}

\section{Solution to Average-Constrained Network} \label{sec:average_solution}

In this section, we take advantage of the \ac{ldp} approach to address $\mathscr{P}_2^{\text{e}}$ and guide the design of a fully distributed, near-optimal randomized algorithm for $\mathscr{P}_1$ (by Corollary \ref{coro:opt_solution}).

\subsection{Optimal Virtual Flow} \label{sec:ldp_alg}

We first present the \ac{ldp}-based algorithm to solve $\mathscr{P}_2^{\,\text{e}}$ given by \eqref{eq:problem_2_eq}. Define the Lyapunov function as $L(t) = \|\V{U}(t)\|_2^2 \big/ 2$, and the Lyapunov drift $\Delta(\V{U}(t)) = L(t+1) - L(t)$. The \ac{ldp} approach aims to minimize a linear combination of an upper bound of the Lyapunov drift (which can be derived by some standard manipulation \cite{Nee:B10}) and the objective function weighted by a tunable parameter $V$, or
\begin{align}\label{eq:virtual_ub}
\Delta( \V{U}(t) ) + V h( \V{\nu}(t) )
\leq B - \langle \tilde{\V{a}}, \V{U}(t) \rangle - \langle \V{w}(t), \V{\nu}(t) \rangle
\end{align}
where $B$ is a constant, $\tilde{\V{a}} = \{ - \gamma A(t) \} \cup \big\{ a_i^{(\geq l)} : \forall\,i\in \Set{V}\setminus\{d\},\, l\in \Set{L} \big\}$, and the weights $\V{w}(t)$ are given by
\begin{align}\label{eq:weight}
w_{ij}^{(l)}(t) = - V e_{ij} - U_i^{(\leq l)}(t) +
\begin{cases}
U_d(t) & j = d \\
U_j^{(\leq l-1)}(t) & j\ne d
\end{cases}
\end{align}
where the superscript $^{(\leq l)}$ refers to the operation of $\sum_{\ell = 1}^{l}$.

To sum up, at every time slot, the algorithm decides the virtual flow $\V{\nu}(t)$ by addressing the following problem
\begin{align} \label{eq:virtual_opt}
\max_{\V{\nu}(t)} \ \langle \V{w}(t), \V{\nu}(t) \rangle,\ 
\st\ 0\leq \nu_{ij}(t) \leq C_{ij},\ \forall\, (i,j) \in \Set{E}
\end{align}
and the solution to it is in the {\em max-weight} fashion. More concretely, for each link $(i, j)$, we first find the {\em best} lifetime $l^\star$ with the {\em largest weight}, and devote all the transmission resource to serve packets of this lifetime if the weight is positive. Therefore, the optimal virtual flow assignment is
\begin{align} \label{eq:opt_virtual_flow}
\nu_{ij}^{(l)}(t) = C_{ij}\,\mathbb{I}\big\{ l = l^\star, w_{ij}^{(l^\star)}(t) > 0 \big\}
\end{align}
where $l^\star = \argmax_{l\in \Set{L}}\ w_{ij}^{(l)}(t)$, $\mathbb{I}\{ \cdot \}$ is the indicator function.

To implement the above algorithm, at each time slot, a considered node exchanges the virtual queue information with its neighbor nodes (to calculate the weight of each lifetime by \eqref{eq:weight}), and decides the virtual flow according to \eqref{eq:opt_virtual_flow}, which can be completed in a fully distributed manner; the computational complexity at node $i$ is given by $\mathcal{O}(L |\delta_i^+|)$.

\subsection{Flow Matching}

The algorithm developed above can provide a near-optimal (will be proved in next subsection) solution $\{\V{\nu}(t): t\geq 0\}$ to $\mathscr{P}_2$, from which  we will design a feasible, near-optimal policy for $\mathscr{P}_1$ in this section. The decided (actual) flow is denoted by $\V{\mu}(t)$, to distinguish it from the virtual flow $\V{\nu}(t)$.

We will design an {\em admissible} policy for $\mathscr{P}_1$ (i.e., satisfying \eqref{eq:p1_capacity} -- \eqref{eq:p1_availability}) to pursue the goal of {\em flow matching}, i.e.,
\begin{align} \label{eq:flow_matching}
\avg{ \V{\mu}(t) } = \avg{ \V{\nu}(t) }.
\end{align}
The reason to set the above goal is two-fold:
(i) it ensures that the designed policy can attain the same throughput and cost performance (recall that both metrics are linear functions of the flow assignment) as the virtual flow, which is feasible (satisfying the reliability constraint) and achieves near-optimal cost performance,
(ii) the existence of the policy is guaranteed (as a result of identical flow spaces); actually, given the feasible solution $\{\V{\nu}(t)\}$, Corollary \ref{coro:opt_solution} presents a construction procedure of a feasible policy for $\mathscr{P}_1$ to realize the goal.

Corollary \ref{coro:opt_solution} requires the exact values of $\avg{ \V{\nu}(t) }$ and $\V{\lambda}$ as input, which are not available in practice. As an alternative, we employ the corresponding empirical values, i.e., the finite-horizon average of the virtual flow and the arrival rate
\begin{align} \label{eq:virtual_avg}
\bar{\V{\nu}}(t) = \frac{1}{t} \sum_{\tau = 0}^{t-1}{ \V{\nu}(\tau) },\ 
\hat{\V{\lambda}}(t) = \frac{1}{t} \sum_{\tau = 0}^{t-1}{ \V{a}(\tau) }
\end{align}
to calculate the probability values in \eqref{eq:p1_randomize}, by which we decide the outgoing flow at time slot $t$. Since the above empirical values are updated at every time slot, it leads to a time-varying randomized policy; as $\bar{\V{\nu}}(t) \to \avg{ \V{\nu}(t) }$ and $\hat{\V{\lambda}}(t) \to \V{\lambda}$ asymptotically, the policy gradually converges.%
\footnote{
	It is possible that $\bar{\V{\nu}}(t)$ can violate \eqref{eq:flow_conserve} at some time slot, which is not qualified to construct a valid randomized policy. However, as $t\to\infty$, $\bar{\V{\nu}}(t)$ converges to $\avg{\V{\nu}(t)}$, which satisfies the constraints. With this asymptotic guarantee, when such violation occurs, we can choose not to update the control policy at that time slot.
}

\begin{algorithm}[t]
\caption{Randomized Flow-Matching Algorithm}\label{alg:p1}
\begin{algorithmic}[1]
\FOR{ $t \geq 0$ and $i\in \Set{V}$ }
\STATE{ Solve the virtual flow $\V{\nu}(t)$ from \eqref{eq:virtual_opt}; }
\STATE{ Update the empirical averages $\bar{\V{\nu}}(t)$ and $\hat{\V{\lambda}}(t)$ by \eqref{eq:virtual_avg}; }
\STATE{ Update probability values $\big\{ \hat{\alpha}_{i}^{(l)}(j): j\in \delta_i^+ \big\}_{l\in \Set{L}}$ by \eqref{eq:p1_randomize} (using the above empirical averages); }
\FOR{ $l \in L$ }
\STATE{
For each packet in $Q_{i}^{(l)}(t)$, decide its outgoing link according to pdf $\big\{ \hat{\alpha}_{i}^{(l)}(j): j\in \delta_i^+ \big\}$;
}
\ENDFOR
\ENDFOR
\end{algorithmic}
\end{algorithm}

The proposed control policy is summarized in Algorithm \ref{alg:p1}, and we emphasize that
(i) at a given time slot, the policy in Corollary \ref{coro:opt_solution} makes i.i.d. decisions for packets with the same lifetime (i.e., fix the lifetime $l$, the \ac{pdf} $\big\{ \hat{\alpha}_{i}^{(l)}(j): j\in \delta_i^+ \big\}$ to determine the routing decision for each packet is the same). It is equivalent to make flow-level decisions based on packets' lifetime, by generating multinomial random variables with parameter $Q_i^{(l)}(t)$ and the common \ac{pdf};
(ii) in addition to deciding the virtual flow, the developed randomized policy requires each node to update the empirical averages \eqref{eq:virtual_avg}, calculate the \ac{pdf} $\hat{\V{\alpha}}$, and make the decisions at a complexity of $\mathcal{O}(L|\delta_i^+|)$.

\begin{remark}
For the studied packet routing problem (where flow scaling is not relevant), under the widely used assumption of Poisson arrivals, we can show (see Appendix \ref{apdx:poisson}) that the instantaneous flow size $x_{ij}(t), \forall\, (i, j),\,t,$ follows a Poisson distribution, which enjoys good concentration bounds.
\end{remark}

\begin{remark}
In \cite{singh2018delay}, a subproblem of $\mathscr{P}_1$ is studied, which involves constraints on average capacity and timely throughput, while leaving out the aspect of operational cost. The formulated \ac{cmdp} problem is solved by a dynamic programming algorithm, which can also be addressed following the same procedure presented in this section, at a lower complexity.
\end{remark}

\subsection{Performance Analysis}

In this section, we first prove that the \ac{ldp}-based algorithm (for $\mathscr{P}_2$) stabilizes the virtual queues (and consequently, the timely throughput satisfies the reliability constraint \eqref{eq:p2_reliability}) and attains near-optimal cost performance; then we show that the flow matching-based randomized policy (for $\mathscr{P}_1$) achieves the same throughput and cost performance as the previous algorithm. The effects of parameter $V$ are also analyzed.

\subsubsection{Virtual Network}

In addition to proving that the algorithm stabilizes the virtual queues, we analyze the effect of $V$ on the $\varepsilon$-convergence time defined as follows.

\begin{definition}[$\varepsilon$-Convergence Time]
The $\varepsilon$-convergence time $t_\varepsilon$ is the running time for the average solution to achieve a reliability within a margin of $\varepsilon$ from the desired value, i.e.,
\begin{align}
\hspace{-.1in}t_\varepsilon \triangleq \min_\tau \Big\{ \sup_{s\geq \tau} \Big[\gamma \| \V{\lambda} \|_1 -  \sum_{t=0}^{s-1} \frac{\E{ \nu_{\to d}(t) }}{s} \Big] \leq \varepsilon \Big\}.
\end{align}
\end{definition}

The existence of $t_\varepsilon$ (under the proposed algorithm) is shown in Appendix \ref{apdx:converge_cost_2} for any $\varepsilon > 0$.

\begin{proposition}\label{prop:cost_V}
For any point in the interior of the stability region, the virtual queues are mean rate stable under the proposed algorithm with a convergence time $t_\varepsilon \sim \mathcal{O}(V)$ for any $\varepsilon > 0$, and the achieved cost performance satisfies
\begin{align} \label{eq:achieved_cost}
\avg{ \E{ h(\V{\nu}(t)) } } \leq h_2^\star( \V{\lambda}, \gamma ) + \frac{B}{V}
\end{align}
where $h_2^\star( \V{\lambda}, \gamma )$ denotes the optimal cost performance that can be achieved under $( \V{\lambda}, \gamma )$ in $\mathscr{P}_2$.
\end{proposition}

\begin{IEEEproof}
See Appendix \ref{apdx:converge_cost} in supplementary material.
\end{IEEEproof}

We make the following clarifications about the above proposition,
(i) for a finite horizon, the reliability \eqref{eq:p2_reliability} and causality \eqref{eq:p2_causality} constraints might not be satisfied;
(ii) the virtual queues are stabilized, implying that the two constraints hold asymptotically;
(iii) by pushing the parameter $V\to \infty$, the achieved cost performance approaches the optimal cost (since the gap $B/V$ vanishes), by compromising the convergence time.

\subsubsection{Performance of Algorithm \ref{alg:p1}}

\begin{proposition} \label{prop:flow_matching}
For any point in the interior of the stability region, Algorithm \ref{alg:p1} is feasible for $\mathscr{P}_1$, while achieving the near-optimal cost performance of $h( \avg{\V{\nu}(t) } )$.
\end{proposition}

\begin{IEEEproof}
Algorithm \ref{alg:p1} makes decisions for the packets in the queuing system, and thus satisfying the constraints \eqref{eq:p1_availability}. Besides, as $\lim_{t\to\infty} \bar{\V{\nu}}(t) = \avg{ \V{\nu}(t) }$ and $\lim_{t\to\infty} \hat{\V{\lambda}}(t) = \V{\lambda}$, the instantaneous policy converges to a fixed policy constructed from $\avg{ \V{\nu}(t) }$ and $\V{\lambda}$, which achieves the same flow assignment $\avg{ \V{\mu}(t) } = \avg{ \V{\nu}(t) }$ as is proved in Corollary \ref{coro:opt_solution}, leading to identical throughput and cost performance.
\end{IEEEproof}

\begin{remark}
We note that Algorithm \ref{alg:p1} relies on the knowledge of the arrival rate (via \eqref{eq:p1_randomize} in step 4), and the empirical estimate \eqref{eq:virtual_avg} we use for implementation may be subject to estimation errors that can impact the attained cost performance. As shown in Appendix \ref{apdx:estimation_error}, in some extreme cases, the estimation error can lead to a considerable performance loss, driven by the Lagrangian multiplier (or shadow price) associated with the constraints involving $\lambda$. However, under i.i.d. arrivals, the estimated rate converges to the true value, and Algorithm \ref{alg:p1} is guaranteed to achieve near-optimal {\em asymptotic} performance.
\end{remark}

\section{Solution to Peak-Constrained Network} \label{sec:peak_solution}

In this section, we aim to address the original problem $\mathscr{P}_0$ (with peak-capacity constraint), leveraging the flow matching technique we develop in the previous section.

There are two problems we need to address:
\begin{itemize}
	\item[(i)] the actual flow (decided by the randomized policy) can violate the peak capacity constraint \eqref{eq:p0_capacity};
	\item[(ii)] the actual and virtual flow spaces are not identical, i.e., $\Gamma_0 \subset \Gamma_2$ (while in the average-constrained case, $\Gamma_1 = \Gamma_2$).
\end{itemize}

To address problem (i), we propose a request queue stability approach in order to constrain instantaneous transmission rates. For problem (ii), we introduce an auxiliary variable $\epsilon_{ij}, \forall (i, j)$ to represent the gap in flow spaces, leading to the following optimization problem over $\{ \V{\nu}(t), \V{\mu}(t), \V{\epsilon} \}$:
\begin{subequations} \begin{align}
\mathscr{P}_3: \ & \min\ \avg{\E{h(\V{\nu}(t))}} \\
& \st\ \nu_{ij}(t) \leq C_{ij} - \epsilon_{ij},
\text{(11b), (11d), (11e),} \\
& \hspace{24 pt} \avg{\E{\mu_{ij}(t)}} = \avg{\E{\nu_{ij}(t)}},
\text{(8c) -- (8f),} \\
& \hspace{24 pt} 0 \preceq \V{\epsilon} \triangleq \{ \epsilon_{ij} \} \preceq \{ C_{ij} \}.
\end{align} \end{subequations}

While solving $\mathscr{P}_{3}$ in a joint manner is difficult,
we propose an iterative optimization approach:
\begin{itemize}
	\item[i)] fix $\V{\epsilon}$ and $\V{\mu}(t)$:
	find $\V{\nu}(t)$ by LDP control (19), and derive the virtual flow assignment with optimal operational cost;
	\item[ii)] fix $\V{\epsilon}$ and $\V{\nu}(t)$:
	find $\V{\mu}(t)$ with the goal of flow matching (i.e., by stabilizing the request queues);
	\item[iii)] fix $\V{\nu}(t)$ and $\V{\mu}(t)$: update $\V{\epsilon}$ based on the gap between optimal and achievable rates, i.e., $ \avg{\E{\V{\nu}(t)}} - \avg{\E{\V{\mu}(t)}}$, which is non-zero if (25c) is violated.
\end{itemize}
Due to the randomness of network states, we introduce a {\em time frame} structure: step i) and ii) are executed on a per-slot basis, while step iii) on a per-frame basis (to obtain better rate estimates). The developed algorithm, referred to as {\em reliable cloud network control} (RCNC), is described in Algorithm \ref{alg:p0}.

\subsection{Request Queue} \label{sec:requst_queue}

We propose to achieve \eqref{eq:flow_matching} by making admissible flow decisions (i.e., satisfying \eqref{eq:p0_capacity} -- \eqref{eq:p0_queue}) to stabilize the {\em request queues} $\V{R}(t) = \{R_{ij}^{(l)}(t): \forall (i,j)\in \Set{E}, l\in \Set{L}\}$, defined as
\begin{align}\label{eq:request_q}
R_{ij}^{(l)}(t+1) = R_{ij}^{(l)}(t) + \bar{\nu}_{ij}^{(l)}(t) - \mu_{ij}^{(l)}(t)
\end{align}
where $\bar{\V{\nu}}(t)$ is given by \eqref{eq:virtual_avg}, and we still adopt the notation $\V{\mu}(t)$ to refer to the actual flow decided in $\mathscr{P}_0$, without causing ambiguity ($\mathscr{P}_1$ is not relevant in this section).

We consider the {\em $n$-slot look-ahead} scheme, under which the current decision is made together with $n-1$ (anticipated) future decisions. Such a scheme is employed since it creates flexibility for a packet to change its lifetime by delaying transmission, in favor of relieving the burden of the request queue with the heaviest backlog, as well as balancing the transmission load. From a formal point of view, we make decisions for $n$ time slots (starting from the current time slot) to optimize the {\em multi-slot Lyapunov drift}, defined as
\begin{align}
\Delta_n ( \V{R}(t) ) \triangleq \frac{ \| \V{R}(t+n-1) \|_2^2 - \| \V{R}(t) \|_2^2 }{2}.
\end{align}
An upper bound for the multi-slot drift is derived in the following. We apply telescope sum on the queuing dynamics \eqref{eq:request_q} for the period $t, \cdots, t+(n-1)$, which leads to
\begin{align}\label{eq:request_q_multi}
R_{ij}^{(l)}(t+n-1) = R_{ij}^{(l)}(t) + \sum_{\tau = t}^{t + n-1} \big[ \bar{\nu}_{ij}^{(l)}(\tau) - \mu_{ij}^{(l)}(\tau) \big].
\end{align}
Following the same procedure as in Section \ref{sec:virtual_network}, we obtain
\begin{align} \label{eq:multi_slot_drift}
\Delta_n(\V{R}(t)) \leq B'(t) - \sum\nolimits_{(i,j)\in \Set{E}} \langle \V{R}_{ij}(t), \V{M}_{ij} \V{1} \rangle
\end{align}
where $\V{M}_{ij}$ is a $L\times n$ matrix associated with link $(i, j)$, with column $\tau\ (0 \leq \tau < n)$ representing the vector $\V{\mu}_{ij}(t+\tau)$, and $B'(t)$ gathers all the uncontrollable terms.

The goal is to minimize the bound for multi-slot drift, by making admissible flow decisions for the $n$ time slots. By applying the queuing dynamics \eqref{eq:queue_dynamics_1} recursively, the availability constraint \eqref{eq:p0_availability} at the $(t+\tau)$-th time slot can be cast as
\begin{align}\label{eq:delay_compact}
\sum_{j\in \delta_i^+} g_{\tau+1}( \V{M}_{ij} ) - \V{D} \sum_{j\in \delta_i^-} g_\tau( \V{M}_{ji} ) \leq g_{\tau+1}( \V{A}_i ),\ \forall i
\end{align}
where $\V{A}_i = [ \V{Q}_i(t), \V{a}_i(t), \cdots, \V{a}_i(t+n-1) ]$ is the arrival matrix, with the columns $\V{a}_i(t + \tau) = \{ a^{(l)}_i(t + \tau): l\in \Set{L} \}$ representing the exogenous arrivals in the future; $g$ denotes the delay function, given by
\begin{align}
g_{\tau+1}(\V{X}) = \sum\nolimits_{s=0}^{\tau}{ \V{D}^{\tau-s} \V{X}[\,:\,, s] }
\end{align}
in which $\V{D}$ is the delay matrix of order $n$, and $\V{X}[\,:\,, \tau]$ is the $\tau$-th column of matrix $\V{X}$. We clarify that $\V{a}_i(t + \tau)$ are {\em random} vectors, leading to a complex stochastic optimization problem. We simplify the problem by replacing the random vectors with their estimated averages, i.e., empirical arrival rates $\hat{\V{\lambda}}_i$. This reduces $\V{A}_i$ to a deterministic matrix
\begin{align}\label{eq:emp_arrival}
\V{A}_i^{\text{emp}} = [ \V{Q}_i(t), \hat{\V{\lambda}}_i, \cdots, \hat{\V{\lambda}}_i ],
\end{align}
and the problem to a common \ac{lp} that can be addressed by standard solvers.

To sum up, at each slot, the proposed RCNC algorithm solves the following \ac{lp} to determine the transmission flow
\begin{subequations}
\label{eq:flow_matching_peak}
\begin{align} 
\mathscr{H}: \quad & \max_{ \Set{M} }\ \sum_{(i,j)\in \Set{E}} \langle \V{R}_{ij}(t), \V{M}_{ij} \V{1} \rangle \\
& \st\ \V{1}^\T \V{M}_{ij} \preceq C_{ij},\ \forall\,(i, j)\in \Set{E} \label{eq:req_capacity} \\
& \hspace{.33in} \eqref{eq:delay_compact} \text{ with } \V{A}_i = \V{A}_i^{\text{emp}},\ \forall\, i,\, 0\leq \tau < n \label{eq:req_delay} \\
& \hspace{.33in} \V{M}_{ij} \succeq 0,\ \forall\, (i,j)\in \Set{E}
\end{align}\end{subequations}
where $\Set{M} = \cup_{i\in \Set{V}} \, \Set{M}_i \triangleq \{ \V{M}_{ij}: j\in \delta_i^+ \}$, and \eqref{eq:req_capacity} is the peak capacity constraint. Note that the above problem involves all the flow variables of the entire network ($nL| \Set{E} |$ in total); and due to \eqref{eq:req_delay}, the decisions of the nodes are dependent on each other, which are determined in a centralized manner.

After the optimal solution $\V{M}_{ij}^{\star}$ is obtained, its first column $\V{\mu}_{ij}^{\star}(t) = \V{M}_{ij}^{\star}[\,:\,, 0]$ will be used as the decided flow for the current time slot. The rest of its columns are discarded, and the procedure repeats at the next time slot based on the updated information to make the corresponding decision.

\begin{remark}[Choice of $n$] \label{remark:n}
An intuitive choice is $n = L$, since the packets of the largest lifetime $L$ will be outdated after $L$ time slots, and we ignore the effects of the current decision at the distant time slots in the future. Another choice is $n = 1$, which simplifies the formulation by considering only the current time slot, and optimizes the \ac{ldp} greedily; the solution does not involve any (estimated) future information, which can be implemented in a {\em distributed} manner.
\end{remark}

\begin{remark}[Distributed RCNC] \label{remark:distributed}
To develop a distributed algorithm, we assume that future arrivals from neighbor nodes $\mu_{ji}^{(l)}(t+\tau)$ are estimated by their empirical average $\hat{u}_{ji}^{(l)}$. This leads to an \ac{lp} formulation that is the same as \eqref{eq:flow_matching_peak}, only to replace \eqref{eq:req_capacity} by $\sum_{j\in \delta_i^+} g_{\tau+1}( \V{M}_{ij} ) \leq g_{\tau+1}( \tilde{\V{A}}_i^{\text{emp}} )$, with $\tilde{\V{A}}_i^{\text{emp}} = [ \V{Q}_i(t), \hat{\V{\lambda}}_i + \hat{\V{u}}_{\to i}, \cdots, \hat{\V{\lambda}}_i + \hat{\V{u}}_{\to i}]$. However, numerical results suggest that this formulation does not outperform the simple algorithm using $n = 1$ (see Section \ref{sec:apx_alg}).
\end{remark}

\begin{algorithm}[t]
\caption{RCNC}\label{alg:p0}
\begin{algorithmic}[1]
\FOR{ each frame $k \geq 0$ }
\FOR{ $t = 0: K-1$ }
\STATE{ Solve the virtual flow $\V{\nu}(t)$ from \eqref{eq:virtual_opt}; }
\STATE{ Solve the actual flow $\V{\mu}(t)$ from \eqref{eq:flow_matching_peak}; }
\STATE{ Update the request queue $\V{R}(t)$ by \eqref{eq:request_q} (using the virtual and actual flows derived above); }
\ENDFOR
\STATE{ Update the transmission capacities of the links in the virtual network by \eqref{eq:capacity_growing_rate} -- \eqref{eq:capacity_delta}; }
\ENDFOR
\end{algorithmic}
\end{algorithm}

\begin{remark}[Complexity] \label{remark:complexity}
At every time slot, the centralized algorithm requires solving an LP problem with $nL|\Set{V}| + n|\Set{E}|$ constraints and $nL|\Set{E}|$ variables, and the time complexity is $\Set{O}(n^2 L^2 |\Set{E}|^2)$ (at the centralized controller). For the distributed algorithm, the complexity reduces to $\Set{O}(n^2 L^2 |\delta_i^+|^2)$ at node $i$. Intuitively, the centralized algorithm with $n = L$ can achieve a better performance; however, its complexity is $\Set{O}(L^4 |\Set{E}|^2)$, which can become prohibitive in practice. By selecting $n = 1$, we can obtain the most efficient algorithm (with some performance loss), at a complexity of $\Set{O}(L^2 |\delta_i^+|^2)$.
\end{remark}

\begin{remark}
In practice, we can select $L$ as tens of time slots (e.g., $L = 10$), based on following considerations. On one hand, it is on the order of network diameter and sufficient to support packet transmission within the network (note that the network diameter -- representing the hop-distance of the longest path -- of a hierarchical edge computing network is $\sim \mathcal{O}(\log(|\Set{V}|)$). On the other hand, it falls into the regime in which RCNC can run efficiently.

Accordingly, we can select appropriate time slot lengths based on the delay requirement of the supported applications,%
\footnote{
	Note that network slicing allows customizing (virtualized) networks for applications with similar delay requirements.
}
to achieve a value of $L$ as marked above. For delay-sensitive applications, such as VR ($7$--$20$ ms) \cite{cuervo2018creating} and real-time gaming ($50$ ms) \cite{anton2014machine}, a choice of $L$ on tens of time slots would result in time slot length of around $1$ ms. A larger slot length can be considered for applications with higher delay budgets, e.g., it can be selected as $15$ ms for live streaming ($150$ ms) \cite{anton2014machine}.
\end{remark}

\subsection{Capacity Iteration}

The flow matching technique proposed in the previous section assumes that the virtual flow assignment is achievable. This is not necessarily true since there is no guarantee for the equivalence of the flow spaces, as opposed to the average-constrained case. In fact, when deciding the flow assignment in the virtual network, the network controller prefers to transmit the packets along the low-cost routes, leading to a considerable amount of {\em bottleneck} links (i.e., for whom the assigned rate equals the transmission capacity), especially in the high-congestion regime. However, the achieved rate on the bottleneck link is sensitive to the dynamic input, which is usually strictly lower than the link capacity due to truncation (recall {\bf Example} in Section \ref{sec:approximate}). This motivates us to reduce the assigned virtual flow on these links, which can be realized by decreasing the corresponding link capacity in the virtual network.

In practice, a sign of unsuccessful flow matching is the instability of the request queues (i.e., $R_{ij}^{(l)}(t)$ grows linearly). In addition to the reason of {\em overestimating} the transmission capacity of the bottleneck link, in a multi-hop network, the request queue of link $(i, j)$ can also exhibit unstable behavior when its source $i$ receives {\em insufficient} packets from the neighbors compared to the virtual flow assignment. Both factors will be considered when updating the parameters (i.e., link capacities) of the virtual network.

To sum up, the parameters of the virtual network are updated on a larger timescale unit, we referred to as frames. Each frame $k$ consists of $K$ time slots, during which the algorithm developed in the previous subsection is performed in an attempt to stabilize the request queues. At the end of the frame, the increasing rate of the request queue for each link $(i, j)\in \Set{E}$ is calculated by
\begin{align} \label{eq:capacity_growing_rate}
r_{ij} = \sum_{l\in \Set{L}} r_{ij}^{(l)},\text{ with } r_{ij}^{(l)} \triangleq \max \Big\{ 0, \frac{1}{K} R_{ij}^{(l)}(K) \Big\}
\end{align}
and its link capacity is updated by
\begin{align}\label{eq:capacity_update}
C_{ij}(k+1) = \big[ (1-\kappa) \left[ C_{ij}(k) - \epsilon_{ij}^{(k)} \right] + \kappa C_{ij} \big]_{0}^{C_{ij}}
\end{align}
in which
\begin{align}\label{eq:capacity_delta}
\epsilon_{ij}^{(k)} = r_{ij} - r_{\to i} \left( \bar{\nu}_{ij} / \bar{\nu}_{i\to} \right)
\end{align}
where $\kappa\in (0, 1)$ is a constant, $[ z ]_0^{C_{ij}} \triangleq \min\{ \max\{ 0, z \}, C_{ij} \}$.
The update rule is explained as follows. First, the second term in \eqref{eq:capacity_delta} results from insufficient input, where $r_{\to i}$ is the total amount of insufficient input to node $i$, and $\bar{\nu}_{ij} / \bar{\nu}_{i\to}$ is the percentage that link $(i, j)$ takes up among all the outgoing interfaces. Second, note that even if the request queue is stabilized, it is possible for $\epsilon_{ij}^{(k)}$ to be positive due to random arrival, leading to too conservative flow assignment; therefore, we add $\kappa \, C_{ij}$ in \eqref{eq:capacity_update} to avoid such situation, which explores the possibility to increase the assigned flow rate in the considered link $(i, j)$.

\section{Numerical Experiments} \label{sec:experiments}

\begin{figure}[t]
	\centering
	\subfloat[Illustrative network.]{
		\includegraphics[width = 0.4\linewidth]{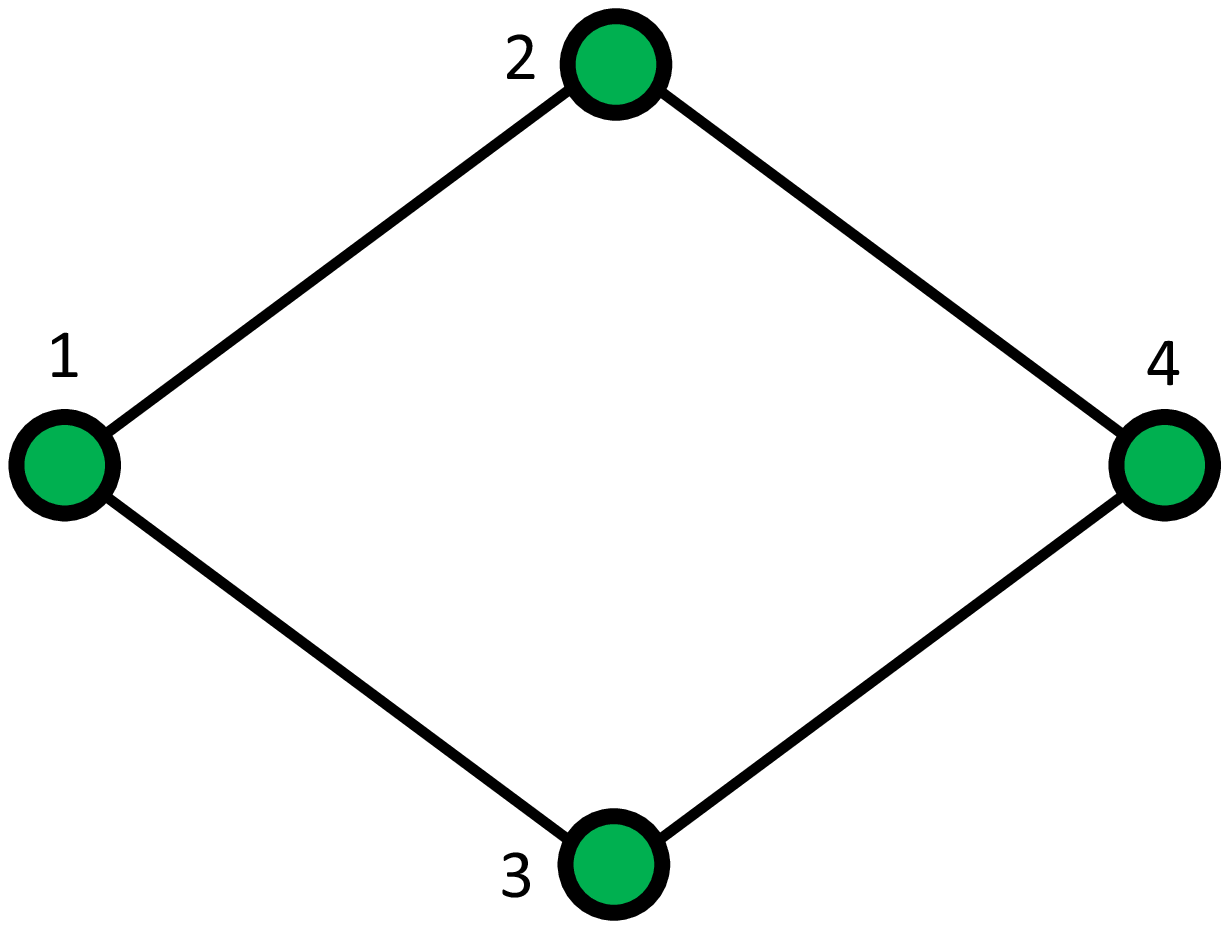}
		\label{fig:network1}
	}
	\hfill
	\subfloat[Mesh cloud network.]{
		\includegraphics[width = 0.45\linewidth]{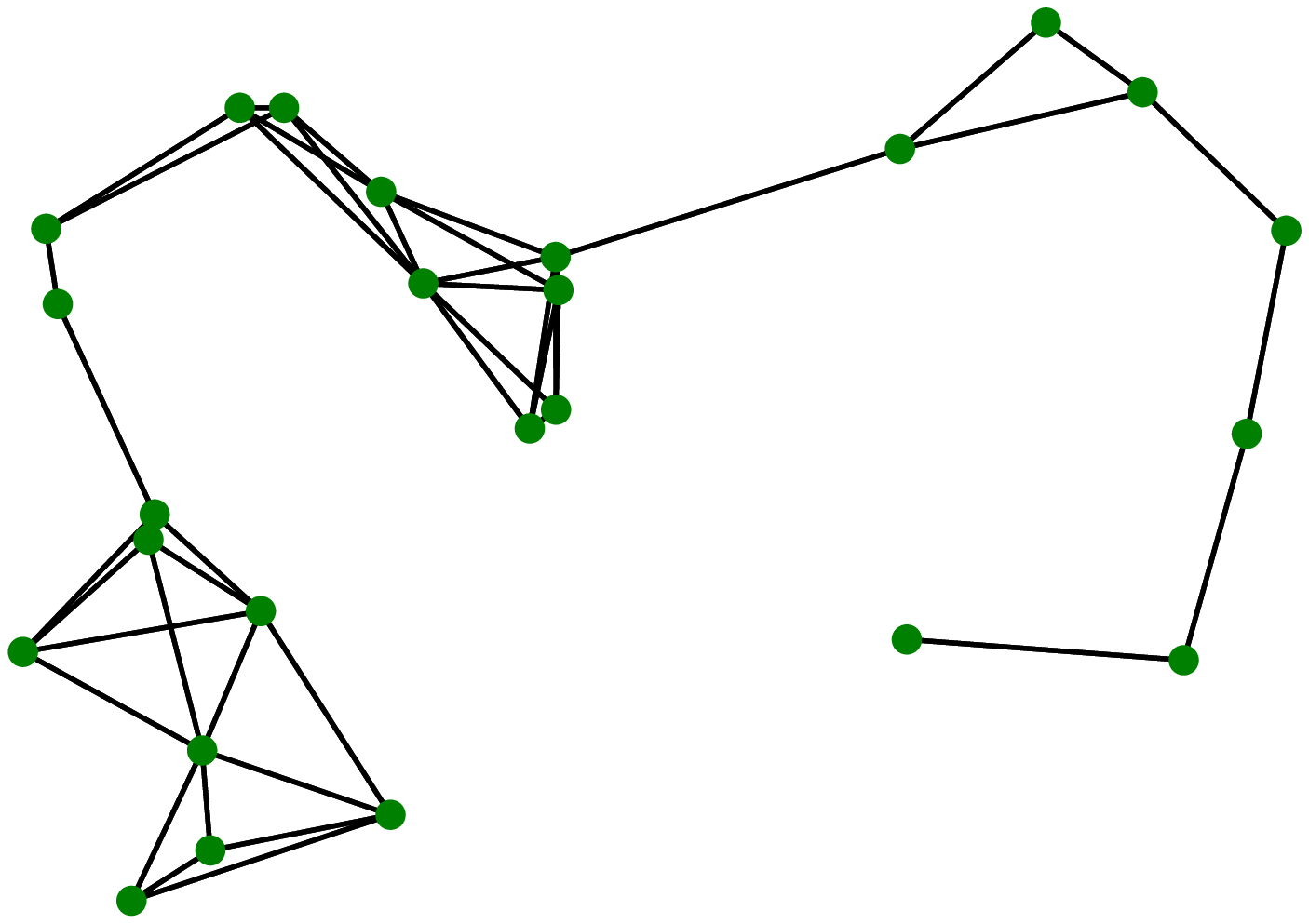}
		\label{fig:network2}
	}
	\hfill
	\subfloat[Hierarchical cloud network (devices of the same type have the same configuration).]{
		\includegraphics[width = .95\columnwidth]{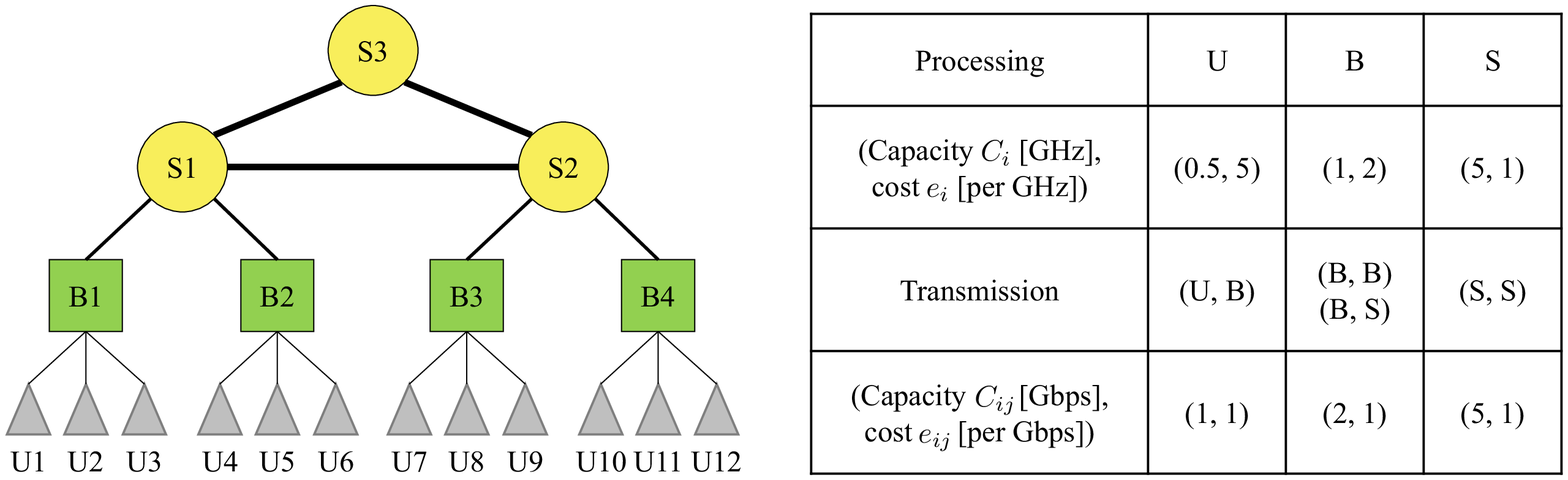}
		\label{fig:network3}
	}
	\caption{The studied networks.}
	\label{fig:networks}
\end{figure}

In this section, we carry out numerical experiments to evaluate the performance of the proposed design. We start with an illustrative example (Fig. \ref{fig:network1}), in which we explain the related concepts, as well as showing some intermediate results. After that, a more realistic scenario of edge computing network (Fig. \ref{fig:network2}) is studied. Both average- and peak-constrained networks are considered, and the term ``link capacity'' should be interpreted either way. We set $n = L$ and $K = 2\times 10^3$ as the default setting for the proposed RCNC algorithm.

Some key observations are listed as follows:
1) the analytical results (e.g., Proposition 3) are validated;
2) there is a performance gap between average- and peak-constrained problems, which vanishes as we reduce the arrival dynamics;
3) the throughput and cost performance improve with longer admissible lifetimes;
4) the distributed algorithm with $n = 1$ can achieve a comparable performance (especially in low-congestion regimes) with much lower complexity.

\subsection{Illustrative Example}
\label{sec:illustrate_experiment}

We study the packet routing problem based on the illustrative network in Fig. \ref{fig:network1}, which consists of $4$ nodes and $4$ undirected links. The links exhibit homogeneous transmission capacity of $C_{ij} = 5$ for $\forall\, (i,j)\in \Set{E}$, with different costs given by: $e_{12} = e_{24} = 1$, $e_{13} = e_{34} = 5$.

A single commodity is considered, where the packets of interest emerge at node $1$ (the source node), and are desired by node $4$ (the destination node). Each packet is of maximum lifetime of $L = 2$ at birth, which implies that it can not be delayed for even one single time slot in order to be effective. The packet arrival process follows a Poisson distribution with parameter $\lambda = 6$, and a reliability level of $\gamma = 90\%$ is demanded by the application.

\subsubsection{Effects of Parameter $V$}

\begin{figure}[t]
	\centering
	\subfloat[Operational cost.]{
	\includegraphics[width = .45 \linewidth]{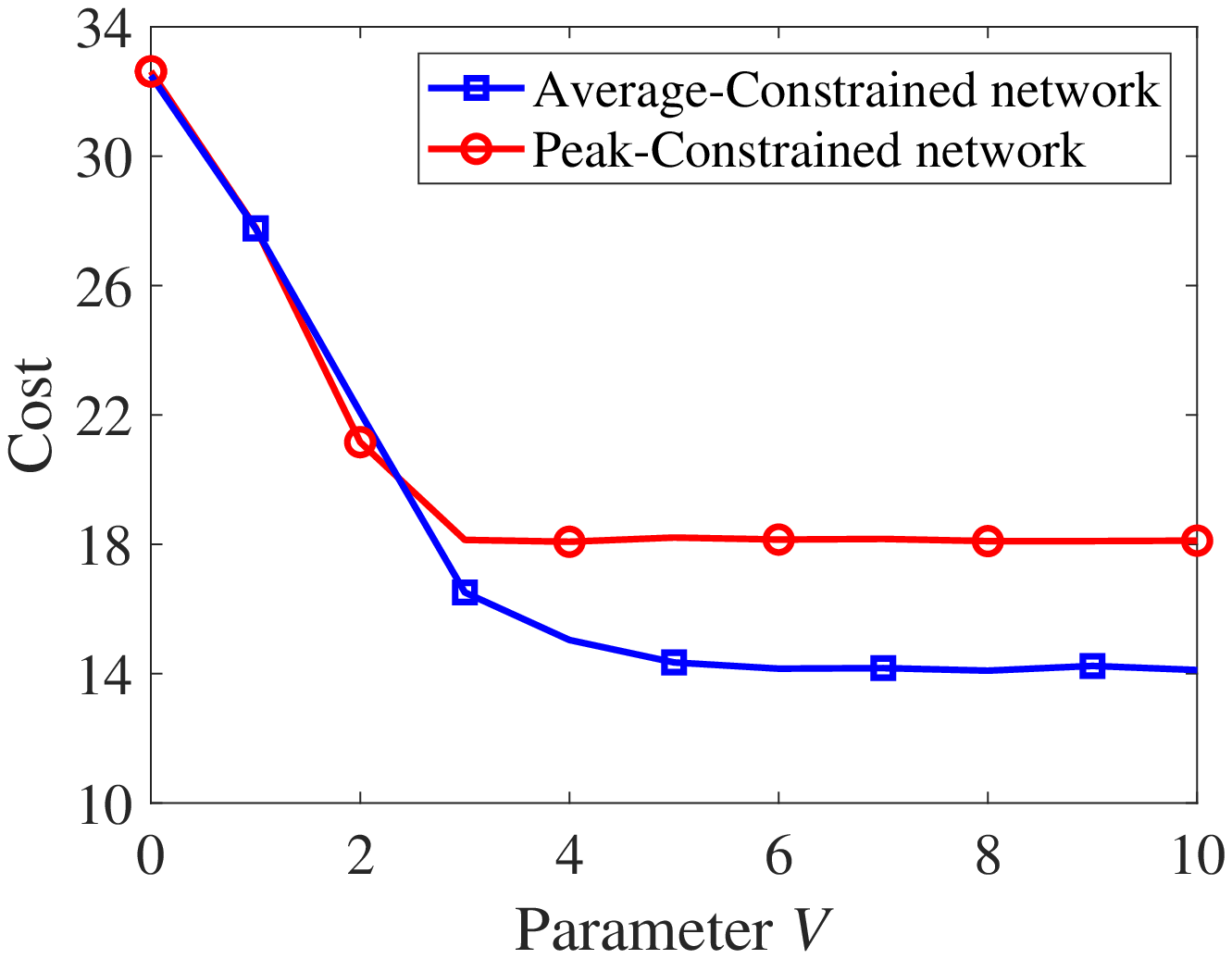}
	\label{fig:V_cost}
	}
	\hfill
	\subfloat[Capacity iteration of link $(1,2)$.]{
	\includegraphics[width = .45 \linewidth]{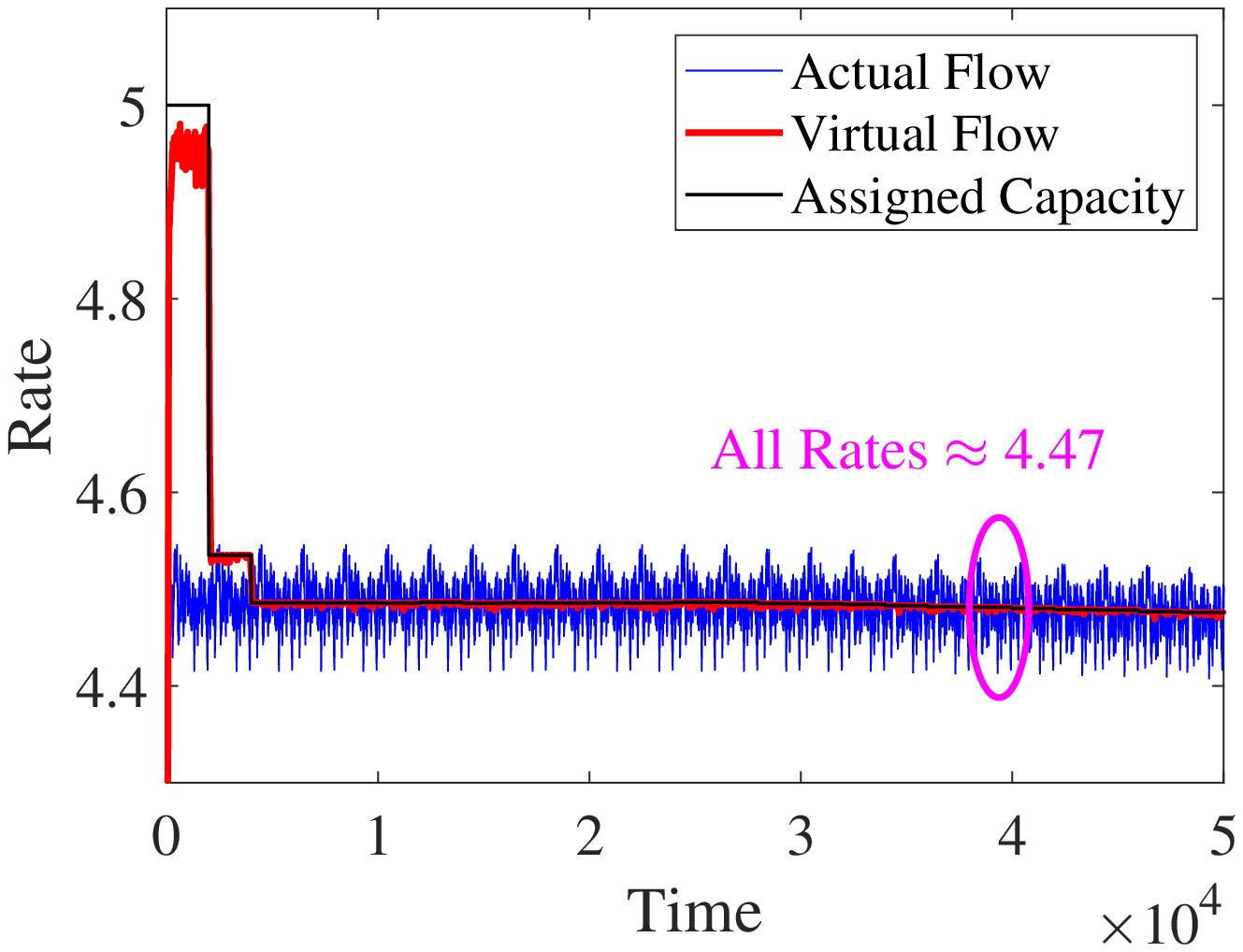}
	\label{fig:assignment}
	}
	\caption{The effect of $V$ on the achieved operational cost, and the flow assignment for peak-constrained network under $V = 5$.}
	\label{fig:V_cost_1}
\end{figure}

\begin{figure}[t]
	\centering
	\subfloat[$.01$-convergence time.]{
	\includegraphics[width = .45 \linewidth]{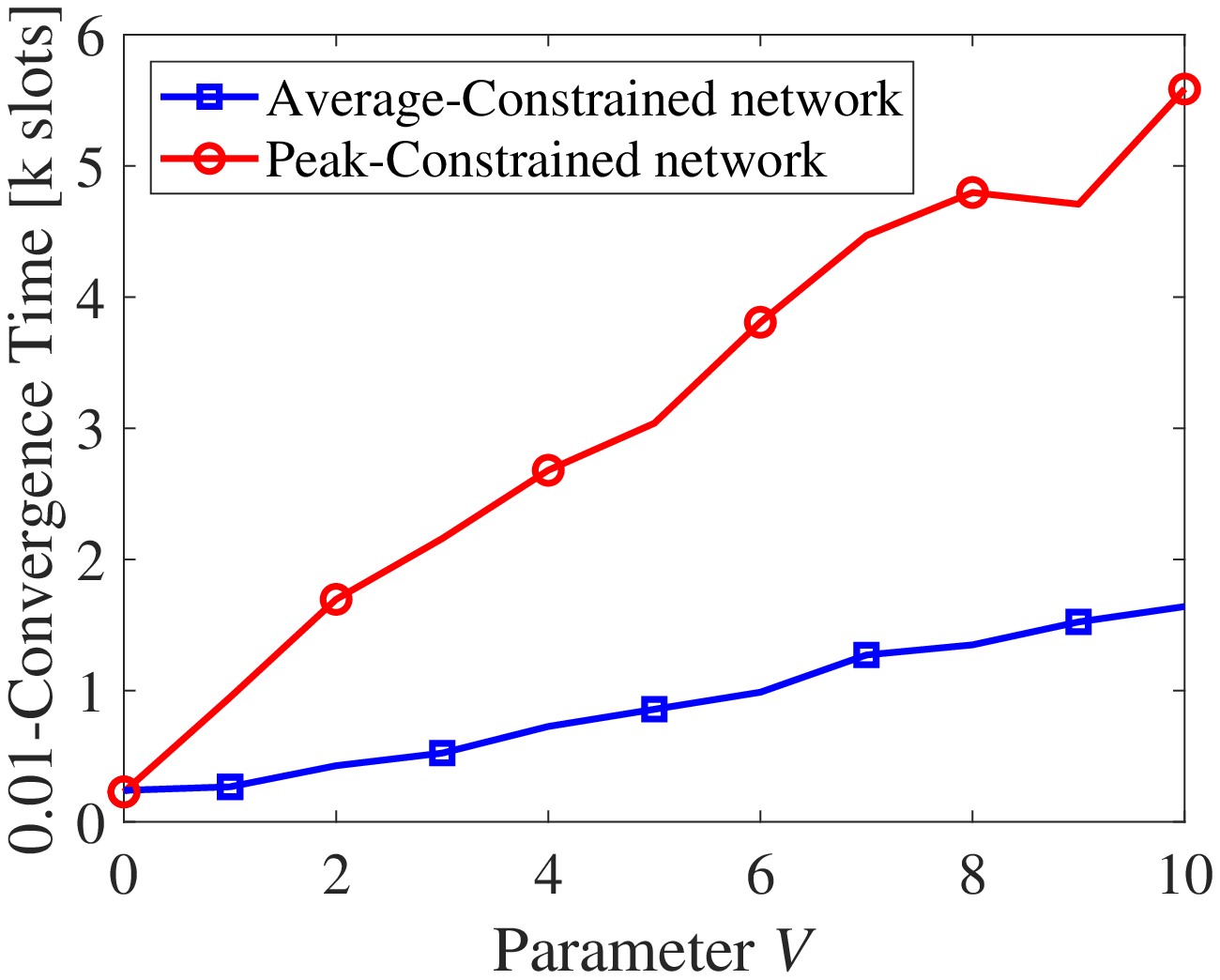}
	\label{fig:V_time}
	}
	\hfill
	\subfloat[The achieved reliability level.]{
	\includegraphics[width = .45 \linewidth]{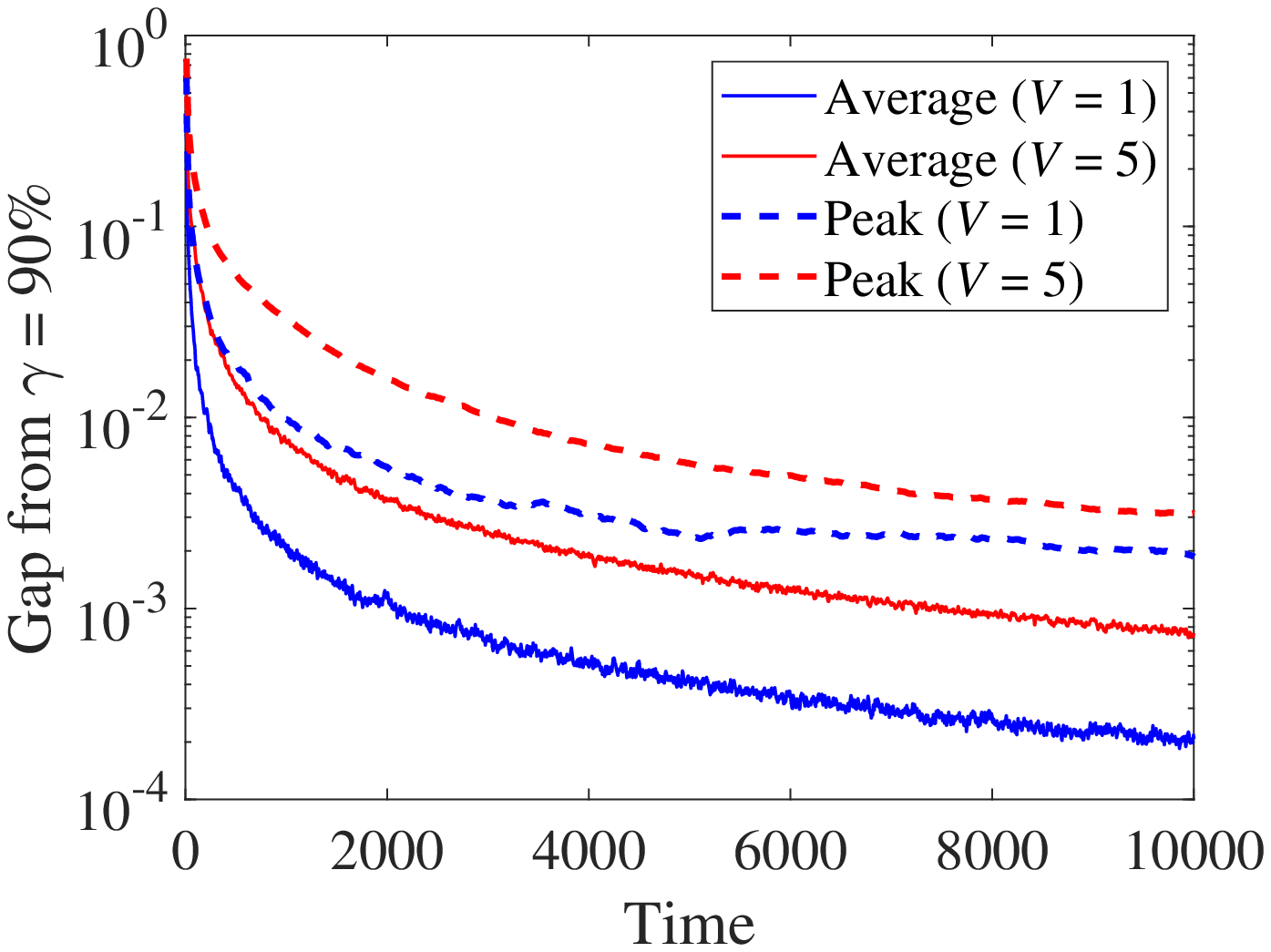}
	\label{fig:reliability}
	}
	\caption{The effect of $V$ on the $\varepsilon$-convergence time (with $\varepsilon = .01$), and the achieved reliability level over time under various settings.}
	\label{fig:V_time_1}
\end{figure}

In this experiment, we study the tradeoff between the convergence time and operational cost controlled by parameter $V$. We implement and run the control algorithm using various parameters $V \in \{0, 1, \cdots, 10\}$. For each $V$ value, we carry out $100$ experiments, and observe the system for $T = 1 \times 10^6$ time slots. The results are depicted in Fig. \ref{fig:V_cost_1} and \ref{fig:V_time_1}, and we make the following observations.

First, for the average-constrained network, the operational cost reduces with $V$ (Fig. \ref{fig:V_cost}), which is in accordance with the analytical results presented in Proposition \ref{prop:cost_V} and \ref{prop:flow_matching} (it also implies that flow matching is achieved). By intuition, we find that in this two-route example, the cheap route $1\to 2\to 4$ (with cost $e_{12} + e_{24} = 2$) is preferable when transmitting the packets to benefit the cost performance, which should be exploited to the largest extent; while to satisfy the throughput constraint, some packets still need to be pushed through the expensive route $1\to 3\to 4$ (with cost $e_{13} + e_{34} = 10$). More concretely, the flow assignment of the entire network is $x_{12} = x_{24} = \min\{ C_{12}, C_{24} \} = 5$ (the corresponding link capacity), $x_{13} = x_{34} = \gamma \lambda - x_{12} = 5.4 - 5 = 0.4$, leading to a cost performance of $h_1^\star = 5\times 2 + 0.4\times 10 = 14$. As we can observe in Fig. \ref{fig:V_cost}, the blue curve converges to the value of $14$ as $V$ increases, which agrees with the above result.

On the other hand, for the peak-constrained network, the principle to prioritize the cheap route also applies when transmitting the packets. However, due to the dynamics of the arrival process $a(t)$ and the truncation, the amount of packets that can be scheduled for this route is $x_{12}(t) = \min\{ a(t), C_{12} \}$ at every time slot. Under the assumption of i.i.d. Poisson arrival, it can be calculated that $\avg{ x_{12}(t) } \approx 4.47$. Therefore, the optimal flow assignment for the peak-constrained network is $x_{12} = x_{24} = 4.47$, and $x_{13} = x_{34} = \gamma \lambda - x_{12} = 0.93$, leading to a cost of $h_2^\star = 4.47\times 2 + 0.93 \times 10 = 18.24$. The proposed RCNC algorithm finds the flow assignment by trial and exploration, as shown in Fig. \ref{fig:assignment} (with $V = 5$): we use the link capacity $C_{12} = 5$ as the initial guess for the achievable flow rate, which overestimates the transmission capacity of the link; then its link capacity in the virtual network is reduced (on a frame basis), which gears the corresponding virtual flow; finally, flow matching is achieved when the $\text{link capacity}\approx\text{achieved rate}\approx 4.47$.

Finally, we emphasize that by increasing the value of $V$, it takes longer to converge to the desired reliability level. Fig. \ref{fig:reliability} shows the gap between the achieved and the desired reliability level over time, for two particular values $V = 1$ and $V = 5$. We find that (i) all the gap curves reduce over time, implying convergence to the desired value, (ii) it takes longer for the peak-constrained network to converge than the corresponding averaged case (under the same $V$), which is due to the additional procedure of capacity iteration to find the feasible flow assignment, and (iii) the gap grows with $V$ at a fixed time point for both average- and peak-constrained networks, i.e., a larger $V$ results in slower convergence. In particular, we study the $\varepsilon$-convergence time with $\varepsilon = .01$, and the result is plotted in Fig. \ref{fig:V_time}. For the average-constrained network, the convergence time grows linearly with $V$, which supports the analytical result of $\mathcal{O}(V)$ in Proposition \ref{prop:cost_V}; similar result is observed from the peak-constrained network.

\subsubsection{Effects of Lifetime and Arrival Model}

Next, we study the effects of the maximum lifetime $L$ and the statistics of the arrival process. The considered lifetime $L$ ranges from $2$ to $10$, and we try different models for the arrival process, including uniform $\mathcal{U}([0, 2\lambda])$, Poisson $\text{Pois}(\lambda)$, binomial $\mathcal{B}(2\lambda, 1/2)$, as well as the constant arrival $a(t) = \lambda$. The four distributions are of the same mean value $\lambda$, but decreasing dynamic (the corresponding variances are $\lambda^2/3 > \lambda > \lambda/2 > 0$ if we assume $\lambda > 3$). The reliability level is set as $\gamma = 90 \%$, and $V = 10$ is chosen to optimize the operational cost.

Two performance metrics are studied for each settings. One is the achieved operational cost, and the other is the stability region.%
\footnote{
	We recall that the stability region of $\mathscr{P}_0$ is defined w.r.t. the \ac{pdf} of the arrival process $f_{\V{a}}$. In the experiment, as the model of the arrival process is fixed, we only need to specify $\lambda$ to determine the \ac{pdf}; in other words, the maximum arrival rate $\lambda$ can represent the stability region under each model.
}
In the first part of the experiment, we assume $\lambda = 6$ as in the previous experiments.

\begin{figure}[t]
	\centering
	\subfloat[Operational cost.]{
		\includegraphics[width = .45 \linewidth]{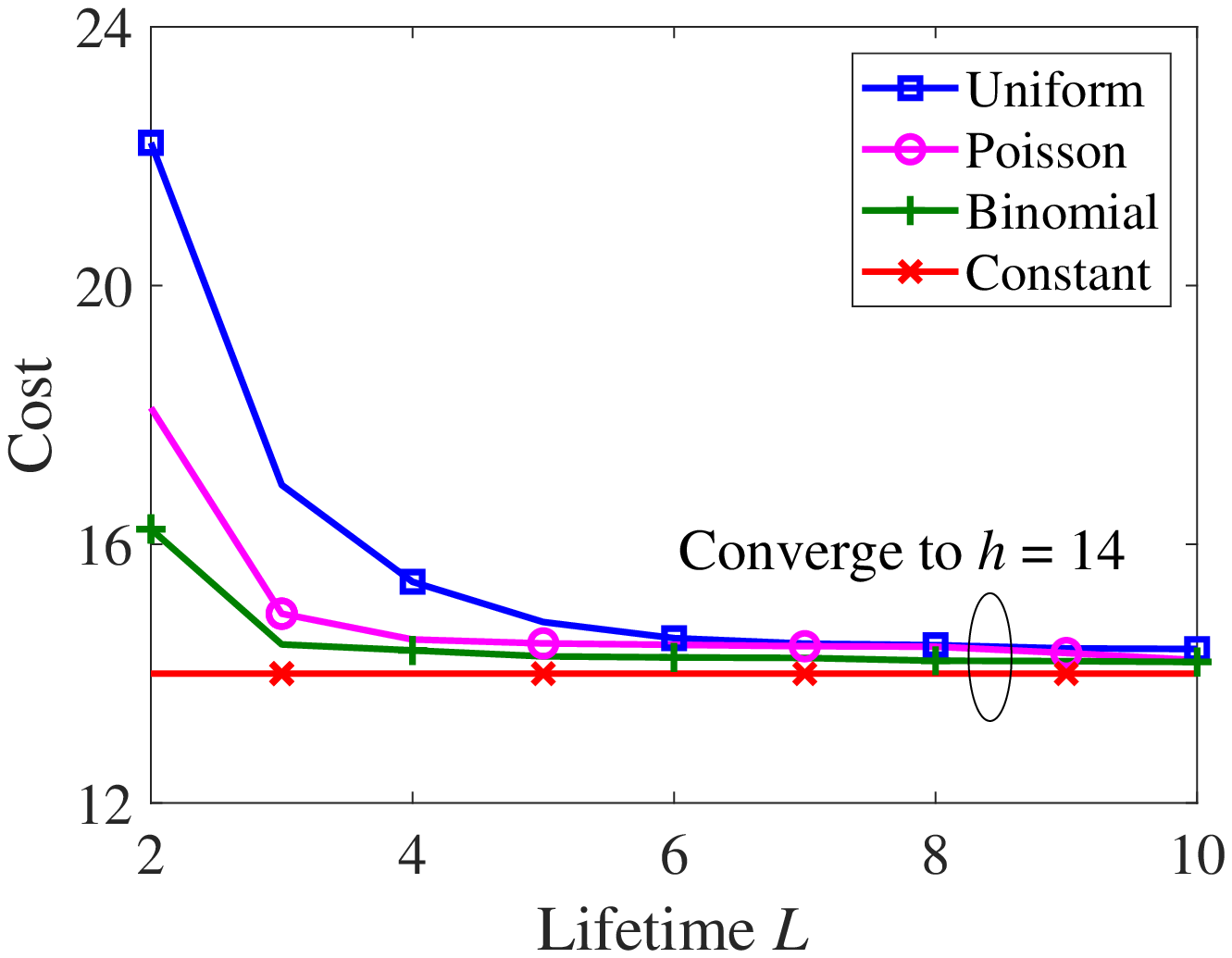}
		\label{fig:L_pdf_cost}
	}
	\hfill
	\subfloat[Stability region.]{
		\includegraphics[width = .45 \linewidth]{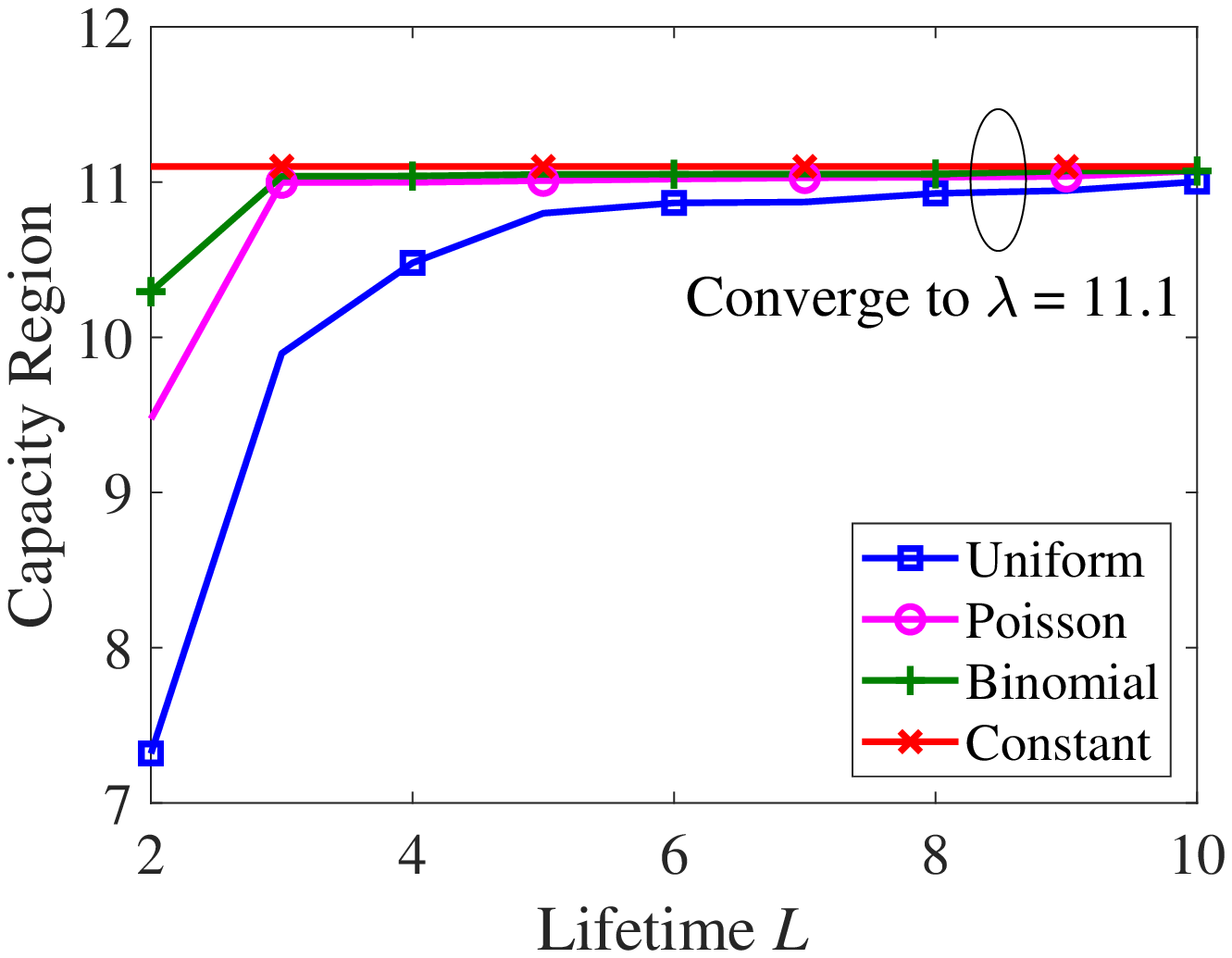}
		\label{fig:L_pdf_cr}
	}
	\caption{
	The effect of maximum lifetime $L$ and the arrival model on the achieved operational cost and stability region.
	}
	\label{fig:L_pdf}
\end{figure}

The results for the peak-constrained networks are shown in Fig. \ref{fig:L_pdf}. As we can observe, for any arrival model, as the maximum lifetime $L$ grows, the operational cost attained by RCNC reduces, while the stability region enlarges. The result agrees with our intuition, that as the initial lifetime grows, the packets are more likely to arrive at the destination while effective, and furthermore, through the cheap route (when possible). In this example, node $1$ can withhold the packets in its queuing system at bursting time slots, leaving them for future transmission through $1\to 2\to 4$ to optimize the cost. As $L$ increases, the problem reduces to the traditional packet routing problem, where packet lifetime is not relevant, and the attained operational cost and stability region {\em converge} to the corresponding optimal results.%
\footnote{
	However, we stress that RCNC does not guarantee convergence to the optimal performance in all cases. As we compare the uniform arrival with other models, there is gap in terms of both metrics, which is probably due to the sub-optimality of the flow matching technique in this case.
}
We also find that under constant arrival, the maximum lifetime does not impact the performance, which approximately equals the optimal values of the traditional problem.%
\footnote{
	In addition, for average-constrained networks, the various settings (including maximum lifetime and arrival model) do not impact the performance, either, which is the same as the plotted results for constant arrival.
}

\begin{figure*}[t]
	\centering
	\subfloat[Effect of lifetime ({\em mesh}).]{
		\includegraphics[width = .23 \linewidth]{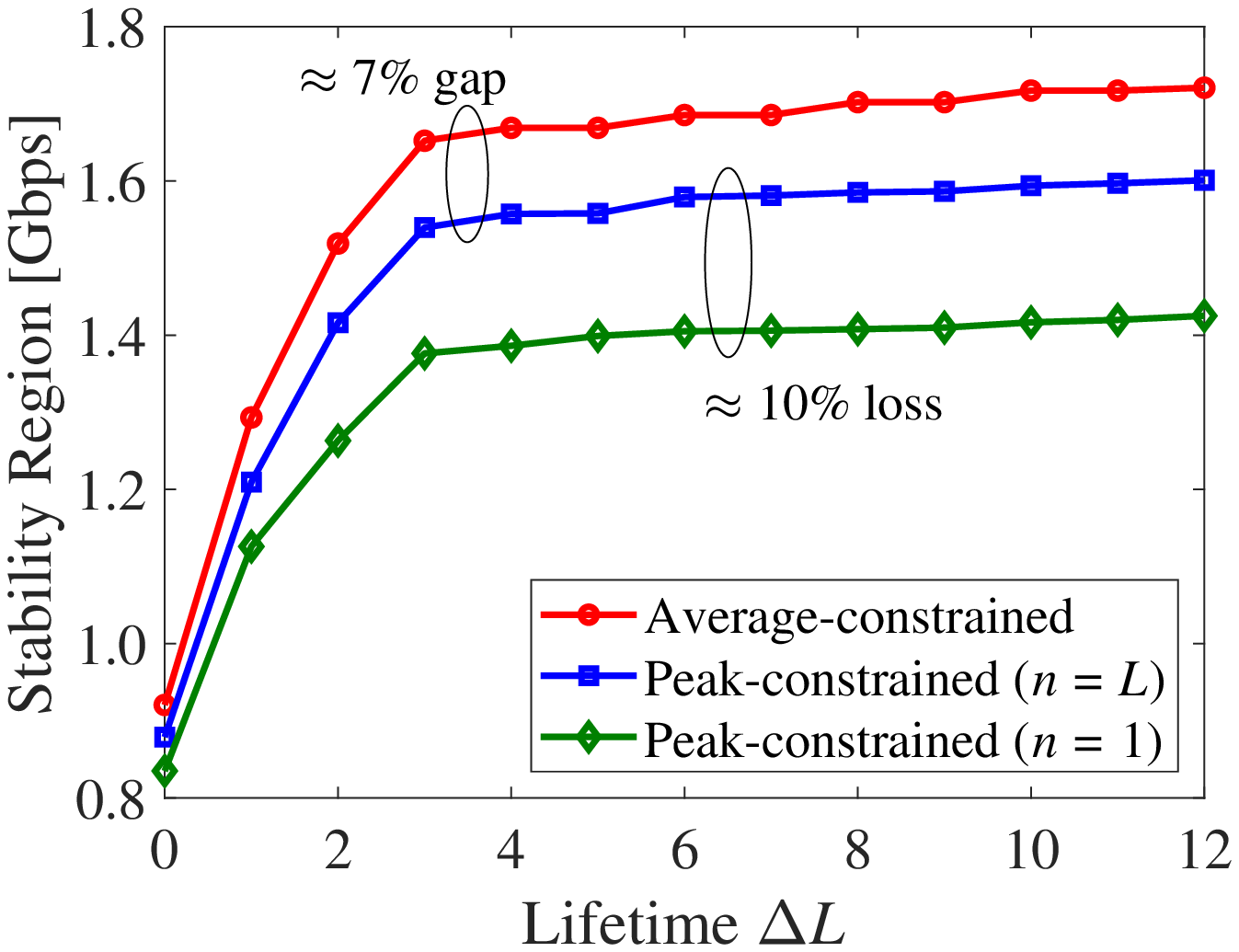}
		\label{fig:nsr_lifetime_edge}
	}\hfill
	\subfloat[Effect of time slot length ({\em mesh}).]{
		\includegraphics[width = .23 \linewidth]{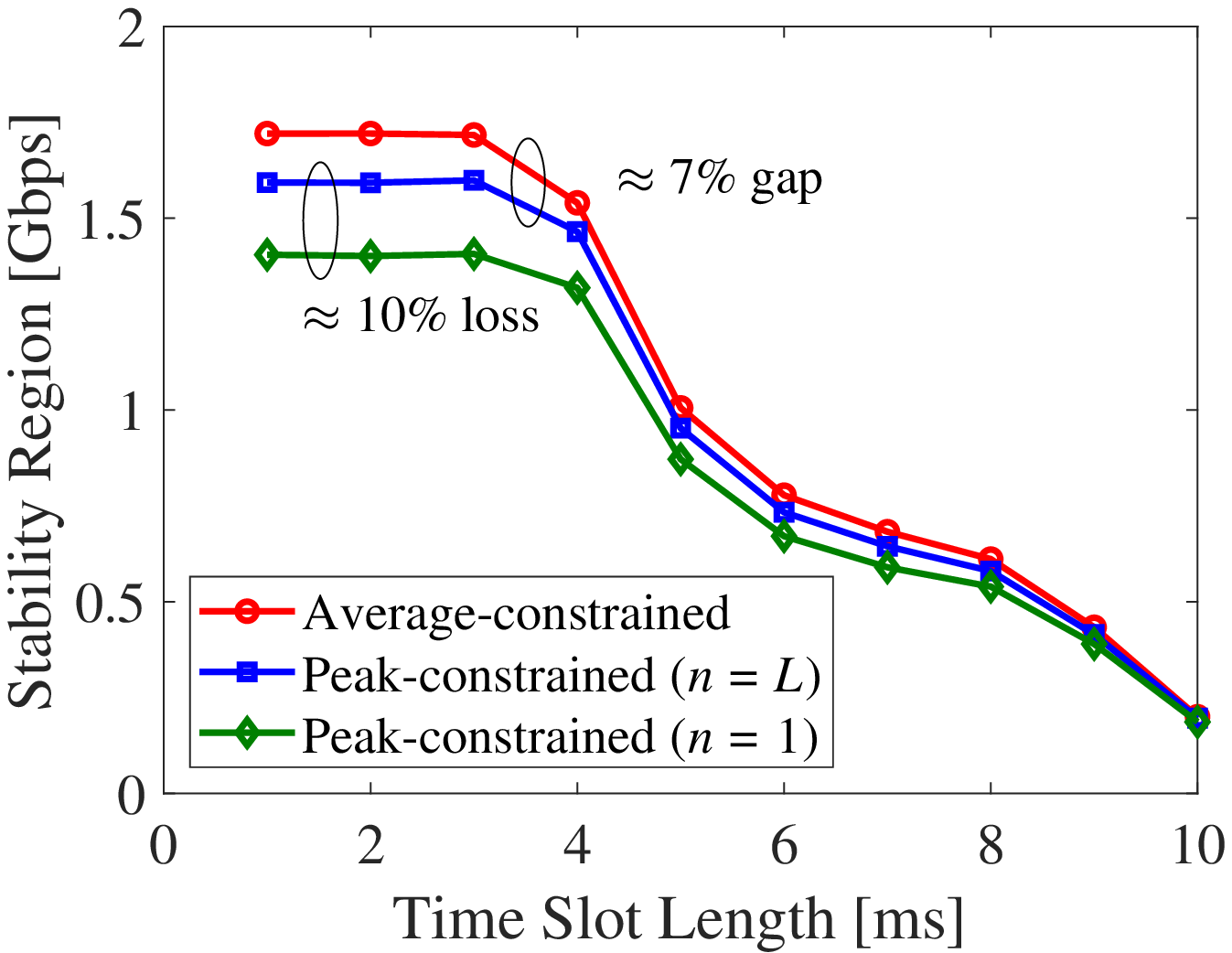}
		\label{fig:nsr_slotlength_edge}
	}\hfill
	\subfloat[Effect of lifetime ({\em hierarchical}).]{
		\includegraphics[width = .23 \linewidth]{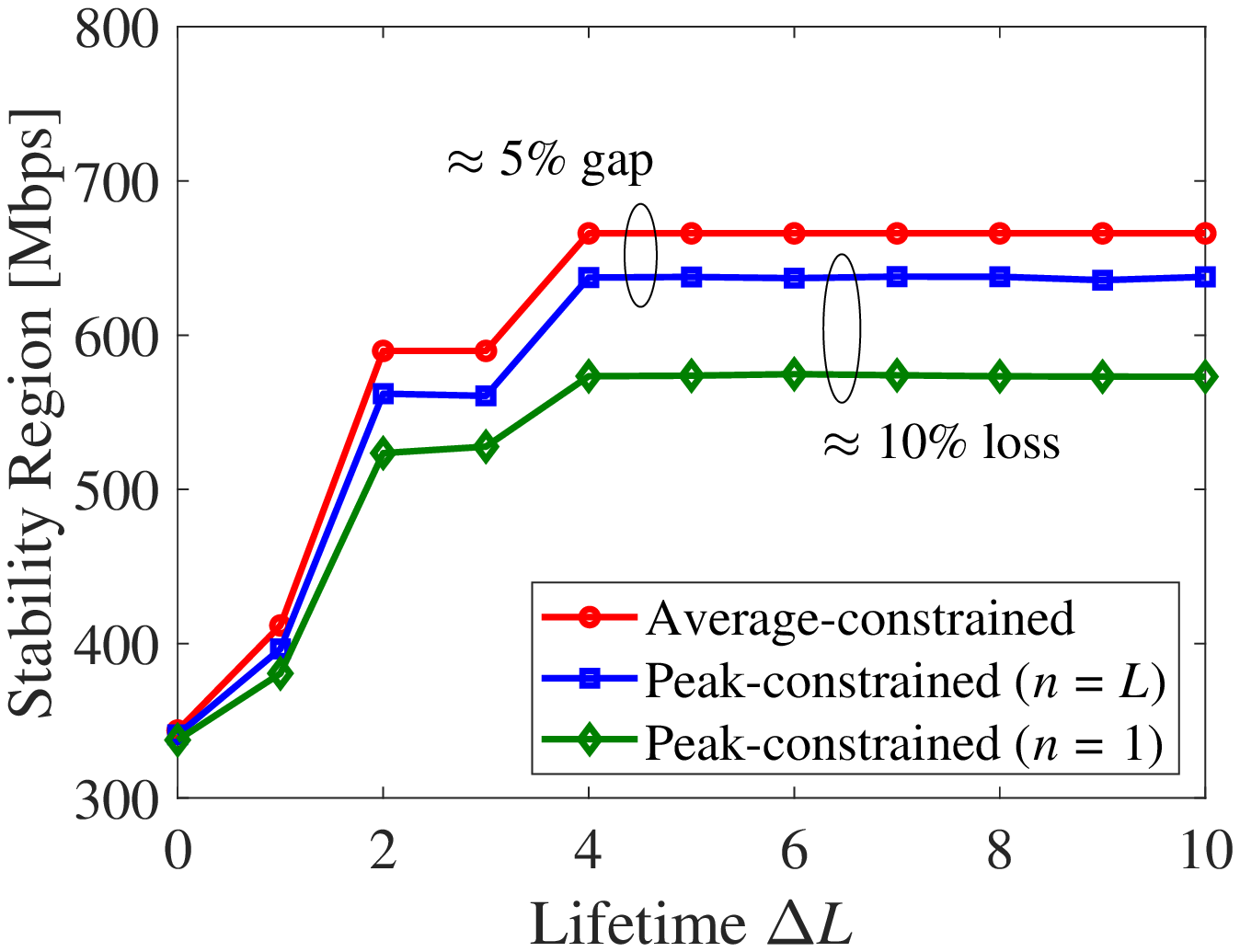}
		\label{fig:nsr_lifetime_fog}
	}\hfill
	\subfloat[Effect of time slot length ({\em hierarchical}).]{
		\includegraphics[width = .23 \linewidth]{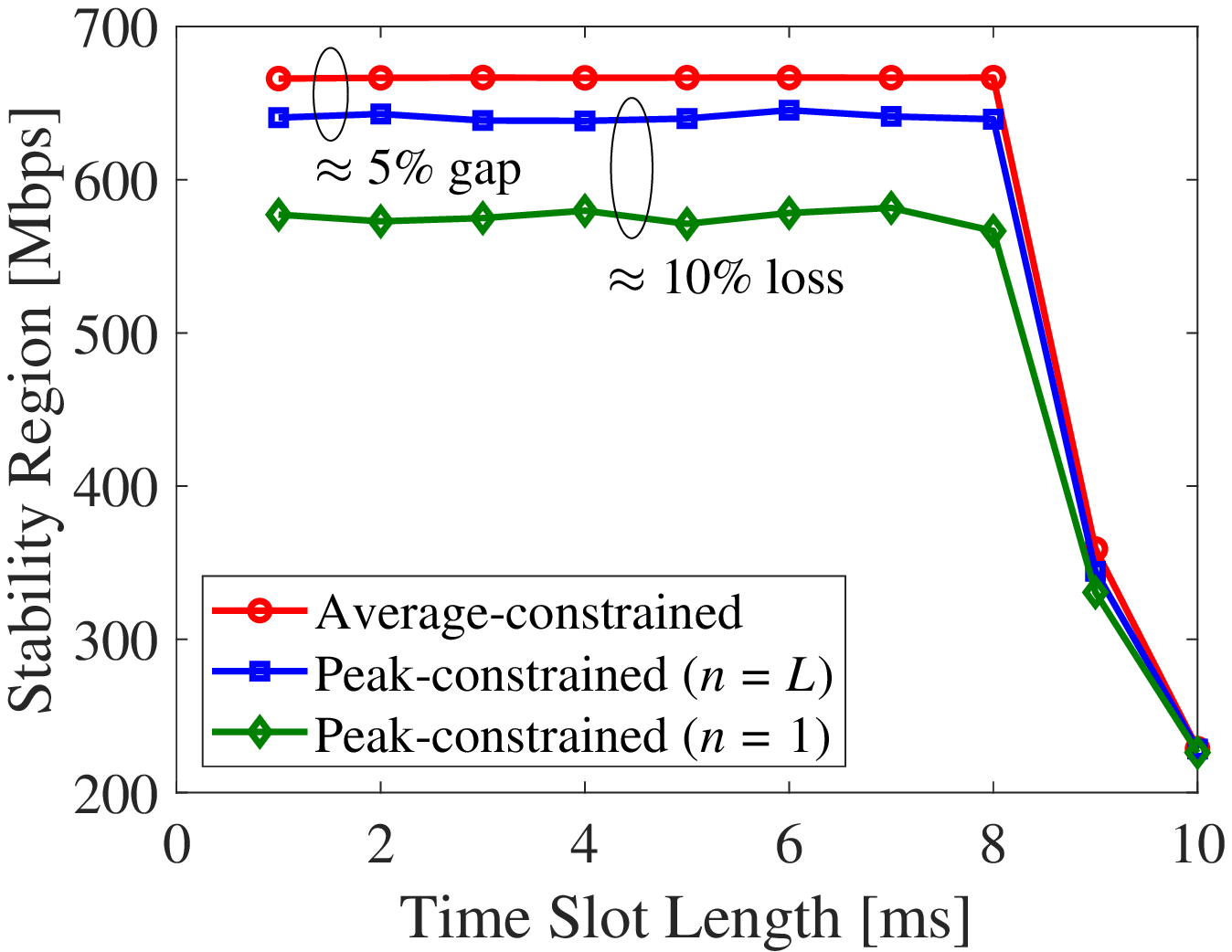}
		\label{fig:nsr_slotlength_fog}
	}
	\caption{Stability regions of Algorithm \ref{alg:p1} and RCNC (for average- and peak-constrained cases, respectively), under different lifetimes and time slot lengths.}
	\label{fig:nsr}
\end{figure*}

Finally, we explain the effect of the arrival model. By fixing the maximum lifetime (e.g., $L = 2$), we compare different arrival models, and find that a higher dynamic of the arrival process can increase the operational cost, while shrinking the stability region, both due to the truncation effect of the peak-constrained links. With a given mean rate $\lambda$, it is more likely for a high-dynamic arrival process to exceed the transmission capacity, and more packets must be delayed for transmission, which can possibly lead to packet outdatedness, and thus reducing the achievable output rate. This is true for any link in general, and in particular, the links lying in the cheap routes; as a result, a worsened performance of stability region and operational cost can be expected.%
\footnote{
	The results suggest that when 1) the packet lifetime is abundant for transmission, or 2) the arrival process is of low-dynamic, $\mathscr{P}_1$ makes a good approximation for the original problem $\mathscr{P}_0$. However, the gap can be large in some extreme cases, e.g., uniform arrival with $L = 2$ in the experiment.
}

\subsection{Practical Scenarios}

In this section, we demonstrate the performance of the proposed RCNC algorithm in two representative network scenarios:
\begin{itemize}
	\item {\em mesh}: a mesh edge computing network including $25$ servers that are randomly placed in an $1\,\text{km}\times 1\,\text{km}$ square area, with links established between any pair of servers within a distance of $250$\,m, as shown in Fig. \ref{fig:network2}, which is representative of a generic unstructured scenario. 
	\item {\em hierarchical}: a hierarchical edge computing network  \cite{yeh2021deco} composed of core, edge, access, and user layers, as shown in Fig. \ref{fig:network3}, which represents envisioned practical MEC systems.
\end{itemize}

In the {\em mesh} network, each link has a transmission capacity of $C_{ij} = 1$ Gbps with a cost of $e_{ij} = 1\,/$Gbps, and each server has a processing capacity of $C_i = 2$ GHz with a random cost $e_i \in \{5, 7.5, 10\}\,/$GHz, which accounts for the heterogeneity of the computing devices; the parameters of the {\em hierarchical} network are summarized in the table in Fig. \ref{fig:network3}. The default time slot length is $1$ ms in both networks.

We adopt the \ac{agi} service model used in \cite{FenLloTulMol:J18a}. The \ac{agi} service $\phi$ is modeled by a sequence of ordered functions, through which incoming packets must be processed to produce consumable results. The service functions can be executed at different network locations, and we assume that each network location can host all the service functions. Each function (say the $m$-th function of service $\phi$) is specified by two parameters: (i) $\xi_\phi^{(m)}$: scaling factor, i.e., the output flow units per input flow unit. (ii) $r_\phi^{(m)}$: workload, i.e., the required computational resource per input flow unit.

In this experiment, we consider two \ac{agi} services including $2$ functions, with parameters given by (the workload $r_\phi^{(m)}$ is in GHz/Gbps):
\begin{align*}
\text{Service }1 &:\ \xi_1^{(1)} = 1,\ \xi_1^{(2)} = 2;\ r_1^{(1)} = \frac{1}{300},\ r_1^{(2)} = \frac{1}{400}, \\
\text{Service }2 &:\ \xi_2^{(1)} = \frac{1}{3},\ \xi_2^{(2)} = \frac{1}{2};\ r_2^{(1)} = \frac{1}{200},\ r_2^{(2)} = \frac{1}{100}.
\end{align*}

Each service has an i.i.d. Poisson arrival process (for packets with maximum lifetime), with $\lambda_1 = \lambda_2 = \lambda$ Mbps, and requires a reliability level of $\gamma_1 = \gamma_2 = 90\%$.%
\footnote{
	The service chain can expand or compress the size of the input flow, and we calculate the throughput on the basis of the input flow size. See Appendix \ref{apdx:agi} for detailed explanation.
}
The source-destination pair of each service is selected at random (with the shortest distance between them denoted by $\sigma$). The maximum lifetime is then chosen as $L = \sigma + 2 + \Delta L$, where $\sigma + 2$ is the least lifetime for packet delivery (``2'' account for two processing slots), and $\Delta L \geq 0$ denotes some allowable relaxation slots.

\begin{figure*}[t]
	\centering
	\subfloat[$V$ ({\em mesh}, with $\Delta L = 2$).]{
		\includegraphics[width = .23 \linewidth]{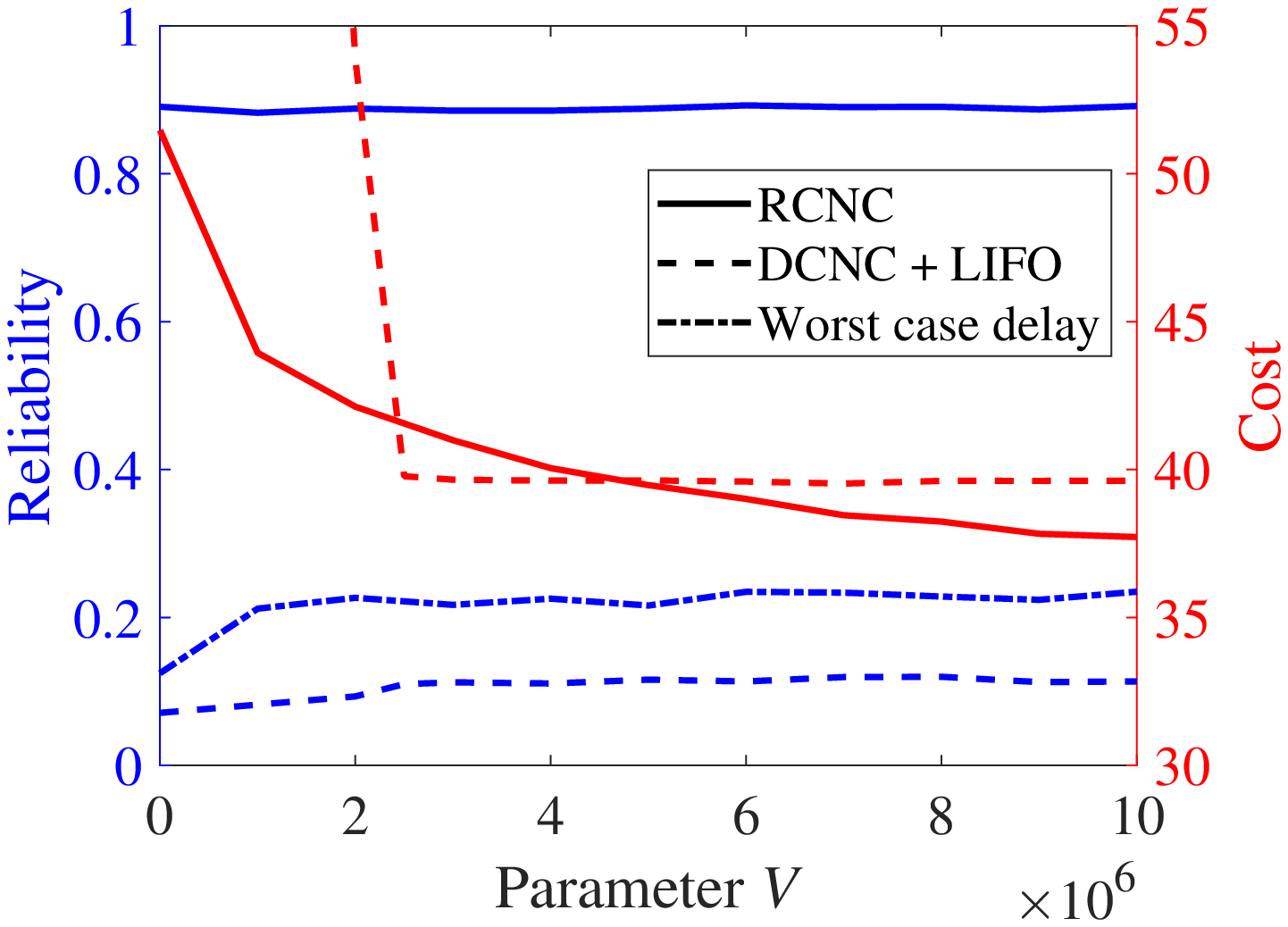}
		\label{fig:V_2}
	}
	\hfill
	\subfloat[$\Delta L$ ({\em mesh}, with $V = 1\times 10^8$).]{
		\includegraphics[width = .23 \linewidth]{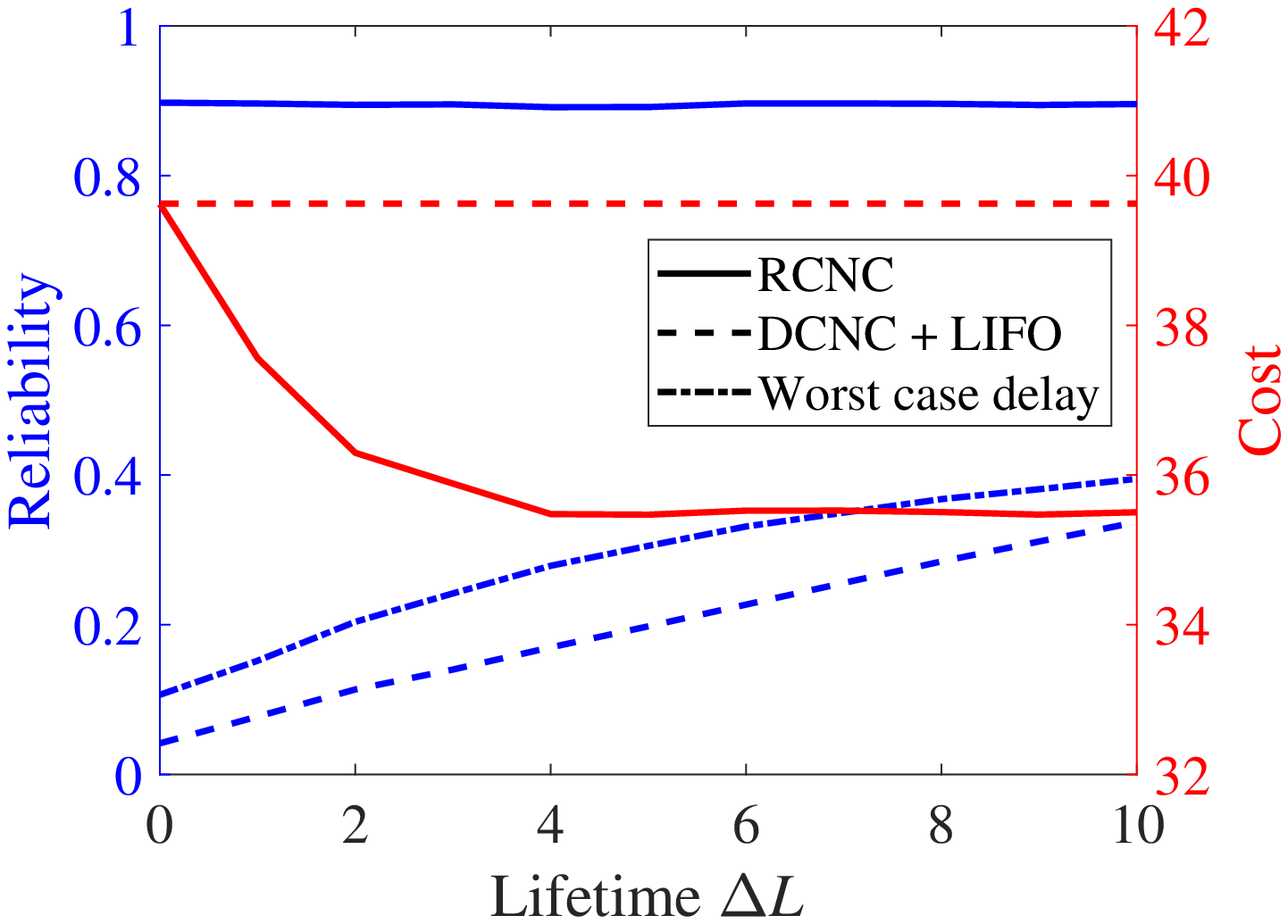}
		\label{fig:L_2}
	}
	\hfill
	\subfloat[$V$ ({\em hierarchical}, with $\Delta L = 2$).]{
		\includegraphics[width = .23 \linewidth]{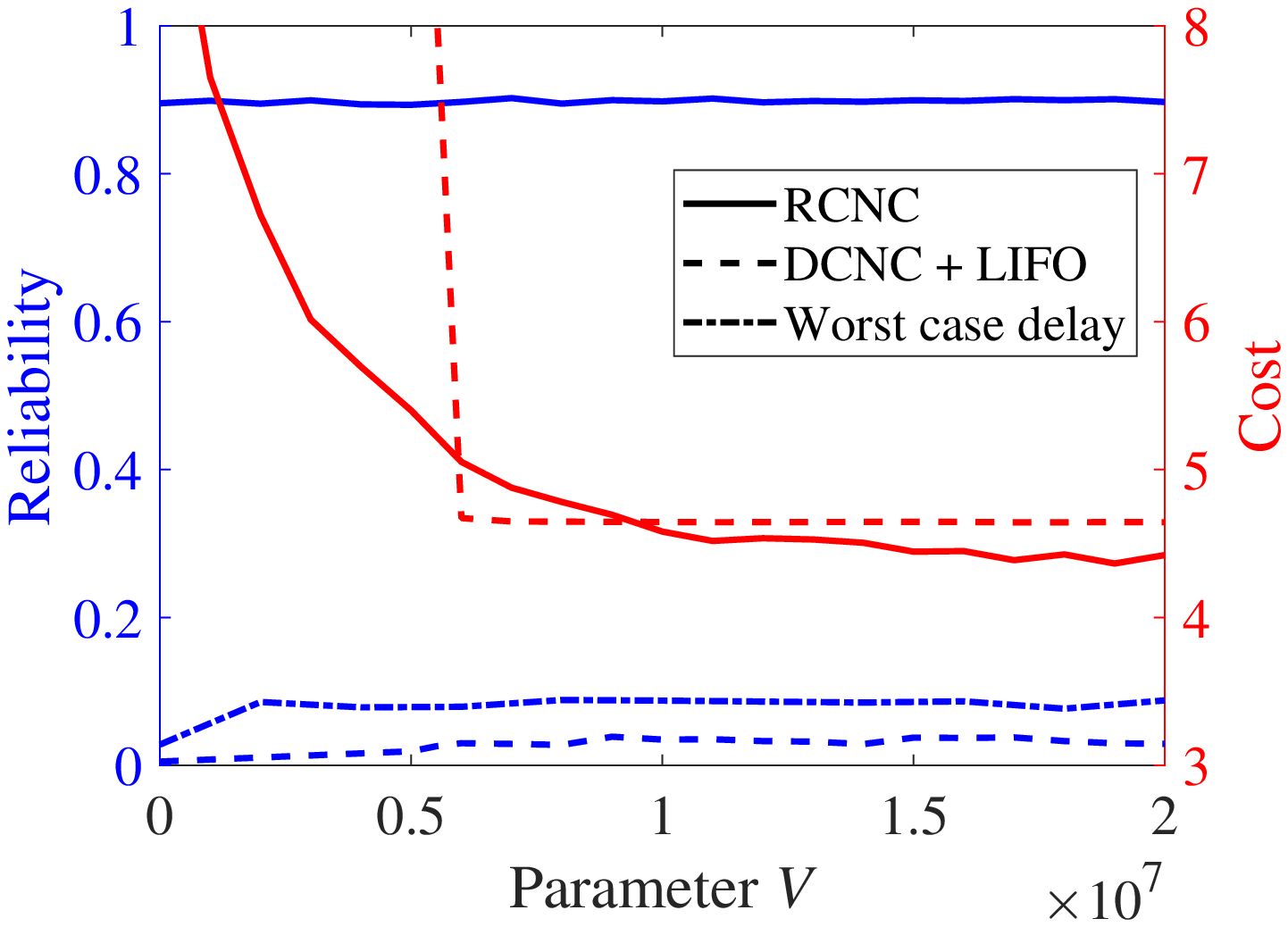}
		\label{fig:V_2_fog}
	}
	\hfill
	\subfloat[$\Delta L$ ({\em hierarchical}, with $V = 1\times 10^8$).]{
		\includegraphics[width = .23 \linewidth]{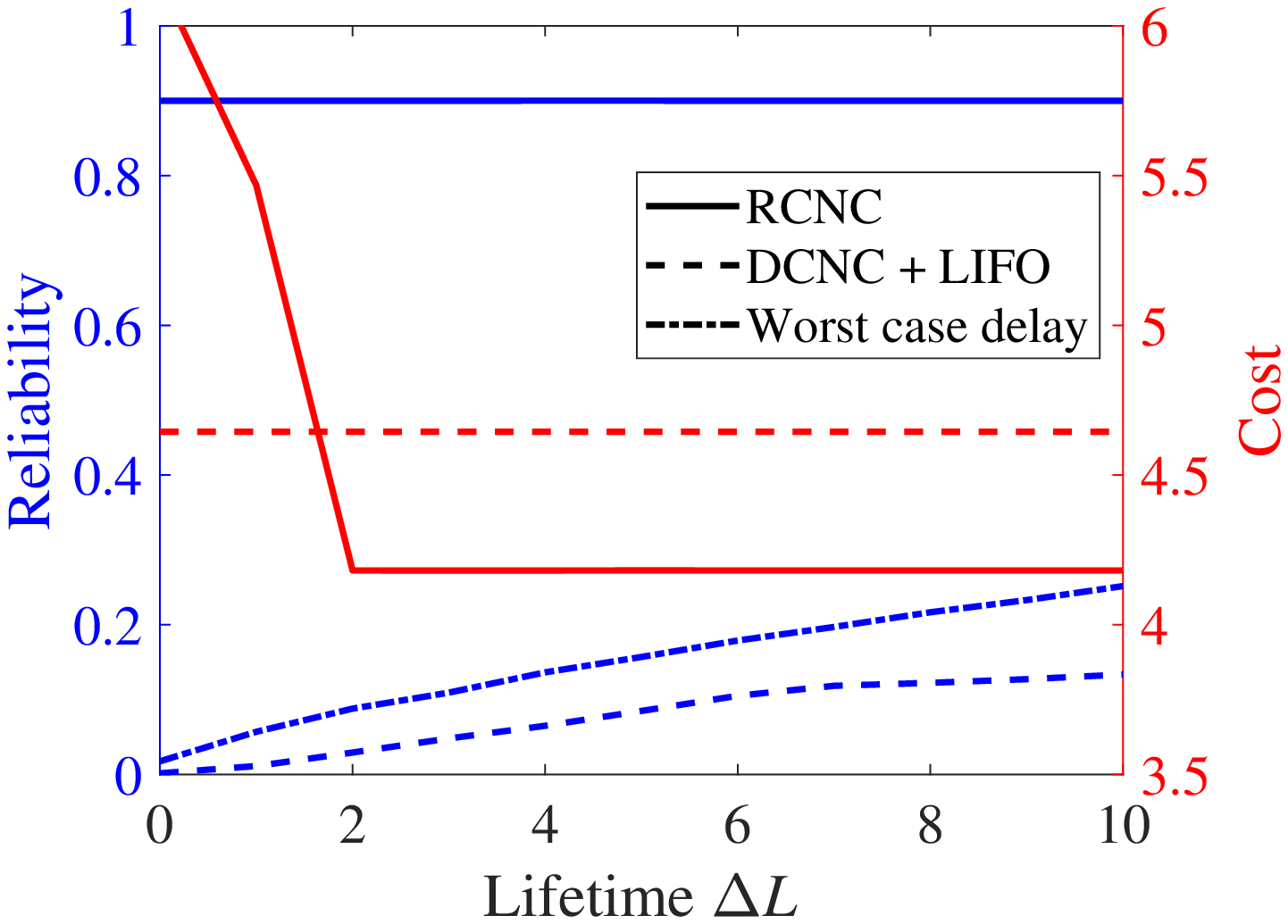}
		\label{fig:L_2_fog}
	}
	\caption{
		Throughput and cost achieved by DCNC (with LIFO) \cite{FenLloTulMol:J18a}, worst-case delay \cite{Nee:C11} and RCNC.
	}
	\label{fig:compare_bp}
\end{figure*}

\begin{figure*}[t]
	\centering
	\subfloat[Low-congestion regime ({\em mesh}).]{
		\includegraphics[width = .23 \linewidth]{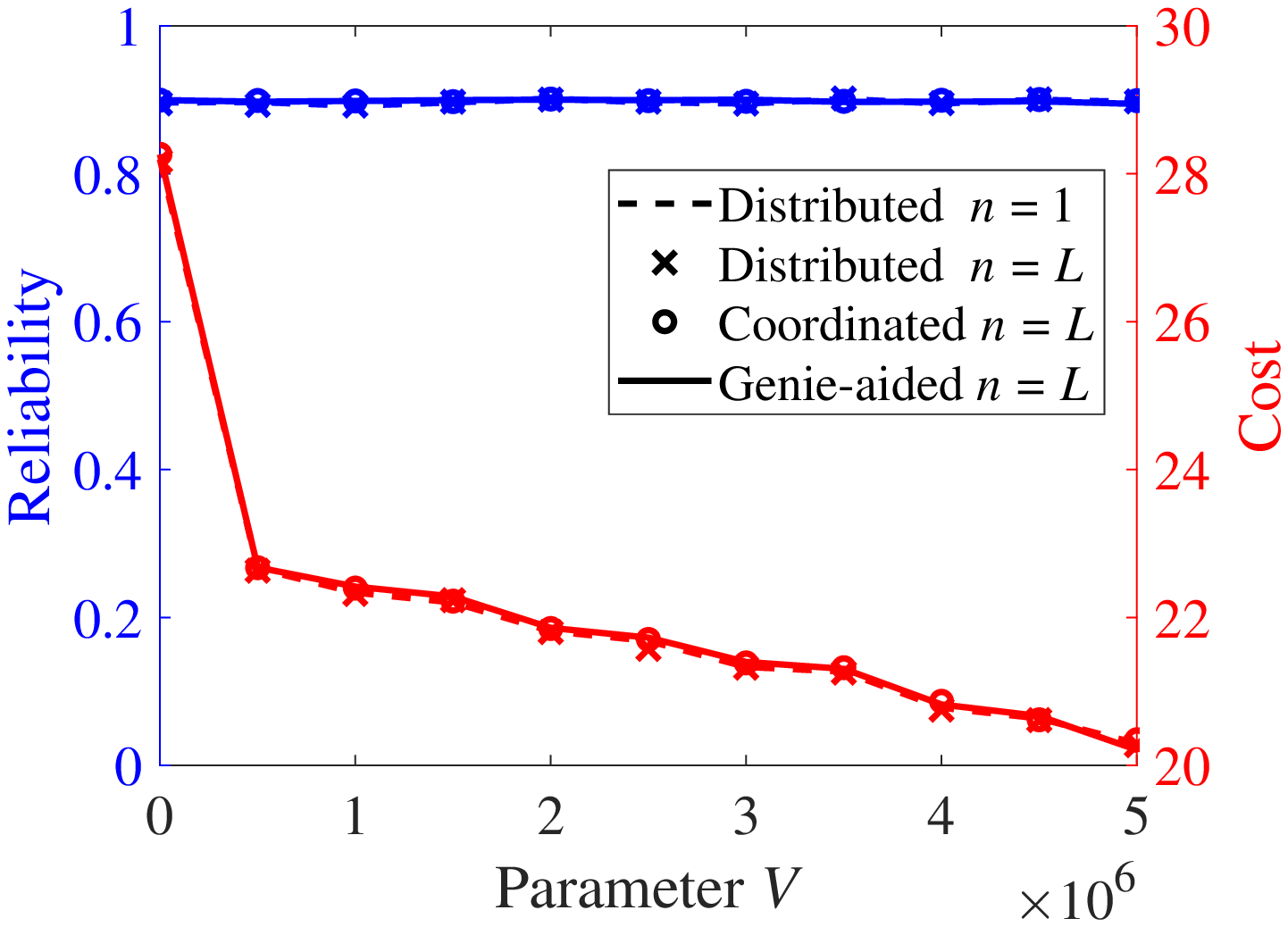}
		\label{fig:approx_low_edge}
	}
	\hfill
	\subfloat[High-congestion regime ({\em mesh}).]{
		\includegraphics[width = .23 \linewidth]{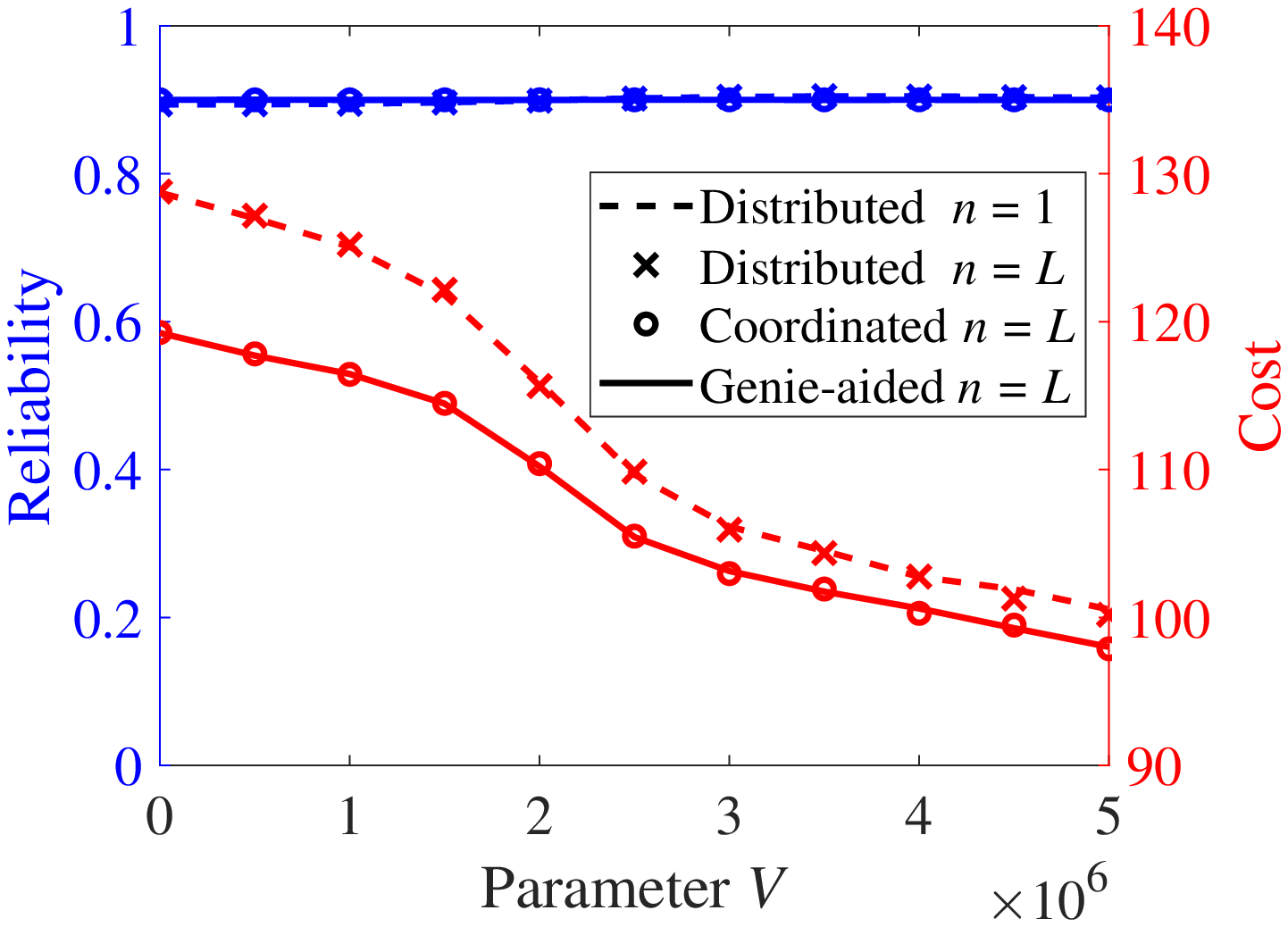}
		\label{fig:approx_high_edge}
	}
	\hfill
	\subfloat[Low-congestion regime ({\em hierarchical}).]{
		\includegraphics[width = .23 \linewidth]{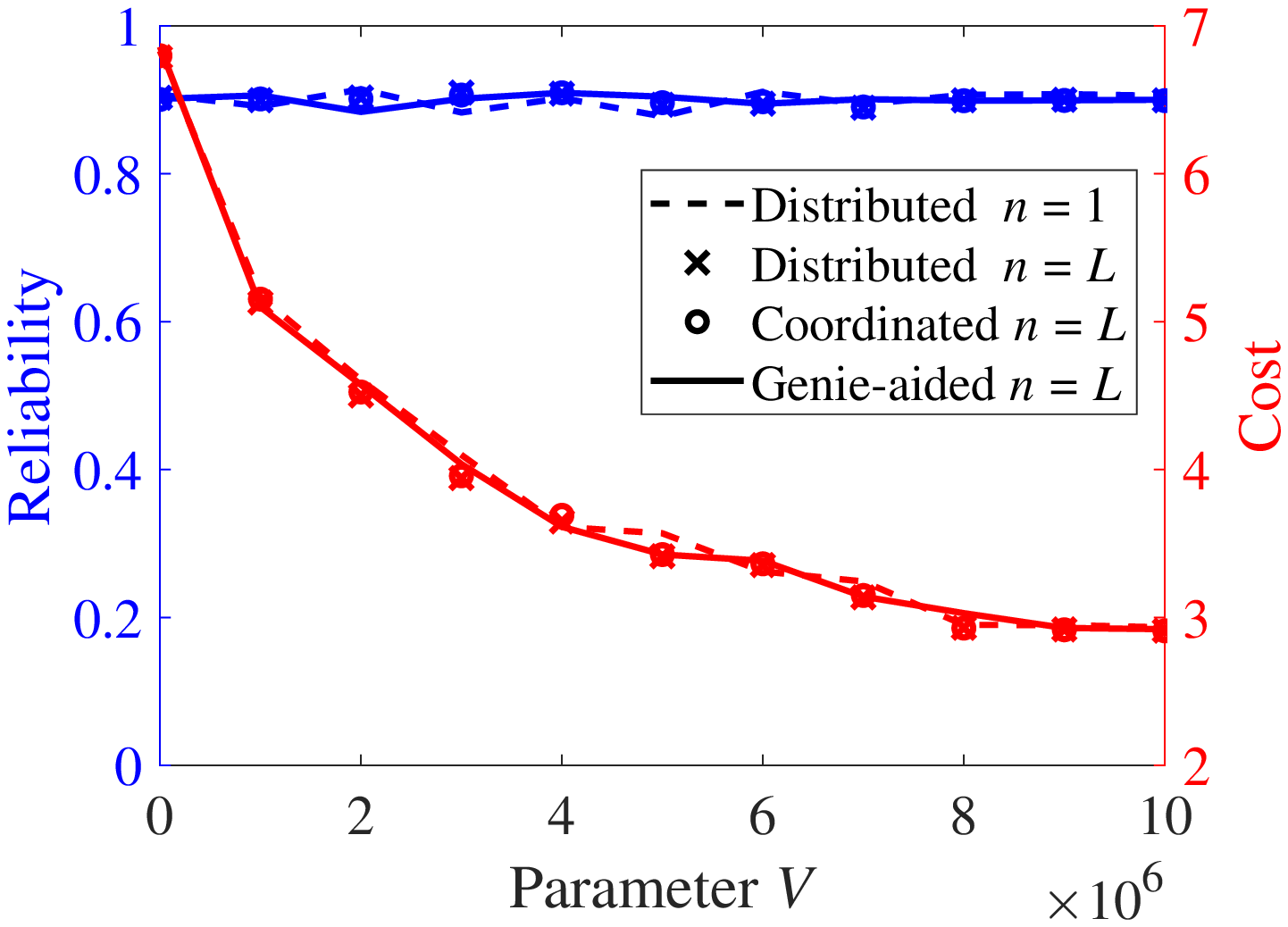}
		\label{fig:approx_low_fog}
	}
	\hfill
	\subfloat[High-congestion regime ({\em hierarchical}).]{
		\includegraphics[width = .23 \linewidth]{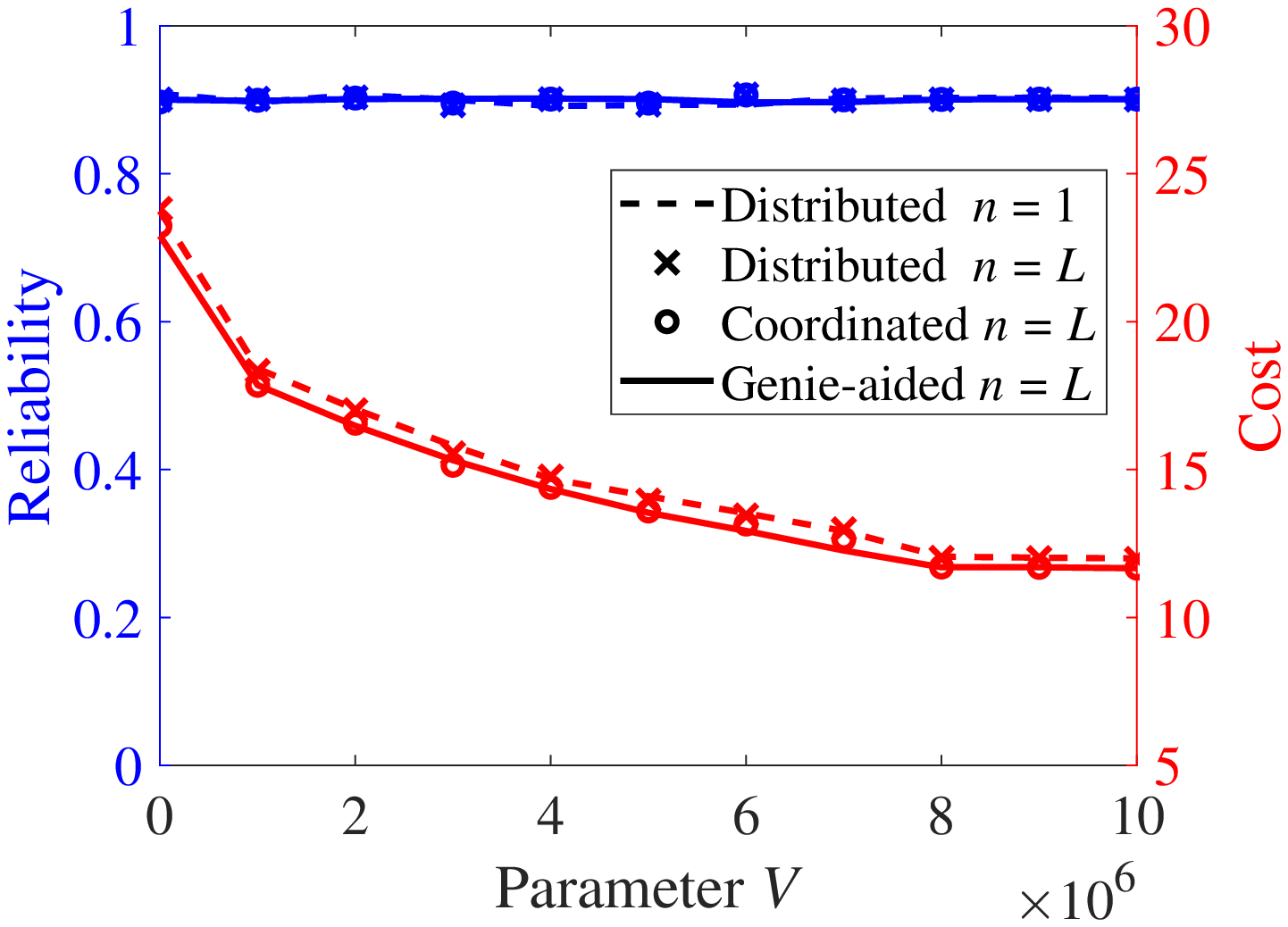}
		\label{fig:approx_high_fog}
	}
	\caption{
		Performances attained by RCNC under different configurations.
	}
	\label{fig:approx}
\end{figure*}

\subsubsection{Stability Region}
\label{sec:agi_stability_region}

In this section, we present the network stability regions achieved by the proposed algorithms, under different lifetime constraints and time slot lengths. We use a slot length of $1$ ms when conducting experiments for lifetime (Fig. \ref{fig:nsr_lifetime_edge} and \ref{fig:nsr_lifetime_fog}), and fix the delay constraint as $L = 50$ ms when studying the effect of slot length (Fig. \ref{fig:nsr_slotlength_edge} and \ref{fig:nsr_slotlength_fog}).

Fig. \ref{fig:nsr_lifetime_edge} and \ref{fig:nsr_lifetime_fog} depict the effect of lifetime, and we make following observations. First, the stability region enlarges with more available lifetimes, since packets can explore more network locations for additional computational resource; in particular, Fig. \ref{fig:nsr_lifetime_fog} saturates at $\Delta L = 4$ because the bottleneck links are constrained by the transmission limits. Second, the gap between the stability regions of average- and peak-constrained networks, is not significant (around $7\%$ for {\em mesh} and $5\%$ for {\em hierarchical}).%
\footnote{
	We emphasize that RCNC does NOT guarantee to achieve the entire stability region in the peak-constrained case. In other words, the exact stability region in the peak-constrained case lies between the blue and red curves.
}
Finally, by comparing $n = 1$ and $n = L$, we find that including more look-ahead slots can benefit the throughput performance (by around $10\%$), while resulting in a higher complexity of $\mathcal{O}(n^2 L^2) = \mathcal{O}(L^4)$.

Next, we tune the time slot length to study its impact on the stability region, with the results shown in Fig. \ref{fig:nsr_slotlength_edge} and \ref{fig:nsr_slotlength_fog}. As we increase the slot length, the attained throughput in {\em mesh} starts to degrade when it exceeds $3$ ms, since the resulting maximum lifetime $L\ (\leq 12)$ is not admissible to support the delivery of some services; while the throughput remains unchanged until a slot length of $8$ ms in {\em hierarchical} due to its simpler topology. On the other hand, a larger slot length can accelerate the algorithm: for $n = 1$, as we increase the slot length from $1$ to $10$ ms, the running time for decision making reduces: $52.8$, $13.5$, $6.6$, $4.1$, $2.7$, $2.2$, $1.8$, $1.5$, $1.2$, $0.94$ in {\em mesh}, and $27.2$, $7.3$, $3.7$, $2.4$, $1.6$, $1.4$, $1.2$, $0.99$, $0.83$, $0.68$ in {\em hierarchical} (in milliseconds).%
\footnote{
	Results are obtained using MATLAB 2021a running on a $3.2$ GHz computer, which leaves room for improvement, e.g., using a commercial solver and/or a faster processor.
}

\subsubsection{Throughput and Cost}

In this experiment, we compare the timely throughput and operational cost attained by RCNC with two benchmark algorithms:
\begin{itemize}
	\item DCNC \cite{FenLloTulMol:J18a}, which is shown to achieve optimal throughput and (near-optimal) cost performances, combined with last-in-first-out (LIFO) scheduling \cite{HuaMoeNeeKri:J13};
	\item an opportunistic scheduling algorithm \cite{Nee:C11} that provides worst-case delay guarantees for hop-count-limited transmissions, using the following parameters (notations are in line with \cite{Nee:C11}): $g_m(x) = x$, $\beta = \nu_m = 1$, $A_m^{\max} = 1.25 \lambda$, $D_n^{(m), \max} = \max\{ \epsilon, 1_n^{(m)} A_m^{\max} + \mu_{n}^{\max, in} \}$, with $\epsilon$ found by grid search to optimize the timely throughput.
\end{itemize}
We assume $\lambda = 500$ Mbps for {\em mesh} and $200$ Mbps for {\em hierarchical}, and Fig. \ref{fig:compare_bp} depicts the achieved throughput and cost under different lifetimes $\Delta L$ and $V$ values.

First, we focus on the reliability (or timely throughput) attained by the algorithms. The proposed RCNC algorithm can achieve a reliability level of $90\%$ under any $\Delta L$ and $V$ that meets the requirement of the services (and the reliability constraint \eqref{eq:p0_reliability} holds with equality). For DCNC, although it proves to be throughput optimal, the attained {\em timely} throughput is much lower, which increases with $\Delta L$ since more packets are counted as effective under a more relaxed lifetime constraint.%
\footnote{
	As $\Delta L \to \infty$, timely throughput converges to throughput, and DCNC can achieve a reliability level of $100\%$ since it is throughput optimal.
}
The worst-case delay algorithm \cite{Nee:C11} behaves slightly better than DCNC; however, there is still a considerable gap between the attained reliability and the imposed requirement (always guaranteed by the proposed RCNC algorithm), especially when the deadline constraint is stringent.

Next, we compare the operational cost of RCNC and DCNC.%
\footnote{
	The worst-case delay algorithm \cite{Nee:C11} is excluded from the comparison, because (i) it does not optimize the operational cost (and the attained costs are around 70 for {\em mesh} and 30 for {\em hierarchical}), and (ii) in contrast to the other two algorithms, the parameter $V$ has an essentially different interpretation.
}
As shown in Fig. \ref{fig:V_2} and \ref{fig:V_2_fog}, the operational costs of both algorithms reduce as $V$ increases. When $V$ is small, the cost of DCNC is significantly higher, since it might deliver packets through cyclic routes; while RCNC can reduce the number of extra transmissions due to the deadline constraint. Second, in Fig. \ref{fig:L_2} and \ref{fig:L_2_fog}, as we relax the deadline constraint (or increase $\Delta L$), RCNC can achieve better cost performance, since packets can ``detour'' to cheaper network locations for processing; while packet lifetime is not relevant in DCNC, and its cost performance stays constant (which is near-optimal). Last but not least, we note that the operational cost of RCNC is lower than DCNC, because DCNC delivers {\em all} the packets to the destination; while RCNC only delivers {\em effective} packets to meet the reliability requirement.

\subsubsection{Performance of RCNC}
\label{sec:apx_alg}

Finally, we compare the performance of RCNC with different implementations: using $n = 1$ or $L$ look-ahead slots (see Remark \ref{remark:n}), centralized or distributed decision making (see Remark \ref{remark:distributed}). A genie-aided algorithm serves as the benchmark, which can (by assumption) use {\em accurate} future arrival information to calculate $\V{A}_i$ \eqref{eq:delay_compact} for $n = L$ look-ahead slots. Assume $\Delta L = 2$.

As we can observe from Fig. \ref{fig:approx}, in the low-congestion regime ($\lambda = 20\%$ of the stability region, Fig. \ref{fig:approx_low_edge} and \ref{fig:approx_low_fog}), the four implementations achieve comparable performance in both {\em mesh} and {\em hierarchical} scenarios. However, when the network traffic becomes heavier ($\lambda = 80\%$ of the stability region, Fig. \ref{fig:approx_high_edge} and \ref{fig:approx_high_fog}), the two distributed algorithms (using $n=1$ or $L$) achieve sub-optimal cost performance (while still satisfying the reliability constraint); in contrast, the centralized algorithm remains robust, and the difference of its cost performance compared to the genie-aided algorithm is negligible. The result shows the importance of coordinated decision making among the nodes in the high-congestion regime to preserve the optimality of the solution; yet, it also motivates the use of the simplified algorithm ($n = 1$) in practical systems, especially for networks with simpler topologies (such as {\em hierarchical}), which can achieve sub-optimal performance with greatly reduced computational complexity.

\section{Extensions}
\label{sec:extensions}

In this section, we briefly discuss flexible extensions to the proposed approach in order to handle scenarios of practical relevance.

\subsection{Mixed Deadline-Constrained and Unconstrained Users}

It is flexible to combine the proposed approach with existing queuing techniques \cite{Nee:B10} in order to treat hybrid scenarios that include both deadline-constrained and unconstrained users \cite{neely2013MDPdelay}. To be specific, we can establish a queuing system that is a hybrid of the proposed lifetime queues for the constrained users, and standard queues (i.e., without lifetime structure) for unconstrained users. As shown in Appendix \ref{apdx:hybrid_queues}, the decisions for the two groups of users are {\em loosely} coupled, where unconstrained users follow the {\em max-weight} rule \cite{Nee:B10} for scheduling, and interact with constrained users via {\em one} additional variable for each link and look-ahead slot that represents the entire group, regardless of the number of unconstrained users.

\subsection{Time-Varying Slot Length}

While the technique developed in this paper assumes a fixed slot length, it is also possible to adopt a varying slot length via ``lifetime mapping''. In principle, when the slot length changes, we can construct a new queuing system, assign packets to queues of corresponding lifetimes, and map the decisions produced by the original policy to suite the new slot length. For example, if the slot length changes from $1$ ms to $2$ ms, we can add up the decisions (i.e., transmitted flows) for lifetime $2k-1$ and $2k$ packets to obtain the decision for lifetime $k$ packets based on the new slot length. The design of the mapping functions and associated performance loss analysis are topics worth further investigation.

\section{Conclusions} \label{sec:conclusion}

In this paper, we investigated the delay-constrained least cost dynamic network control problem. We established a new queuing system to keep track of data packets' lifetime, based on which we formalized the problem $\mathscr{P}_0$. To find an efficient approximate solution to this challenging problem, we first derived a relaxed problem with average capacity constraints $\mathscr{P}_1$ and designed a fully distributed, near-optimal solution that matches the \ac{ldp} assigned flow on an equivalent virtual network $\mathscr{P}_2$. The methodology was then extended to solve $\mathscr{P}_0$, where we proposed a two-way optimization approach in order to use the assigned flow in $\mathscr{P}_2$ to guide the flow solution to $\mathscr{P}_0$. Extensive numerical results were presented to validate the analytical results, illustrate the performance gain, and guide the system configuration.

\ifCLASSOPTIONcompsoc
	\section*{Acknowledgments}
\else
	\section*{Acknowledgment}
	The authors would like to thank Prof. Michael J. Neely for helpful discussions.
\fi

\ifCLASSOPTIONcaptionsoff
  \newpage
\fi






\bibliographystyle{IEEEtran}
\bibliography{IEEE_abrv,AgI}

%










\cleardoublepage

\appendices

\section{Proof for Proposition \ref{prop:causality_constraint}}\label{apdx:proof_causality}

Consider any sample path (the randomness comes from both the arrival process and the policy). Denote by $\V{x}(t)$ the decisions made by the policy, which satisfies the availability constraint \eqref{eq:p1_availability}, i.e.,
\begin{align}
x_{i\to}^{(l)}(t) \leq Q_i^{(l)}(t),\ \forall i \in \Set{V},\ l\in \Set{L}.
\end{align}

Specially, we focus on the intermediate nodes $i\in \Set{V}\setminus\{d\}$. Recall the queuing dynamics \eqref{eq:queue_dynamics_1}, and sum up the equations of $\ell = l,\cdots, L$, which leads to
\begin{align}\begin{split} \label{eq:q_geq_1}
Q_i^{(\geq l)}(t+1) & = Q_i^{(\geq l+1)}(t) \\
& \quad + x_{\to i}^{(\geq l+1)}(t) - x_{i\to}^{(\geq l+1)}(t) + a_i^{(\geq l)}(t)
\end{split}\end{align}
where $Q_i^{(\geq l)}(t) \triangleq \sum_{\ell = l}^{L}{ Q_i^{(\ell)}(t) }$ (and the other terms are defined in the same way). By definition
\begin{align} \label{eq:q_geq_2}
Q_i^{(\geq l)}(t+1) = Q_i^{(l)}(t+1) + Q_i^{(\geq l+1)}(t+1).
\end{align}

By availability constraint \eqref{eq:p0_availability}, we know
\begin{align}
Q_i^{(l)}(t+1) \geq x_{i\to}^{(l)}(t+1).
\end{align}
Applying the above relationship to \eqref{eq:q_geq_2}, combining with \eqref{eq:q_geq_1} and rearranging the terms, we can obtain
\begin{align}\begin{split}
\hspace{-.2cm} Q_i^{(\geq l+1)}(t+1) - Q_i^{(\geq l+1)}(t) \leq \big[ x_{\to i}^{(\geq l + 1)}(t) + a_i^{(\geq l)}(t) \big] \\
- \big[ x_{i\to}^{(l)}(t+1) + x_{i\to}^{(\geq l + 1)}(t) \big].
\end{split}\end{align}

The telescoping sum is then applied to the above inequality \cite{Nee:B10}. Fix some $T$, sum the inequalities of $t = 0,\cdots, T-1$, and the result is (w.l.o.g., assume $\V{Q}(0) = \V{0}$)
\begin{align} \label{eq:telescope}
\begin{split}
Q_i^{(\geq l+1)}(T) & \leq \sum_{t=0}^{T-1}\big[ x_{\to i}^{(\geq l + 1)}(t) + a_i^{(\geq l)}(t) \big] \\
& \quad - \sum_{t=0}^{T-1}\big[ x_{i\to}^{(l)}(t+1) + x_{i\to}^{(\geq l + 1)}(t) \big]
\end{split}
\end{align}
Some standard operations are performed, including taking expectation (since the above inequality holds for each sample path), dividing by $T$, and pushing $T\to\infty$. Use the fact that $\V{Q}(t)$ is bounded (due to the assumptions of bounded arrival and packet drop), and thus stable, and we can obtain
\begin{align}
\begin{split}
0 & \leq \avg{ \E{ x_{\to i}^{(\geq l + 1)}(t) } } + \lambda_i^{(\geq l)}(t) \\
& \quad - \avg{ \E{ x_{i\to}^{(l)}(t+1) + x_{i\to}^{(\geq l + 1)}(t) } }.
\end{split}
\end{align}
Furthermore, we can replace $x_{i\to}^{(l)}(t+1)$ by $x_{i\to}^{(l)}(t)$ in the second line because the two expressions lead to identical long-term average (just one time slot shift), and the two terms on the right-hand-side can be combined, i.e.,
\begin{align}
\avg{ \E{ x_{\to i}^{(\geq l + 1)}(t) } } +  \lambda_i^{(\geq l)}
\geq \avg{ \E{ x_{i\to}^{(\geq l)}(t) } }
\end{align}
which is true for all $l = 1,\cdots, L$. It is also true for $l = 0$, but the relationship is trivial (which is implied by that of $l = 1$). This is the causality constraint \eqref{eq:p2_causality}.

\section{Stability region of $\mathscr{P}_1$}\label{apdx:cap_p1}

\subsection{Necessity} \label{apdx:p1_necessity}

Suppose $(\V{\lambda},\gamma) \in \Lambda_1$. By definition there exists a policy under which the constraints \eqref{eq:p1_reliability} -- \eqref{eq:p1_queue} are satisfied, while achieving the optimal cost. Denote by $X_{ij}^{(l)}(t) \geq 0$ the number of successfully delivered packets within the first $t$ time slots, which are transmitted on link $(i, j)$ at lifetime $l$, and define
\begin{align}
X_{ij}(t) = \sum_{l\in \Set{L}} X_{ij}^{(l)}(t).
\end{align}
In addition, denote by $A_i^{(l)}(t)$ the total number of lifetime $l$ packets that arrives at node $i$ during the first $t$ time slots.

The basic facts are listed in the following
\begin{subequations}
\begin{align}
\sum_{j\in \delta_d^-} X_{jd}(t) + A_d(t) & \geq \gamma \sum_{i\in \Set{V}} A_i(t) \\
\lim_{t\to\infty} \frac{X_{ij}(t)}{t} & \leq C_{ij} \label{eq:avg_cap} \\
\sum_{j\in \delta_i^-}{ X_{ji}^{(\geq l+1)} (t) } + A_i^{(\geq l)}(t) & \geq \sum_{j\in \delta_i^+}{ X_{ij}^{(\geq l)}(t) } \label{eq:causality_1} \\
X_{ij}^{(0)}(t) = X_{dj}^{(l)}(t) & = 0
\end{align}
\end{subequations}
where \eqref{eq:avg_cap} holds due to the average capacity constraint, and \eqref{eq:causality_1} is implied by \eqref{eq:telescope} (using the fact $Q_i^{(\geq l+1)}(T) \geq 0$).
Divide by $t$, take the limit $t\to\infty$, and note that
\begin{align}
x_{ij} \triangleq \lim_{t\to\infty} \frac{X_{ij}(t)}{t},\ \lim_{t\to\infty} \frac{A_i^{(l)}(t)}{t} = \lambda_i^{(l)}.
\end{align}
Then we obtain the characterization \eqref{eq:capacity_region}.

\subsection{Sufficiency}

For a given pair $(\V{\lambda},\gamma)$, suppose there exists a flow assignment $\V{x}$ that satisfies \eqref{eq:capacity_region}. We will prove $(\V{\lambda},\gamma) \in \Lambda_1$ by constructing a randomized policy, and showing that the achieved flow assignment equals $\V{x}$.

\subsubsection{Randomized Policy} \label{sec:random_1}

The policy is construct as follows. For any node $i\in \Set{V}$ and lifetime $l\in \Set{L}$, we define a set of probability values $\big\{ \alpha_i^{(l)}(j) : j\in \delta_i^+ \big\}$ with
\begin{align}\label{eq:prob_1}
\alpha_i^{(l)}(j) = \Big( \frac{ x_{i\to}^{(l)} }{ \tilde{x}_{\to i}^{(\geq l)} - x_{i\to}^{(\geq l+1)} } \Big) \Big( \frac{ x_{ij}^{(l)} }{ x_{i\to}^{(l)} } \Big)
\end{align}
(the expression can be simplified, while we preserve this form for ease of exposition), where
\begin{align}
\tilde{x}_{\to i}^{(l)} = x_{\to i}^{(l+1)} + \lambda_i^{(l)},\text{ and }
\tilde{x}_{\to i}^{(\geq l)} = x_{\to i}^{(\geq l+1)} + \lambda_i^{(\geq l)}
\end{align}
We claim that this is a set of valid probability values, since by \eqref{eq:flow_conserve}, we have $\tilde{x}_{\to i}^{(\geq l)} \geq x_{i\to}^{(\geq l)} = x_{i\to}^{(l)} + x_{i\to}^{(\geq l+1)}$, and thus%
\footnote{
	If the sum is strictly less than $1$, the complementary part corresponds to the {\em idle} operation (i.e., the packet is not scheduled for transmission at the current time slot).
}
\begin{align}\label{eq:aggregate_prob_1}
\alpha_i^{(l)} \triangleq \sum_{j\in \delta_i^+} \alpha_i^{(l)}(j)
= \frac{ x_{i\to}^{(l)} }{ \tilde{x}_{\to i}^{(\geq l)} - x_{i\to}^{(\geq l+1)} } 
\leq 1.
\end{align}

The developed policy $*$ operates in the following way: at every time slot, node $i$ makes independent transmission decisions for each packet of lifetime $l$ (i.e., to transmit or not; if yes, through which outgoing interface) according to the \ac{pdf} $\big\{ \alpha_i^{(l)}(j) : j \in \delta_i^+ \big\} \cup \{1 - \alpha_i^{(l)}\}$, where the complement part $1 - \alpha_i^{(l)}$ accounts for the idle operation.

\subsubsection{Validate the Constraints} \label{apdx:flow_1}

Next, we show that the decisions made by this policy satisfy \eqref{eq:problem_1}.

Since the policy makes independent decision for each packet in the queue, the availability constraint \eqref{eq:p1_availability} is satisfied. Next, we prove that the decisions $\V{\mu}(t)$ made by this policy satisfy the remaining constraints, by showing $\avg{ \E{ \V{\mu}(t) } } = \V{x}$.

The decision of policy $*$ can be decomposed into two steps, including 1) whether to transmit the packet or not, 2) if yes, through which interface $j$. For the first step, define $\alpha_i^{(\ell, l)}$ as the probability that a packet of lifetime $\ell$ (when first loaded to the queuing system at node $i$) gets transmitted at the lifetime of $l$. By definition, $\alpha_i^{(\ell, l)} = \alpha_i^{(\ell)}$ when $l = \ell$.

We first present some useful results for the following proofs. The transition probability follows the recurrent formula
\begin{align}
\alpha_i^{(\ell, l)} = \big( 1 - \alpha_i^{(\ell)} \big) \alpha_i^{(\ell - 1, l)};
\end{align}
on the other hand, for $\forall\, k \leq \ell - 2$
\begin{align}\begin{split}
\alpha_i^{(\ell, k)} & = \prod_{l=k+1}^{\ell }\big( 1 - \alpha_i^{(l)} \big) \alpha_i^{(k)} \\
& = \bigg[ \prod_{l=k+2}^{\ell}\big( 1 - \alpha_i^{(l)} \big) \alpha_i^{(k+1)} \Big( \frac{1 - \alpha_i^{(k+1)}}{\alpha_i^{(k+1)}} \Big) \bigg] \alpha_i^{(k)} \\
& = \alpha_i^{(\ell, k+1)} \Big( \frac{ 1 - \alpha_i^{(k+1)} }{ \alpha_i^{(k+1)} } \Big) \alpha_i^{(k)}
\end{split}\end{align}
and the above relationship also holds for $k = l-1$.

\begin{lemma}
The probability $\V{\alpha}$ defined as \eqref{eq:prob_1} leads to the following relationship
\begin{align}\label{eq:average_flow_1}
x_{i\to}^{(l)} = \sum_{\ell \geq l}{ \alpha_i^{(\ell, l)} \tilde{x}_{\to i}^{(\ell)} }.
\end{align}
\end{lemma}

\begin{IEEEproof}
We give a proof by induction on $l$.

{\em Base case:} we show the result holds for $l = L$. By \eqref{eq:aggregate_prob_1},
\begin{align}
\sum_{\ell \geq L}{ \alpha_i^{(\ell, L)} \tilde{x}_{\to i}^{(\ell)} } = \alpha_i^{(L)} \tilde{x}_{\to i}^{(L)}
= \frac{ x_{i\to}^{(L)} }{ \tilde{x}_{\to i}^{(\geq L)} - 0 } \, \tilde{x}_{\to i}^{(L)}
= x_{i\to}^{(L)}.
\end{align}

{\em Inductive step:} Assume that the result holds for $l = k+1$, and next we show it also holds for $l = k$. Note that
\begin{subequations}
\begin{align}
& \quad \sum_{\ell \geq k}{ \alpha_i^{(\ell, k)} \tilde{x}_{\to i}^{(\ell)} }
= \alpha_i^{(k)} \tilde{x}_{\to i}^{(k)} + \sum_{\ell \geq k+1}{ \alpha_i^{(\ell, k)} \tilde{x}_{\to i}^{(\ell)} } \\
& = \alpha_i^{(k)} \Big[ \tilde{x}_{\to i}^{(k)} + \Big( \frac{ 1 - \alpha_i^{(k+1)} }{ \alpha_i^{(k+1)} } \Big) \sum_{\ell \geq k+1}{ \alpha_i^{(\ell, k+1)}  \tilde{x}_{\to i}^{(\ell)} } \Big] \label{eq:induction_1} \\
& = \alpha_i^{(k)} \Big[ \tilde{x}_{\to i}^{(k)} + \Big( \frac{ \tilde{x}_{\to i}^{(\geq k+1)} - x_{i\to}^{(\geq k+1)} }{ x_{i\to}^{(k+1)} } \Big) x_{i\to}^{(k+1)} \Big] \label{eq:induction_2} \\
& = \alpha_i^{(k)} \big( \tilde{x}_{\to i}^{(\geq k)} - x_{i\to}^{(\geq k+1)} \big) = x_{i\to}^{(k)}
\end{align}
\end{subequations}
where we use the result of $l = k+1$, and substitute the definition \eqref{eq:aggregate_prob_1} into \eqref{eq:induction_1} to derive \eqref{eq:induction_2}.

To sum up, by mathematical induction, the expression \eqref{eq:average_flow_1} holds, and this concludes the proof.
\end{IEEEproof}

On the other hand, an equivalent way to describe the randomized policy $*$ is, as soon as each packet (of lifetime $\ell$) arrives, we make the decision of {\em how many time slots to wait before transmission} (and through which interface); more concretely, the probability for the wait time of $\tau$ is $\alpha_i^{(\ell, \ell - \tau)}$. By exploiting this interpretation, for node $i$, the amount of outgoing packets of lifetime $l$ at time slot $t$ can be those of lifetime $\ell + 1\ (\ell\geq l)$ at $t' = t - (\ell + 1 - l)$ (where $\ell$ is the lifetime of the packet when it is first loaded into the queuing system of node $i$), leading to the following relationship
\begin{align}\begin{split}
\E{\mu_{i\to}^{(l)}(t)} & = \mathbb{E} \Big\{ \sum_{\ell \geq l} \alpha_i^{(\ell, l)} \Big( \mu_{\to i}^{(\ell + 1)}(t') + a_i^{(\ell)}(t') \Big) \Big\} \\
& = \sum_{\ell \geq l} \alpha_i^{(\ell, l)} \Big( \E{ \mu_{\to i}^{(\ell + 1)}(t') } + \lambda_i^{(\ell)} \Big).
\end{split}\end{align}
By taking long-term average, the effect of finite (at most $L$) time slot-shift vanishes, and we obtain
\begin{align}\label{eq:average_flow_2}
\hspace{-.2 cm}\avg{ \E{\mu_{i\to}^{(l)}(t)} } = \sum_{\ell \geq l} \alpha_i^{(\ell, l)} \Big( \avg{ \E{ \mu_{\to i}^{(\ell + 1)}(t) } } + \lambda_i^{(\ell)} \Big).
\end{align}
By comparing \eqref{eq:average_flow_1} (where $\tilde{x}_{\to i}^{(\ell)} = x_{\to i}^{(\ell+1)} + \lambda_i^{(\ell)}$) and \eqref{eq:average_flow_2}, we find that $\avg{ \E{ \V{\mu}(t) } }$ and $\V{x}$ are the solutions to the same linear system, which implies
\begin{align} \label{eq:flow_matching_1}
\avg{ \E{\mu_{ij}^{(l)}(t)} } = x_{ij}^{(l)}.
\end{align}

Thus we can replace $\V{x}$ by $\avg{ \E{ \V{\mu}(t) } }$, in all the constraints \eqref{eq:capacity_region} it satisfies (especially, \eqref{eq:cr_reliability} and \eqref{eq:cr_capacity}). Therefore, we can conclude that policy $*$ is an admissible policy for $\mathscr{P}_1$.

\subsubsection{Optimality} \label{apdx:optimality}

For any point within the stability region, assume $\V{x}$ is the flow assignment corresponding to the cost-optimal policy. We can construct the randomized policy by the procedure introduced in Appendix \ref{sec:random_1}, which leads to \eqref{eq:flow_matching_1}. Since the operational cost is a linear function of the flow assignment, the designed randomized policy achieves the same objective value as the cost-optimal policy.

\section{Stability region of $\mathscr{P}_2$}\label{apdx:cap_p2}

\subsection{Necessity}

Suppose $(\V{\lambda},\gamma) \in \Lambda_2$. By definition there exists a policy under which the constraints \eqref{eq:p1_reliability} -- \eqref{eq:p1_queue} are satisfied, while achieving the optimal cost. Define $\V{X}(t)$ as in Appendix \ref{apdx:p1_necessity}.

The basic facts are listed in the following
\begin{subequations}
\begin{align}
\sum_{j\in \delta_d^-} X_{jd}(t) + A_d(t) & \geq \gamma \sum_{i\in \Set{V}} A_i(t) \\
X_{ij}(t) & \leq C_{ij} t \\
\lim_{t\to \infty} \bigg[ \sum_{j\in \delta_i^-}{ \frac{X_{ji}^{(\geq l+1)}(t)}{t} } + \frac{A_i^{(\geq l)}(t)}{t} \bigg] \hspace{-.4in} \nonumber\\
& \geq \lim_{t\to \infty} \sum_{j\in \delta_i^+}{ \frac{X_{ij}^{(\geq l)}(t)}{t} } \label{eq:causality_2} \\
X_{ij}^{(0)}(t) = X_{dj}^{(l)}(t) & = 0
\end{align}
\end{subequations}
where \eqref{eq:causality_2} holds due to the causality constraint. Divide by $t$, take the limit $t\to\infty$, and note that
\begin{align}
x_{ij} \triangleq \lim_{t\to\infty} \frac{X_{ij}(t)}{t},\ \lim_{t\to\infty} \frac{A_i^{(l)}(t)}{t} = \lambda_i^{(l)}.
\end{align}
Then we obtain the characterization \eqref{eq:capacity_region}.

\subsection{Sufficiency}

For a given pair $(\V{\lambda},\gamma)$, suppose there exists a flow assignment $\V{x}$ that satisfies \eqref{eq:capacity_region}. We will prove $(\V{\lambda},\gamma) \in \Lambda_2$ by constructing a stationary randomized policy, and show that the achieved flow assignment equals $\V{x}$.

\subsubsection{Randomized Policy} \label{sec:random_2}

For any link $(i, j)\in \Set{E}$, define a set of probability values $\big\{ \alpha_{ij}^{(l)} : l\in \Set{L} \big\}$, with
\begin{align}
\alpha_{ij}^{(l)} \triangleq x_{ij}^{(l)} \big/ C_{ij}.
\end{align}
We claim that this is a set of valid probability values, since they sum up $\leq 1$ due to \eqref{eq:cr_capacity}.%
\footnote{
	If the sum is strictly less than $1$, the complementary part corresponds to the {\em idle} operation (i.e., no packet is transmitted).
}

The policy $*$ operates in the following way: at every time slot, for each link $(i, j)$, choose a lifetime $l^*$ according to the probability values $\big\{ \alpha_{ij}^{(l)} : l \in \Set{L} \big\}$ independently. The assigned flow is given by $\nu_{ij}^{(l)} = C_{ij} \, \mathbb{I}\{ l \equiv l^* \}$, i.e., we borrow $C_{ij}$ lifetime $l^*$ packets from the reservoir to transmit on link $(i, j)$.

\subsubsection{Validate the Constraints}

Next, we show that this policy makes decisions satisfying \eqref{eq:problem_2}.

Since the decisions are made in an i.i.d. manner over time slots, for any time slot, we have
\begin{align}\label{proof:2}
\E{\nu_{ij}^{(l)}(t)} = \alpha_{ij}^{(l)} C_{ij} = x_{ij}^{(l)}.
\end{align}
The above equation still holds when we take long-term average of it (since it is true for all time slots), i.e.,
\begin{align}
\avg{ \E{\nu_{ij}^{(l)}(t)} } = x_{ij}^{(l)}.
\end{align}
Thus we can replace $\V{x}$ by $\avg{ \E{\V{\nu}(t)} }$, in all the constraints \eqref{eq:capacity_region} it satisfies (specially, \eqref{eq:cr_reliability} and \eqref{eq:flow_conserve}). Therefore, we can conclude that policy $*$ is an admissible policy for $\mathscr{P}_2$.

\subsubsection{Optimality}

The argument is the same as Appendix \ref{apdx:optimality} (the only difference is to construct the randomized policy according to Appendix \ref{sec:random_2}).

\section{Distribution of $x_{ij}(t)$} \label{apdx:poisson}

In the studied packet routing problem (where flow scaling is not relevant), we make the following assumptions:
(i) the arrival process of packets with any lifetime is Poisson,
(ii) the randomized policies of the nodes are not time varying.
We will show: under the above assumptions, $x_{ij}(t)$ follows Poisson distribution for $\forall\,(i, j),\, t$.

We present two facts that will be used in the derivation.
{\bf Fact A}: Define $X = \sum_{k=1}^{N} X_k$, where $N$ is a Poisson random variable (r.v.), and $X_k$ are i.i.d. Bernoulli r.v.s, then $X$ is a Poisson r.v..
{\bf Fact B}: Suppose $X_k$'s are i.i.d. Poisson r.v.s, then $X = \sum_{k=1}^{n} X_k$ (where $n$ is a constant) is a Poisson r.v..

Note that the total flow size is the sum over flows of individual lifetimes:
\begin{align} \label{eq:outgoing_flow}
	x_{ij}(t) & = \sum_{l = 1}^{L} x_{ij}^{(l)}(t)
	= \sum_{l = 1}^{L} \Big[ \sum_{s\in \Set{V}} \sum_{l_0 = l}^{L} y_s^{(l_0)}( t - (l_0 - l) ) \Big] \nonumber \\
	& = \sum_{s\in \Set{V}} \sum_{l_0 = 1}^L \sum_{l = 1}^{l_0} y_s^{(l_0)}( t - (l_0 - l) )
\end{align}
where we exchange the order of summations in the last equation, and define
\begin{align} \label{eq:total_flow}
	y_s^{(l_0)}( t - (l_0 - l) ) = \sum\nolimits_{k=1}^{ a_s^{(l_0)}( t - (l_0 - l) ) } y_k.
\end{align}
In the above expression, we define event $\mathscr{A}_{ij}(s, l_0, l)$: a packet, which is of initial lifetime $l_0$ when arriving at node $s$ at time slot $t-(l_0 - l)$, crosses link $(i, j)$ at time slot $t$ (when its lifetime is $l$); and $p_{ij}(s, l_0, l)$ as its probability (which is fixed given the randomized policies of all the nodes). Let $y_k$'s be Bernoulli r.v.s indicating whether event $\mathscr{A}_{ij}(s, l_0, l)$ is true for packet $k \in \{ 1,\cdots, a_s^{(l_0)}( t - (l_0 - l) ) \}$, and they are i.i.d. because the decisions for each packet are made independently.

Apply {\bf Fact A} to \eqref{eq:total_flow}, and we obtain that $y_s^{(l_0)}( t - (l_0 - l) )$ is a Poisson r.v.. Next, for different $s, l, l_0$: $a_s^{(l_0)}( t - (l_0 - l) ) $'s are independent by assumption, and thus $y_s^{(l_0)}( t - (l_0 - l) ) $'s are also independent; thus, $x_{ij}(t)$ is a Poisson r.v. by {\bf Fact B}, concluding the proof.

\section{Proof for Proposition \ref{prop:cost_V}} \label{apdx:converge_cost}

In this section, we analyze the performance of the proposed {\em max-weight} algorithm for the virtual network. Note that the goal is to find the decision $\V{\nu}(t)$ that minimizes the upper bound of the \ac{ldp} at any time slot $t$, and thus
\begin{align}\begin{split}
\Delta( \V{U}(t) ) + V h( \V{\nu}(t) )
& \leq B - \langle \tilde{\V{a}}, \V{U}(t) \rangle - \langle \V{w}(t), \V{\nu}(t) \rangle \\
& \leq B - \langle \tilde{\V{a}}, \V{U}(t) \rangle - \langle \V{w}(t), \V{\varphi}(t) \rangle
\end{split}\end{align}
where $\V{\varphi}(t)$ is the decision chosen by any other feasible policy; specifically, we assume that it is the randomized policy introduced in Appendix \ref{sec:random_2} for the point $(\V{\lambda} + (\epsilon' / \gamma) \V{1}, \gamma)$. The existence of $\epsilon' > 0$ is guaranteed by the assumption that $(\V{\lambda}, \gamma)$ lies in the interior of the stability region). To recall, the policy makes i.i.d. decisions at each time slot (independent with $\V{U}(t)$), while achieving the optimal cost for that point, denoted by $\E{ h_2(\V{\varphi}(t)) } = h_2^\star(\V{\lambda} + (\epsilon' / \gamma) \V{1}, \gamma) = \tilde{h}_2^\star$.

Take expectation, rearrange the terms on the right-hand-side of the inequality, and we obtain the result in \eqref{eq:ldp_bd}.
\begin{figure*}[ht]
\begin{align}\begin{split}\label{eq:ldp_bd}
\E{\Delta( \V{U}(t) ) + V h( \V{\nu}(t) )} & \leq B + V \E{ h_2(\V{\varphi}(t)) } - \E{ \big[ \varphi_{\to d}(t) - \gamma A(t) \big] U_d(t) } \\
& \quad - \sum_{i\in \Set{V}\setminus \{d\}} \sum_{l\in \Set{L}} \E{ \Big[ \varphi_{\to i}^{(\geq l+1)}(t)  + a_i^{(\geq l)}(t) - \varphi_{i\to}^{(\geq l)}(t) \Big] U_i^{(l)}(t) } \\
& \leq B + V \tilde{h}_2^\star - \big[ \E{ \varphi_{\to d}(t) } - \gamma \|\V{\lambda}\|_1 \big] \E{U_d(t)} \\
& \quad - \sum_{i\in \Set{V}\setminus \{d\}} \sum_{l\in \Set{L}}\Big[ \E{ \varphi_{\to i}^{(\geq l+1)}(t) } + \lambda_i^{(\geq l)} - \E{ \varphi_{i\to}^{(\geq l)}(t) } \Big] \E{ U_i^{(l)}(t) }
\end{split}\end{align}
\rule{\textwidth}{0.5pt}
\end{figure*}
The second inequality is obtained by the fact that $\V{\varphi}(t)$ and $\V{U}(t)$ are independent. Besides, since the policy is admissible, $\V{\varphi}(t)$ satisfies \eqref{eq:p2_reliability} -- \eqref{eq:p2_causality} (specifically, the reliability and causality constraint). It is straightforward to obtain
\begin{align}
\E{\varphi_{\to d}(t)} - \gamma \|\V{\lambda}\|_1 \geq \epsilon', \label{eq:reliability_e} \\
\E{ \varphi_{\to i}^{(\geq l+1)}(t) } + \lambda_i^{(\geq l)} - \E{ \varphi_{i\to}^{(\geq l)}(t) } \geq \epsilon' \label{eq:causality_e}.
\end{align}
Note that in the above expression, we omit the long-term average operator, since $\V{\varphi}(t)$ is i.i.d. at each time slot, and thus the long-term average equals the value of any time slot. Substituting \eqref{eq:reliability_e} and \eqref{eq:causality_e} into \eqref{eq:ldp_bd} leads to
\begin{align}\label{eq:ldp_bd2}
\E{\Delta( \V{U}(t) ) + V h( \V{\nu}(t) )}
\leq B + V \tilde{h}_2^\star - \epsilon' \| \V{U}(t) \|_1.
\end{align}

\subsection{Cost Performance}

Fix some $T > 0$. Apply the telescoping sum to \eqref{eq:ldp_bd2}, and w.l.o.g., assume $\V{U}(0) = \V{0}$. We obtain
\begin{align}\begin{split}
V \sum_{t=0}^{T-1} \E{h( \V{\nu}(t) )} & \leq BT + VT \tilde{h}_2^\star \hspace{1in}\\
& \hspace{-.8in} - \sum_{t=0}^{T-1} \epsilon' \|\V{U}(t)\|_1 - \frac{\E{ \|\V{U}(T)\|_2^2}}{2} + \frac{\|\V{U}(0)\|_2^2}{2} \\
& \leq BT + VT \tilde{h}_2^\star.
\end{split}\end{align}
Divide the inequality by $VT$ and push $T\to\infty$; besides, note that the above inequality holds for any $\epsilon' > 0$, and specially, a sequence $\{ \epsilon_n' \downarrow 0 \}$, which leads to
\begin{align}
\avg{\E{h( \V{\nu}(t) )} } \leq h_2^\star + \frac{B}{V}
\end{align}
which is \eqref{eq:achieved_cost}.

\subsection{$\varepsilon$-Convergence Time} \label{apdx:converge_cost_2}

First, we show that the proposed algorithm stabilizes the virtual queues. Similar to the previous section, we apply the telescoping sum for some fixed $T > 0$, and obtain
\begin{align}\begin{split}
\frac{\E{\| \V{U}(T) \|^2}}{2} & \leq BT + V T \tilde{h}_2^\star + \frac{\E{\| \V{U}(0) \|^2}}{2} \\
& \quad - V \sum_{t=0}^{T-1} \E{h( \V{\nu}(t) )} - \epsilon' \| \V{U}(t) \|_1 \\
& \leq (B + V \tilde{h}_2^\star)T
\end{split}\end{align}
Furthermore, by the definition of the $\ell_2$-norm, the relationship $U_d(t) \leq \|\V{U}(t)\|_2$ always holds, and thus
\begin{align}\begin{split} \label{eq:virtual_q_bound}
\E{ U_d(T) } & \leq \E{ \| \V{U}(T) \|_2 } \leq \sqrt{ \E{ \| \V{U}(T) \|_2^2 } } \\
& \leq \sqrt{ 2(B + V \tilde{h}_2^\star )T }
\end{split}\end{align}
where the second inequality follows from the Cauchy-Schwartz inequality (or the fact that $\E{X}^2 \leq \E{X^2}$ for any random variable $X$). As a result, for any finite $V$,
\begin{align}
\lim_{T\to\infty} \frac{\E{ U_d(T) }}{T} \leq \lim_{T\to\infty} \sqrt{ \frac{2(B + V \tilde{h}_2^\star)}{T} } = 0
\end{align}
which indicates that $U_d(t)$ is mean rate stable, and hence the reliability constraint is satisfied. The same argument also applies to other elements of $\V{U}(t)$, i.e., $U_i^{(l)}(t)$ is also mean rate stable, which implies the causality constraint.

Next, we study the relationship between the $\varepsilon$-convergence time, denoted by $t_\varepsilon$, and parameter $V$. Recall the queuing dynamics of $U_d(t)$, i.e., \eqref{eq:virtual_sink}, which implies that
\begin{align}
U_d(t+1) \geq U_d(t) + \gamma A(t) - \nu_{\to d}(t).
\end{align}
Apply the telescoping sum to the above inequality from $t = 0$ to $T-1$ (for any fixed $T > 0$) and take expectation, we obtain
\begin{align} \label{eq:conv_time_bound}
\gamma \lambda - \frac{1}{T} \sum_{t=0}^{T-1} \E{ \nu_{\to d}(t) } & \leq \frac{\E{U_d(T)}}{T} - \frac{\E{U_d(0)}}{T} \nonumber \\
& \leq \sqrt{ \frac{2(B + V\tilde{h}_2^\star)}{T} }
\end{align}
where the second inequality is obtained from \eqref{eq:virtual_q_bound}.

The left-hand-side can be interpreted as the gap between the desired and achieved reliability level, which is bounded by a non-increasing function converging to $0$. Hence, there must exist some time point, after which the gap is always less than $\varepsilon$, which justifies the definition of the $\varepsilon$-convergence time.

It is difficult to derive an exact form for the $\varepsilon$-convergence time under various $V$ parameters; instead, we can draw an upper bound on it (in the following, we fix the value of $\varepsilon$). When parameter $V$ is used
\begin{align}
T(V):\ \sqrt{ \frac{2(B + V\tilde{h}_2^\star)}{T(V)} } = \varepsilon
\end{align}
can serve as an upper bound for the convergence time $t_\varepsilon(V)$. When we use $V' = \alpha V$ (assuming $\alpha > 1$), and evaluate the bound at $T' = \alpha T(V)$, we find that
\begin{align}
\sqrt{ \frac{ 2(B + V' \tilde{h}_2^\star) }{ T' } } = \sqrt{ \frac{ 2( B/\alpha + V \tilde{h}_2^\star ) }{ T(V) } } < \varepsilon
\end{align}
and therefore, $T(\alpha V) \leq \alpha T(V)$ (since the bound is non-increasing). Therefore, the upper bound $T(V)$ grows sub-linearly, and so is the exact convergence time $t_\varepsilon(V)$; in other words, $t_\varepsilon(V) \sim \mathcal{O}(V)$.

\section{Impacts of Estimation Error}
\label{apdx:estimation_error}

In this section, we analyze the impact of estimation error on the attained cost performance. Assume that the empirical averages of the (virtual and actual) network flows have converged to $x$, which represents the cost-optimal flow assignment obtained using the {\em inaccurate} estimate $\hat{\lambda} = \lambda_0 + \Delta \lambda$, with $\lambda_0$ and $\Delta \lambda$ denoting the true rate and estimation error, respectively. In particular, $x$ can be derived by solving the following LP problem (based on \eqref{eq:problem_2}):%
\footnote{
	We consider one client, and assume that the packets arrive to the network at source node $s$ with maximum lifetime.
}
\begin{subequations} \label{eq:SA} \begin{align}
	& \min_{x, \lambda} \ \langle e, x \rangle \\
	& \st \ x_{\to d} \geq  \gamma \lambda \label{eq:des} \\
	& \hspace{22pt} x_{s \to}^{(\geq l)} - x_{\to s}^{(\geq l+1)} \leq \lambda,\ \forall\,l \in \Set{L} \label{eq:src} \\
	& \hspace{22pt} x_{i \to}^{(\geq l)} - x_{\to i}^{(\geq l+1)} \leq 0\ (i \ne s)   \label{eq:intermediate} \\
	& \hspace{22pt} x_{ij} \leq C_{ij},\ x^{(l)}_{dj} = 0,\ x^{(l)}_{ij} \geq 0 \\
	& \hspace{22pt} \lambda \geq \hat{\lambda} \,\Leftrightarrow\,  \lambda_0 - \lambda \leq - \Delta \lambda \label{eq:auxiliary},
\end{align} \end{subequations}
in which we introduce an auxiliary variable $\lambda$ to represent the estimated rate, and separate \eqref{eq:p2_causality} into \eqref{eq:src} and \eqref{eq:intermediate} for the source node $s$ and intermediate nodes, to clearly indicate the constraints involving $\lambda$ (for illustrative purposes). The optimal (cost) value, denoted by $h^\star(\hat{\lambda})$, is affected by the estimation error via \eqref{eq:auxiliary}, as presented in the following sensitivity analysis.%
\footnote{
	We introduce \eqref{eq:auxiliary} to summarize the effect of the approximation error, which does not change the optimal solution if we replace $\lambda$ by $\hat{\lambda}$ in \eqref{eq:des} and \eqref{eq:src}. To wit, if \eqref{eq:auxiliary} holds with inequality $\hat{\lambda} > \lambda_0$, more network resources will be consumed to handle additional packets.
}

We define an {\em unperturbed} LP problem by setting $\Delta\lambda = 0$ in \eqref{eq:auxiliary}, and denote by $h^\star(\lambda_0)$ its optimal value, i.e., the optimal cost. Consider the optimal solution to its dual problem, and let $y \geq 0$ be the multiplier associated with \eqref{eq:auxiliary}, which depends on network (topology, link capacity) and service (deadline constraint, reliability level) parameters, as well as the true arrival rate $\lambda_0$, in the studied problem.

By the general inequality \cite[(5.57)]{boyd2004convex}, we obtain: for any estimation error $\Delta\lambda$,
\begin{align}
h^\star(\hat{\lambda}) \geq h^\star(\lambda_0) + y \Delta \lambda,
\end{align}
leading to the following qualitative conclusions:
\begin{itemize}
	\item If the multiplier $y\geq 0$ is large and we overestimate the rate (i.e., $\Delta \lambda > 0$), the attained cost is significantly higher than the optimal cost.
	\item If the multiplier $y\geq 0$ is small and we underestimate the rate (i.e., $\Delta \lambda < 0$), the attained cost can be slightly less than the optimal cost.
\end{itemize}
We note that the above analysis does not cover all the cases (e.g., if $y$ is small and we overestimate the rate), under which the impact is indefinite and can vary case by case.

\section{Hybrid Queuing System}
\label{apdx:hybrid_queues}

Consider a scenario including two groups of users, collected in $\Phi$ and $\Psi$, respectively. The users $\phi \in \Phi$ are deadline-constrained, which are treated as presented in the paper: for each user $\phi$, we establish lifetime queues $Q_i^{(\phi, l)}(t)$ at each node $i$, and denote by $\mu_{ij}^{(\phi, l)}(t)$ the actual flows transmitted from node $i$ to $j$. The other users $\psi \in \Psi$ are unconstrained (i.e., without deadline constraints), and for each user $\psi$, we create one queue $\tilde{Q}_{i}^{(\psi)}(t)$ at each node $i$ to accommodate all the packets (regardless of the lifetimes), and the assigned flow transmitted from node $i$ to $j$ is denoted by $\tilde{\mu}_{ij}^{(\psi)}(t)$. In general, the queuing dynamics of unconstrained users are given by \cite{Nee:B10}:
\begin{align} \begin{split}
\tilde{Q}_{i}^{(\psi)}(t+1)
& \leq \max\Big[ \tilde{Q}_{i}^{(\psi)}(t) - \sum_{j \in \delta_i^+} \tilde{\mu}_{ij}^{(\psi)}(t), 0 \Big] \\
& \quad + \sum_{j \in \delta_i^-} \tilde{\mu}_{ji}^{(\psi)}(t) + \tilde{a}_{i}^{(\psi)}(t)
\end{split} \end{align}
where $\tilde{a}_{i}^{(\psi)}(t)$ is the number of arrivals of user $\psi$ at node $i$,
and the overall drift for the queues of unconstrained users, i.e., $\tilde{\V{Q}}(t) = \{ \tilde{Q}_{i}^{(\psi)}(t) : i\in \Set{V}, \psi \in \Psi\}$, can be derived as:%
\footnote{
	We illustrate the approach using one-slot drift as an example, which can be extended to $n$ look-ahead slots using multi-slot-drift \cite[Lemma 4.11]{Nee:B10}.
}
\begin{align} \begin{split}
\Delta(\tilde{\V{Q}}(t))
& \leq B' + \sum_{i\in \Set{V}} \sum_{\psi\in \Psi} \tilde{Q}_{i}^{(\psi)}(t) \tilde{a}_{i}^{(\psi)}(t) \\
& \quad - \sum_{(i, j)\in \Set{E}} \sum_{\psi\in \Psi} \big[ \tilde{Q}_{i}^{(\psi)}(t) - \tilde{Q}_{j}^{(\psi)}(t) \big] \tilde{\mu}_{ij}^{(\psi)}(t)
\end{split} \end{align}
where $B'$ is a constant.

Our goal is to stabilize the entire queuing system, including request queues $\V{R}(t)$ for deadline-constrained users, and $\tilde{\V{Q}}(t)$ for unconstrained ones. We propose to minimize the {\em sum drift} $\Delta(\V{R}(t)) + \Delta(\tilde{\V{Q}}(t))$, leading to the following problem
\begin{subequations} \begin{align}
& \hspace{-2pt} \min_{\mu,\,\tilde{\mu}}\ 
\Delta(\V{R}(t)) - \sum_{(i, j)} \sum_{\psi\in \Psi} \big[ \tilde{Q}_{i}^{(\psi)}(t) - \tilde{Q}_{j}^{(\psi)}(t) \big] \tilde{\mu}_{ij}^{(\psi)}(t) \\ 
& \hspace{-2pt} \st \ \sum_{\phi\in \Phi} \sum_{l\in \Set{L}_{\phi}} \mu_{ij}^{(\phi, l)}(t) + \sum_{\psi\in \Psi} \tilde{\mu}_{ij}^{(\psi)}(t) \leq C_{ij} \label{eq:share_capacity} \\
& \hspace{20pt} \sum_{j \in \delta_i^+} \mu_{ij}^{(\phi, l)}(t) \leq Q_{i}^{(\phi, l)}(t);\ \mu(t),\, \tilde{\mu}(t) \succeq 0
\end{align} \end{subequations}
where \eqref{eq:share_capacity} shows the interaction between the two groups of users in sharing the transmission resource, and $\Delta(\V{R}(t))$ is given by \eqref{eq:multi_slot_drift}. In addition, we find that the optimal decisions for unconstrained users, $\tilde{\mu}(t)$, follow the max-weight rule, and it suffices to focus on the interaction of the selected user (with maximum weight) with the deadline-constrained users.

To sum up, we can solve the problem in two steps. First, find the unconstrained user with the max-weight for $\forall\, (i, j)$:
\begin{align}
\psi_{ij}^\star = \argmax_\psi  \big[ \tilde{Q}_{i}^{(\psi)}(t) - \tilde{Q}_{j}^{(\psi)}(t) \big].
\end{align}
Then, solve the following problem:
\begin{subequations} \label{eq:hybrid} \begin{align}
& \hspace{-2pt} \min_{\mu,\,\tilde{\mu}}\ \Delta(\V{R}(t)) - \sum_{(i, j)} \big[ \tilde{Q}_{i}^{(\psi_{ij}^\star)}(t) - \tilde{Q}_{j}^{(\psi_{ij}^\star)}(t) \big] \tilde{\mu}_{ij}^{(\psi_{ij}^\star)}(t) \\ 
& \hspace{-2pt} \st \ \sum_{\phi\in \Phi} \sum_{l\in \Set{L}_{\phi}} \mu_{ij}^{(\phi, l)}(t) + \tilde{\mu}_{ij}^{(\psi_{ij}^\star )}(t) \leq C_{ij} \\
& \hspace{20pt} \sum_{j \in \delta_i^+} \mu_{ij}^{(\phi, l)}(t) \leq Q_{i}^{(\phi, l)}(t);\ \mu(t),\, \tilde{\mu}(t) \succeq 0
\end{align} \end{subequations}
and other unconstrained users are not served, i.e., $\tilde{\mu}_{ij}^{(\psi)}(t) = 0$ if $\psi \ne \psi_{ij}^\star$.

Compared to the original problem, \eqref{eq:hybrid} includes one additional variable for each link and look-ahead slot (and no new constraints), which represents the entire group of unconstrained users.

\section{The Multi-Commodity \ac{agi} Problem}
\label{apdx:agi}

\subsection{\ac{agi} Service Model}

The cloud network offers a set of \ac{agi} services $\Phi$. Each service $\phi \in \Phi$ is modeled by an ordered chain of $(M_{\phi} - 1)$ service functions, through which incoming packets must be processed to be transformed into results that are consumable by corresponding destination nodes. Service functions can be executed at different network locations. While, for ease of exposition, we assume every cloud node can host any service function, it is straightforward to extend our model to limit the set of functions available at each cloud node. 

There are two parameters associated with each function: for the $m$-th function of service $\phi$, we define
\begin{itemize}
	\item $\xi_\phi^{(m)}$: the scaling factor, i.e., the output data-stream size per unit of input data-stream;
	\item $r_\phi^{(m)}$: the workload, i.e., the required computational resource to process one unit of input data-stream.
\end{itemize}
In addition, the input and output data-streams of the $m$-th function are referred to as the stage $m$ and stage $m+1$ data-streams of the service, respectively.

\subsection{Constructing the Layered Graph}

Denote the topology of the actual network by $\Set{G} = (\Set{V}, \Set{E})$. For a particular service $\phi$ (consisting of $M_{\phi} - 1$ functions), the layered graph $\Set{G}^{(\phi)}$ is constructed by the following steps:
\begin{itemize}
	\item[1)] make $M_{\phi}$ copies of the actual network, indexed as layer $1,\cdots, M_{\phi}$ from top to bottom; specifically, node $i\in \Set{V}$ on the $m$-th layer is denoted by $i_m$;
	\item[2)] add {\em directed} links connecting corresponding nodes between adjacent layers, i.e., $(i_m, i_{m+1})$ for $\forall\, i\in \Set{V}$.
\end{itemize}
To sum up, the layered graph $\Set{G}^{(\phi)} = (\Set{V}^{(\phi)}, \Set{E}^{(\phi)})$ with $\Set{E}^{(\phi)} = \{ \Set{E}^{(\phi)}_{\text{pr}, i}: i\in \Set{V} \} \cup \{ \Set{E}^{(\phi)}_{\text{tr}, (i, j)}: (i, j)\in \Set{E} \}$ is defined as
\begin{subequations}
\begin{align}
\Set{V}^{(\phi)} & = \{ i_m : i\in \Set{V}, 1\leq m\leq M_{\phi} \} \\
\Set{E}^{(\phi)}_{\text{pr}, i} & = \{ (i_m, i_{m+1}) : , 1\leq m\leq M_{\phi}-1 \} \\
\Set{E}^{(\phi)}_{\text{tr}, ij} & = \{ (i_m, j_m) : (i, j)\in \Set{E}, 1\leq m\leq M_{\phi} \}.
\end{align}
\end{subequations}

\begin{figure}[t]
	\centering
	\subfloat[Physical network.]{
		\includegraphics[width = 0.4\columnwidth]{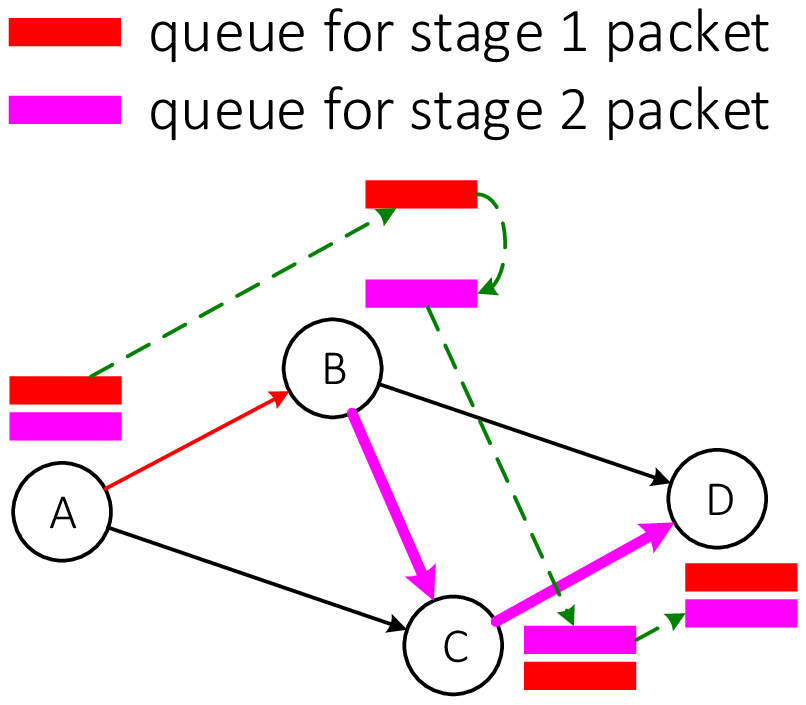}
		\label{fig:physical_network}
	}
	\hspace{10pt}
	\subfloat[Layered graph.]{
		\includegraphics[width = 0.4\columnwidth]{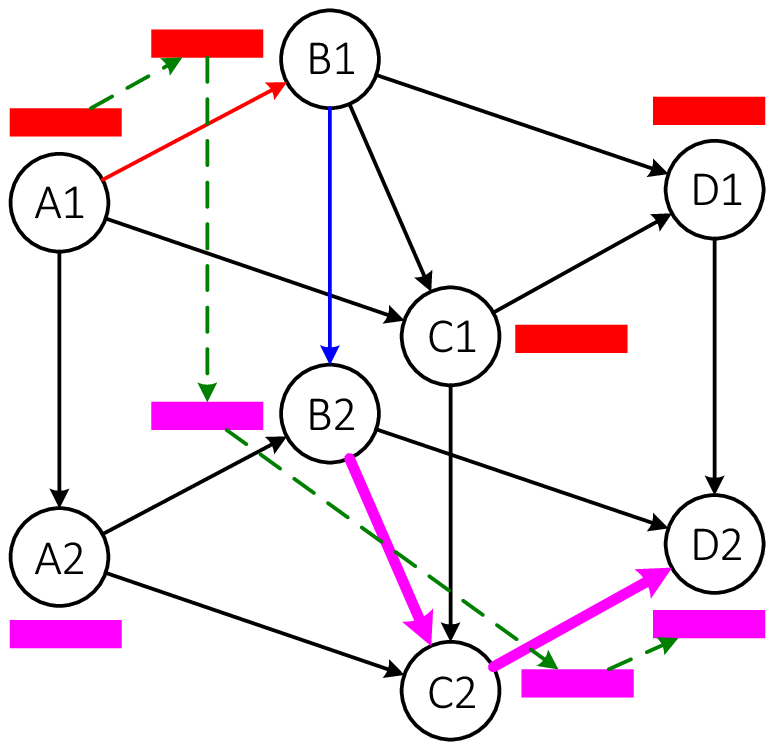}
	}
	\caption{
		An example of delivering a packet, requiring one processing step, over a $4$-node network and its associated layered graph. The stage $1$ packet arrives at the source node $A$ and it is transmitted to node $B$ (along the red path), where it also gets processed. The produced stage $2$ packet is then transmitted to the destination node $D$ (along the pink path). We use red and pink bars to represent queues for packets at different stages, and depict the packet trajectory over the queuing system using green arrows.
	}
	\label{fig:layered_graph}
\end{figure}

Each layer $m$ in $\Set{G}^{(\phi)}$ only deals with packets of specific stage $m$. The edges in $\Set{E}^{(\phi)}_{\text{pr}, i}$ and $\Set{E}^{(\phi)}_{\text{tr}, (i, j)}$ indicate the processing and transmission operations in the actual network, respectively. More concretely, the flow on $(i_m, i_{m+1})$ denotes the processing of stage $m$-packets by function $m$ at node $i$, while $(i_m, j_m)$ denotes the transmission of stage-$m$ packets through the link $(i, j)$. In addition, $d^{(\phi)}_{M_{\phi}}$ is the only destination node (since only stage $M_{\phi}$-packets can be consumed), where $d^{(\phi)}$ is the destination in the actual network.

We define two parameters $(\zeta_{\imath\jmath}^{(\phi)}, \rho_{\imath\jmath}^{(\phi)})$ for each link $(\imath, \jmath)$ in the layered graph
\begin{align}
(\zeta_{\imath\jmath}^{(\phi)}, \rho^{(\phi)}_{\imath\jmath}) = \begin{cases}
(\xi_{\phi}^{(m)}, r_{\phi}^{(m)}) & (\imath, \jmath) = (i_m, i_{m+1}) \\
(1, 1) & (\imath, \jmath) = (i_m, j_m)
\end{cases}
\end{align}
The two parameters can be interpreted as the generalized scaling factor and workload; specifically, for transmission edges (the second case), $\zeta_{\imath\jmath}^{(\phi)} = 1$ since flow is neither expanded or compressed by the transmission operation, and $\rho_{\imath\jmath}^{(\phi)} = 1$ since the flow and the transmission capacity are quantified on the same basis.

\subsubsection{Interpretation}

An example is presented in Fig. \ref{fig:layered_graph}.
In the physical network, two queues are created at each node for packets of different stages. A transmitted packet moves between queues of the same stage (but) at different locations, while a processed packet moves  between queues of different stages at the same node. In the associated layered graph, we create two layers to deal with stage 1 and 2 packets, respectively, and only one queue is created at each node, which accommodates packets of the corresponding stage. Packets traversing nodes in layer $m$ are stage $m$ packets transmitted over the corresponding links in the physical network, and packets crossing from layer $m$ to $m+1$ are stage $m$ packets processed at the corresponding location to create stage $m+1$ packets. For example, $(A_1, B_1)$ indicates that the stage $1$ packet is transmitted over link $(A, B)$, and $(B_1, B_2)$ indicates that the stage $1$ packet is processed at node $B$ into a stage $2$ packet.

\subsection{Relevant Quantities}

\begin{figure*}[ht]
\begin{align}\begin{split}\label{eq:ldp_bd_agi}
\Delta(t) + V h(t)
& \leq B + V \sum_{\phi, (\imath, \jmath), l} e_{\imath \jmath} \rho_{\imath \jmath}^{(\phi)} x_{\imath \jmath}^{(\phi, l)}(t) - \langle \tilde{\V{a}}, \V{\Sigma} \V{U}(t) \rangle \\
& \quad - \sum_{\phi} \beta_d^{(\phi)} U_d^{(\phi)}(t) \sum_{\imath \in \delta_d^-} \zeta_{\imath d}^{(\phi)} x_{\imath d}^{(\phi)}(t)
- \sum_{\phi, \imath, l} \beta_\imath^{(\phi)} U_\imath^{(\phi, l)}(t) \Big[ \sum_{\jmath\in \delta_\imath^-} \zeta_{\jmath \imath}^{(\phi)} x_{\jmath \imath}^{(\phi, \geq l+1)}(t) - x_{\imath\to}^{(\phi, \geq l)}(t) \Big] \\
& = B - \langle \tilde{\V{a}}, \V{\Sigma} \V{U}(t) \rangle \\
& \quad - \sum_{\phi, (\imath, \jmath), l} \underbrace{ \Bigg[ - V e_{\imath \jmath} - \frac{ \beta_\imath^{(\phi)} U_\imath^{(\phi, \leq l)}(t) }{\rho_{\imath \jmath}^{(\phi)}} + \frac{\zeta_{\imath \jmath}^{(\phi)} \beta_\jmath^{(\phi)}}{\rho_{\imath \jmath}^{(\phi)}}
\begin{cases}
U_d^{(\phi)}(t) & \jmath = d_{M_{\phi}}^{(\phi)} \\
U_\jmath^{(\phi, \leq l-1)}(t) & \jmath\in \Set{V^{(\phi)}}\setminus \big\{ d_{M_{\phi}}^{(\phi)} \big\}
\end{cases}
\Bigg] }_{w_{\imath \jmath}^{(\phi, l)}(t)} \underbrace{ \big[\rho_{\imath \jmath}^{(\phi)} x_{\imath \jmath}^{(\phi, l)}(t) \big] }_{ \tilde{x}_{\imath \jmath}^{(\phi, l)}(t) }
\end{split} 
\end{align}
\rule{\textwidth}{0.5pt}
\end{figure*}

The flow variable $x_{\imath\jmath}^{(\phi, l)}(t)$ is defined for link $(\imath, \jmath)$ in the layered graph, which is the amount of packets sent to the corresponding interface. By this definition, for $\forall \phi \in \Phi$ and $\imath \in \Set{G}^{(\phi)}$, the queuing dynamics are modified as
\begin{align}\begin{split}
Q_\imath^{(\phi, l)}(t+1) & = Q_\imath^{(\phi, l+1)}(t) \\
& \hspace{-.5in} - x_{\imath\to}^{(\phi, l+1)}(t) + \sum_{\jmath\in \delta_\imath^-} \zeta_{\jmath\imath}^{(\phi)} x_{\jmath\imath}^{(\phi, l+1)}(t) + a_\imath^{(\phi, l)}(t),
\end{split}\end{align}
where the proposed framework takes the arrival of intermediate stage-packets into account. The causality constraint can be derived as
\begin{align}\label{eq:agi_causality}
\avg{ x_{\imath\to}^{(\phi, \geq l)}(t) } \leq \sum_{\jmath\in \delta_\imath^-} \avg{ \zeta_{\jmath\imath}^{(\phi)} x_{\jmath\imath}^{(\phi,\geq l+1)}(t) } + \lambda_\imath^{(\phi,\geq l)}.
\end{align}

The capacity constraint is given by
\begin{subequations}\label{eq:agi_resource}
\begin{align}
\sum_{\phi\in \Phi} \sum_{(\imath, \jmath)\in \Set{E}_{\text{pr}, i}^{(\phi)}} \sum_{l\in \Set{L}} \rho_{\imath \jmath} x_{\imath \jmath}^{(\phi, l)}(t) & \leq C_{i} \\ 
\sum_{\phi\in \Phi} \sum_{(\imath, \jmath)\in \Set{E}_{\text{tr}, ij}^{(\phi)}} \sum_{l\in \Set{L}} \rho_{\imath \jmath} x_{\imath \jmath}^{(\phi, l)}(t) & \leq C_{ij},
\end{align}
\end{subequations}
and the corresponding operational cost is
\begin{align}
h(t) = \sum_{\phi\in \Phi} \sum_{(\imath, \jmath)\in \Set{E}^{(\phi)}} e_{\imath \jmath} \sum_{l\in \Set{L}} \rho_{\imath \jmath} x_{\imath \jmath}^{(\phi, l)}(t)
\end{align}
where $e_{i_m i_{m+1}} = e_i$ and $e_{i_m j_m} = e_{ij}$, with $C_i$ and $e_i$ denoting the computation capacity and the corresponding cost at each network location $i$, respectively.

Finally, the reliability constraint is given by
\begin{align}\label{eq:agi_reliability}
\frac{1}{\Xi_\phi^{(M_{\phi})}} \sum_{\imath \in \delta_d^-} \avg{\E{ \zeta_{\imath d}^{(\phi)} x^{(\phi)}_{\imath  d}(t) }} \geq \gamma^{(\phi)} \|\V{\lambda}^{(\phi)}\|_1
\end{align}
where the overall scaling factor (for stage $m$ packet of service $\phi$) is defined as
\begin{align}
	\Xi_{\phi}^{(m)} = \prod_{s=1}^{m-1} \xi_{\phi}^{(m)},\text{ and }
	\Xi_{\phi}^{(1)} = 1
\end{align}
and we abuse $d = d_{M_\phi}^{(\phi)}$ for the simplicity of notation. Out of consideration for fairness, in this paper, we calculate the throughput on the basis of input flow size, i.e., the throughput can be interpreted as the rate of {\em served requests}, which is defined as the left-hand-side of \eqref{eq:agi_reliability}.

\subsection{Modifications of Algorithm}

The major difference lies in deriving the solution to the virtual network $\mathscr{P}_2$. The modified constraints \eqref{eq:agi_reliability} and \eqref{eq:agi_causality} lead to the following definition of the virtual queues
\begin{align}
U_d^{(\phi)}(t+1) & = \max\Big\{ 0,\, U_d^{(\phi)}(t) + \Xi_{\phi}^{(M_{\phi})} \gamma^{(\phi)} A^{(\phi)}(t)  \nonumber \\
& \hspace{1in} - \sum_{\imath \in \delta_d^-} \zeta^{(\phi)}_{\imath d} x^{(\phi)}_{\imath d}(t) \Big\}, \\
U_\imath^{(\phi, l)}(t+1) & = \max\Big\{ 0,\, U_\imath^{(\phi, l)}(t) - a_\imath^{(\phi, \geq l)}(t) \nonumber \\
& + x_{\imath\to}^{(\phi, \geq l)}(t) - \sum_{\jmath \in \delta_\imath^-} \zeta^{(\phi)}_{\jmath\imath} x_{\jmath\imath}^{(\phi, \geq l+1)}(t) \Big\}.
\end{align}

The Lyapunov function is defined as
\begin{align}
	L(\V{U}(t)) = \frac{1}{2} \| \V{\Sigma} \V{U}(t) \|_2^2
\end{align}
where $\V{\Sigma} = \diag{ \beta_{d}^{(\phi)}, \beta_{i_m}^{(\phi, l)} }$ is a diagonal matrix with
\begin{align}
	\beta_{d}^{(\phi)} = \frac{1}{ \Xi_{\phi}^{(M_{\phi})} },\ 
	\beta_{i_m}^{(\phi, l)} = \beta_{i_m}^{(\phi)} = \frac{1}{ \Xi_{\phi}^{(m)} }.
\end{align}
The reason to define the coefficients $\V{\beta}$ as above is the following. For any service $\phi$, the virtual queue $U_{i_m}^{(\phi, l)}(t)$ deals with packets of stage $m$ (and thus the virtual queues are of different scale); we define the coefficients to normalize the virtual queues (to the basis of the input flow size).

The \ac{ldp} in this case is given by \eqref{eq:ldp_bd_agi} (where $\tilde{\V{a}}$ is defined in \eqref{eq:virtual_ub}). As a result, the $\min$ \ac{ldp} problem is equivalent to
\begin{subequations}
\begin{align}
& \max\ \sum_{\phi, (\imath, \jmath), l} w_{\imath \jmath}^{(\phi, l)}(t) \tilde{x}_{\imath \jmath}^{(\phi, l)}(t) \\
& \st\ \  \eqref{eq:agi_resource},\ \tilde{\V{x}}(t) \succeq 0.
\end{align}
\end{subequations}
The solution to the problem is in the max-weight fashion. For each transmission link $(i, j)$, we serve the packets of optimal commodity $\phi$, $(\imath, \jmath) \in \Set{E}_{\text{tr}, ij}^{(\phi)}$ and $l$ (with the maximum, positive weight) with all the available resource $C_{ij}$. The processing decisions are made in the same way (note that the solution to the problem $\tilde{\V{x}}$ represents the allocated resource, and the scheduled computation flow equals to the result divided by the corresponding workload parameter).

\end{document}